\documentclass[a4paper,11pt]{article}

\usepackage{amsmath,amssymb,bm,epsf,epsfig,graphicx,dsfont, float,multirow,tensor,url,placeins, subcaption}

\input epsf.sty
\numberwithin{equation}{section}
\numberwithin{figure}{section}
\numberwithin{table}{section}

\def \del {\partial}

\def \rar {\rightarrow}

\def \la {\langle}
\def \ra {\rangle}
\def \rra {\rangle\!\rangle}

\def \ww {|\hspace{-1pt}|}
\def \ps {\phantom{+}}
\def \zb {\bar z}
\def \nn {\nonumber}
\def \Id {\mathds{1}}
\def \inf {\infty}
\def \mr {\multirow}
\def \mc {\multicolumn}

\def \pik {\frac{\pi}{k}}

\def \vv {{\mathcal V}}

\def \mm {{\mathcal M}}
\def \hh {{\mathcal H}}

\def \ee {{\mathcal E}}
\def \CC {{\mathbb C}}
\def \ZZ {{\mathbb Z}}

\def \SU {{SU(2)\ }}
\def \SUk {{SU(2)$_k$\ }}
\def \SL {{SL(2,$\mathbb C$)\ }}
\def \SLk {{SL(2,$\mathbb C$)$_k$\ }}

\def \dexp#1{\times 10^{#1}}

\newcommand{\Cbulk}[3] {C\indices{_#1_#2^#3}}
\newcommand{\Cbound}[6] {C\indices*{^{(#1#2#3)#6}_{#4#5}}}
\newcommand{\Cbb}[3] {\tensor[^{(#1)}]{B}{_#2^#3}}
\newcommand{\qsin}[1] {\lfloor #1 \rfloor}
\newcommand{\qfac}[1] {\lfloor #1 \rfloor!}

\newcommand{\Fmatrix}[6] {
F_{#1#2}
\begin{bmatrix}
 #4 & #5 \\
 #3 & #6 \\
\end{bmatrix}}

\newcommand{\BFmatrix}[6] {
B_{#1#2}
\begin{bmatrix}
 #4 & #5 \\
 #3 & #6 \\
\end{bmatrix}}

\newcommand{\dt}[1] {\frac{#1}{2}}

\begin{document}
\vskip 2.1cm

\centerline{\Large \bf Boundary states in the SU(2)$_k$ WZW model}
\vspace*{3pt}
\centerline{\Large \bf from open string field theory}
\vspace*{8.0ex}

\centerline{\large \rm Mat\v{e}j Kudrna\footnote{Email: {\tt matej.kudrna at email.cz}}}

\vspace*{4.0ex}
\begin{center}
{\it {Institute of Physics of the Czech Academy of Sciences,} \\
{Na Slovance 2, 182 21 Prague 8, Czech Republic}}
\vskip .4cm

\end{center}
\vspace*{6.0ex}

\centerline{\bf Abstract}
\bigskip
We analyze boundary states in the \SUk WZW model using open string field theory in the level truncation approximation. We develop algorithms that allow effective calculation of action in this model and we search for classical solutions of the equations of motion, which are conjectured to describe boundary states. We find three types of solutions. First, there are real solutions that represent maximally symmetric Cardy boundary states and we show that they satisfy certain selection rules regarding their parameters. Next, we find complex solutions that go beyond the \SU model and describe maximally symmetric \SL boundary conditions. Finally, we find exotic solutions that correspond to symmetry-breaking boundary states. Most of real exotic solutions describe the so-called B-brane boundary states, but some may represent yet unknown boundary states.

 \vfill \eject

\baselineskip=16pt

\tableofcontents

\setcounter{footnote}{0}

\section{Introduction}\label{sec:intro}
Wess-Zumino-Witten (WZW) models are one of the most important groups of conformal field theories (CFT). Their defining characteristic is that they do not have just the Virasoro symmetry, but also an affine Lie algebra symmetry, which is based on some Lie group $G$. They are rational with respect to this symmetry, which makes them one of the few types of solvable models. Therefore it is possible to construct certain classes of modular invariant bulk theories and so-called maximally symmetric boundary states, which preserve one half of the bulk symmetry.
However, a boundary conformal field theory is only required to preserve the Virasoro symmetry, which means that there is a more general class of boundary states, which do not have the Lie group symmetry. These are called symmetry-breaking boundary states. Symmetry-breaking boundary theories are generally non-rational and therefore it is difficult to construct them. The known examples are based on techniques like orbifold constructions or conformal embeddings and they do not represent fully generic symmetry-breaking boundary states because they preserve some weaker symmetry or have other special properties.

In the context of string theory, conformal boundary states correspond to D-branes. Therefore classification of boundary states in a given model means understanding of D-branes on the corresponding string background. In WZW models, that means D-branes on the group manifold of the Lie group $G$. The aim of this paper is to investigate boundary states/D-branes in the simplest WZW model, which has the \SU symmetry, using Witten's bosonic open string field theory (OSFT) \cite{WittenSFT}. We will search for both for maximally symmetric boundary states, which can serve as a test of validity of our approach, and for new symmetry-breaking boundary states.

Open string field theory was introduced as a non-perturbative formulation of string theory. As a theory of open strings, it requires a choice of D-brane background, which means that it is formulated using a boundary conformal field theory with some fixed boundary conditions. However, there is a conjecture known as background independence of OSFT \cite{SenBackground1}\cite{SenBackground4}\cite{ErlerMaccaferri3}, which states that the initial background is not important because OSFTs on different backgrounds are related by a field redefinition. These transitions are realized via classical solutions of OSFT, which are conjectured to be in one to one correspondence with conformal boundary states. In other words, by finding classical OSFT solutions, we can move to new backgrounds and find their properties.

Therefore it should theoretically be possible to classify all conformal boundary conditions in a given CFT by analyzing all classical OSFT solutions on a single background. In practice, it is of course not that easy because solving OSFT equations is a difficult task. There are two approaches to OSFT and both have their limitations. One possibility is to find solutions analytically using the $KBc$ algebra or its extensions \cite{AnalyticSolutionSchnabl}\cite{AnalyticSolutionOkawa}\cite{ErlerKBC1}\cite{ErlerKBC2}\cite{FuchsKroyter}\cite{SimpleAnalyticSol}\cite{OkawaLectures}\cite{Multibrane1}\cite{MarginalKOZ}. A remarkable success of this approach are solutions for any background \cite{ErlerMaccaferri}\cite{ErlerMaccaferri2}\cite{ErlerMaccaferri3}, which can be used to prove background independence of OSFT, but their disadvantage is that they are based on boundary condition changing operators. Therefore they require full understanding of the concerned boundary states and they will not help us to find new symmetry-breaking boundary states. The second possibility is to look for solutions numerically, using the level truncation approach \cite{SenZwiebachTV}\cite{RastelliZwiebach}\cite{OkawaLectures}\cite{MSZ lump}\cite{KMS}\cite{GaiottoRastelli}\cite{KudrnaUniversal}\cite{KudrnaThesis}. This approach allows us to find only a limited number of solutions and the results are not exact, but it is possible to reach yet unknown boundary theories. Therefore we choose this approach to analyze the \SUk WZW model.

The \SUk WZW model has been already considered in the context of OSFT in an older work by Michishita \cite{Michishita}, which also looks for classical solutions using the level truncation approach. We take some inspiration from this article, but our paper has a somewhat different focus and we use different computational techniques. We have more efficient numerical algorithms and more computer power, so we can do calculations at much higher level. We also have access to Ellwood invariants \cite{EllwoodInvariants}\cite{KMS}, which allow us to make precise identification of most solutions.

When we compared our numerical results with \cite{Michishita}, we found that they are not the same and we reached a conclusion that the numerical data in \cite{Michishita} are incorrect. The most likely reason is that \cite{Michishita} uses a wrong input for boundary structure constants, see the comment in section \ref{sec:WZW:structure}. We checked which results are correct using the duality of $k=2,3$ \SUk WZW models to Ising and Potts model respectively. We reproduce the correct structure constants in these two minimal models, which does not happen if one uses the structure constants from \cite{Michishita}. The problem with structure constants does not affect much the conclusions of \cite{Michishita}, which seem to be correct qualitatively, although the exact numerical data are not.

The results of this paper can be divided into two different categories:

First, we can learn more about the \SUk WZW model itself. And since the \SU group is a subgroup of all other Lie groups, the results can be also applied to more complex WZW models. We find evidence about existence of new symmetry-breaking boundary states (although there seems to be less of them than we expected), which can be helpful for finding them analytically in the future, and we confirm the existence of the so-called B-brane boundary states from \cite{WZW B-branes}. Next, we observe transitions between Cardy boundary states, which seem to follow some interesting selection rules regarding their parameters (see (\ref{theta 2})). It would be interesting to see whether such rules also apply to RG flows in the \SUk WZW model in general, or whether they are specific to the OSFT approach. By considering complex solutions, it is also possible to explore boundary states in the \SLk WZW model, both maximally symmetric and symmetry-breaking.

Second, our results are interesting purely from the OSFT perspective. We have worked out algorithms for OSFT calculations with three non-commuting currents, which can be extended to other WZW models in the future. We obtained many classical solutions, which help us understand properties of OSFT solutions in general and which can be compared with solutions in other models, like free boson theories and Virasoro minimal models. And since we identified many of our solutions, we get further evidence regarding background independence of OSFT and Ellwood conjecture in the numerical formulation of OSFT.

This paper is organized as follows: Section \ref{sec:WZW} contains a review the \SUk WZW model with focus on topics that are needed for OSFT calculations. In section \ref{sec:SFT}, we discuss properties of the string field in the context of the \SUk WZW model and then we focus on search for OSFT solutions and on their identification. Sections \ref{sec:regular} to \ref{sec:exotic} are used for presentation of our numerical results. Sections \ref{sec:regular} describes solutions describing \SU Cardy boundary states, section \ref{sec:SL} solutions representing \SL Cardy boundary states and finally section \ref{sec:exotic} focuses on exotic solutions, which describe symmetry-breaking boundary states. In section \ref{sec:discussion}, we summarize our results and offer some possible future directions of this research. In appendix \ref{app:complex}, we discuss complex conjugation of \SU primary fields and in appendix \ref{app:Ishibashi}, we construct explicit form of Ishibashi states in the \SUk WZW model. In appendix \ref{app:sewing}, we provide formulas for F-matrices in this model and we use sewing relations to derive structure constants. Appendix \ref{app:Numerics} includes description of some of our numerical methods. Finally, in appendix \ref{app:Data}, we provide additional numerical data regarding several solutions.

\section{Review of the SU(2)$_k$ WZW model}\label{sec:WZW}
In this section, we review some properties of the \SUk WZW model. We focus on aspects that are useful for its implementation in the open string field theory. Basic properties of this model are well-known and they can be found in many books and articles (the traditional reference is the book by di Francesco at al. \cite{DiFrancesco}, other useful references include, for example,  \cite{RecknagelSchomerus}\cite{Blumenhagen}\cite{WZWGoddard}\cite{WZWLectures}\cite{SchomerusLectures}\cite{WaltonWZWIntroduction}). Some topics, like boundary states and structure constants, are less known, so we will discuss them in more detail.

\subsection{SU(2) representations}\label{sec:WZW:SU2}
First, let us review some basic properties of representations of the SU(2) group. The group has three Hermitian generators $J^a$, $a=1,2,3$, but, for most purposes, it is more convenient to replace $J^{1,2}$ by the ladder operators $J^\pm=J^1\pm i J^2$. These generators have commutation relations
\begin{eqnarray}
\left[J^{+},J^{-}\right]&=& 2J^3, \nn\\
\left[J^3,J^{\pm}\right]&=& \pm J^{\pm}.
\end{eqnarray}

Irreducible representations of the SU(2) group are labeled by half-integer spin $j$. States that irreducible representations act on are labeled by the spin $j$ and also by eigenvalues of the operator $J^3$, which are denoted by $m$ and which go from $-j$ to $j$. The group generators act on these states as
\begin{eqnarray}\label{SU2 zero modes}
J^3|j,m\ra&=&m|j,m\ra, \nn\\
J^+|j,m\ra&=&\alpha_{j,m}^+|j,m+1\ra,\\
J^-|j,m\ra&=&\alpha_{j,m}^-|j,m-1\ra, \nn
\end{eqnarray}
where $\alpha_{j,m}^\pm=\sqrt{j(j+1)-m(m\pm1)}$. %Note that $\alpha_{j,m}^\pm=\alpha_{j,-m}^\mp$.

By definition, elements $g$ of the SU(2) group are given by matrices
\begin{equation}\label{SU2 group element abcd}
g=\left(\begin{array}{cc}
a & b \\
c & d
\end{array}\right),
\end{equation}
which satisfy $g^\dagger g=\Id$ and ${\rm{Det}}\, g=1$, which implies $c=-b^*$, $d=a^*$ and $ad-bc=1$. However, it is more convenient for us to parameterize group elements by three angles $\theta\in (0,\pi)$, $\psi\in (0,\pi)$ and $\phi\in (0,2\pi)$ as
\begin{equation}\label{SU2 group element}
g=\left(\begin{array}{cc}
\cos\theta+i \sin\theta \cos\psi & i e^{i \phi} \sin\theta \sin\psi \\
i e^{-i\phi} \sin\theta \sin\psi & \cos\theta-i \sin\theta \cos\psi
\end{array}\right).
\end{equation}
This parameterization also shows that the \SU group is isomorphic to a 3-sphere.

The matrix (\ref{SU2 group element}) forms the fundamental \SU representation of spin $\frac{1}{2}$. A generic spin $j$ irreducible representation of a group element $g$ reads
\begin{equation}
g(|j,m\ra)=\sum_n D^j_{mn}(g)|j,n\ra,
\end{equation}
where $D^j_{mn}(g)$ is the Wigner D-matrix \cite{Hamermesh}
\begin{equation}\label{D gen}
D^j_{mn}(g)=\sum_{l=\max(0,n-m)}^{\min(j-m,j+n)}\frac{\left[(j+m)!\,(j-m)!\,(j+n)!\,(j-n)!\right]^{1/2}}
{(j+n-l)!\,l!\,(m-n+l)!\,(j-m-l)!}\,a^{j+n-l}\,b^{l}\,c^{m-n+l}\,d^{j-m-l}.
\end{equation}

Finally, let us note that the three generators $J^a=(J^-,J^3,J^+)$ form a triplet and transform under the adjoint representation
\begin{equation}
g(J^a)=\Omega^a_{\ b}(g)J^b,
\end{equation}
where the matrix $\Omega$ equals
\begin{equation}\label{Omega}
\Omega(g)=A\,D^1(g)\,A^{-1}.
\end{equation}
This relation includes a diagonal matrix $A=\rm{diag}(\sqrt{2},1,-\sqrt{2})$, which appears because the currents have a different normalization from the usual spin 1 states.

\subsection{\SUk WZW model}\label{sec:WZW:WZW}
Now, let us move from the \SU group to the \SUk WZW model. This model is a conformal field theory which has the \SU group as a symmetry. We will skip the sigma model description of this theory because it is not relevant for this paper and focus purely on its CFT formulation.

The chiral symmetry algebra of the \SUk WZW model includes three currents of dimension one $J^a(z)$, $a=1,2,3$, which generalize the \SU group generators. Their OPE is
\begin{equation}
J^a(z) J^b(w)\sim i\sum_c \epsilon_{abc}\frac{J^c(w)}{z-w}+\frac{k\delta_{ab}}{2(z-w)^2}.
\end{equation}
The OPE includes a parameter $k$, which is called level and which has to be a positive integer in unitary models. This parameter appears in the central extension of the current algebra, in a similar way as the central charge extends the Virasoro algebra.
As before, we define a more convenient linear combinations of the first two currents, $J^\pm (z)=J^1(z) \pm i J^2(z)$. The OPE of the three currents implies that their modes satisfy the following commutation relations
\begin{eqnarray}
\left[J_m^{\pm},J_n^{\pm}\right]&=& 0, \\
\left[J_m^3,J_n^3\,\right]&=& \frac{m k}{2}\delta_{m+n,0}, \\
\left[J_m^{\pm},J_n^{\mp}\right]&=& \pm 2J^3_{m+n}+m k\delta_{m+n,0}, \\
\left[J_m^3,J_n^{\pm}\right]&=& \pm J^{\pm}_{m+n}.
\end{eqnarray}
The zero modes of the currents $J_0^a$ form a subalgebra, which is identical to the Lie algebra of \SU.

The stress-energy tensor of the theory is given by the Sugawara construction
\begin{equation}
T(z)=\frac{1}{2(k+2)}\left(2(J^3J^3)(z)+(J^+J^-)(z)+(J^-J^+)(z)\right)=\frac{1}{2(k+2)}\sum_{a,b}K^{-1}_{ab}(J^a J^b)(z),
\end{equation}
where
\begin{equation}
K^{-1}=\left(\begin{array}{ccc}
0 & 0 & 1 \\
0 & 2 & 0 \\
1 & 0 & 0
\end{array}\right)
\end{equation}
is the inverse of the Killing form. The OPE of the stress-energy tensor with itself implies that the central charge of the theory is
\begin{equation}
c^{SU}=\frac{3k}{k+2}.
\end{equation}

Primary operators\footnote{For our purposes, it is convenient to slightly abuse the terminology and use the term primary operator for all operators $\phi_{j,m}$, while it is more usual to consider only $\phi_{j,j}$ as primary and view the remaining operators with the same $j$ as $J_0^-$ descendants.} in this model are labeled according to \SU irreducible representations by two numbers $(j,m)$, where $j$ is restricted by the level $k$ to $j=0,\dt 1,1,\dots,\dt k$ and $m=-j,\dots,j$ as usual. We denote these primary operators as $\phi_{j,m}(z)$ and their Hilbert space representation as $|j,m\ra=\phi_{j,m}(0)|0\ra$. Conformal weights of these primaries depend only on the label $j$:
\begin{equation}
h_j=\frac{j(j+1)}{k+2}.
\end{equation}
Fusion rules of primary operators are different from the usual spin addition in quantum mechanics because the maximal available spin is restricted by the level $k$:
\begin{equation}
j_1\times j_2=|j_1-j_2|+\dots+\min(j_1+j_2,k-j_1-j_2).
\end{equation}
Zero modes of the currents act on primary states following (\ref{SU2 zero modes}) and positive modes annihilate them
\begin{equation}
J^a_n|j,m\ra=0,\quad n>0.
\end{equation}
Therefore the state space of the \SUk WZW model is spanned by the states
\begin{equation}\label{state space SU}
J_{-n_1}^{a_1}\dots J_{-n_l}^{a_l} |j,m\ra, \quad n_1\geq n_2\geq \ldots\geq n_l \geq 1.
\end{equation}
However, this state space is not irreducible because it contains many null states, especially for low $k$. The basic null state in a given highest weight representation is $(J_{-1}^+)^{k+1-2j}|j,j\ra$ and the rest of the null space is formed by its descendants. We have to remove null states during our string field theory calculations, although we search for them in a different way by analyzing the Gram matrix.

The number of states in an irreducible representation can be determined using characters, which appear for example in \cite{RecknagelSchomerus}. Characters of the \SUk WZW model are defined so that they count both $L_0$ and $J_0^3$ eigenvalues:
\begin{equation}\label{character SU}
\chi_j(z,\tau)={\rm{Tr}}_{\hh_j} \left[q^{L_0-c/24} e^{-2\pi i z J_0^3}\right]=\left(\frac{\Theta_{2j+1}^{(k+2)}-\Theta_{-2j-1}^{(k+2)}}{\Theta_{1}^{(2)}-\Theta_{-1}^{(2)}}\right)(z,\tau),
\end{equation}
where $q=e^{2\pi i \tau}$ as usual and where we introduce the theta functions
\begin{equation}
\Theta_l^{(k)}(z,\tau)=\sum_{m\in \mathbb{Z}+\frac{l}{2k}} q^{k m^2}e^{-2\pi i k m z}.
\end{equation}

So far, we have considered only the chiral theory. However, to construct a complete CFT, which includes both bulk theory and boundary theory, we need to know how to combine the left and the right sector.
In case of bulk theory, there in not much to discuss because our calculations mostly involve boundary CFT. Modular invariants of the \SUk WZW models follow the A-D-E classification and we will work only with the A-series of models, which have diagonal partition function
\begin{equation}
Z=\sum_{j=0}^{k/2}|\chi_j|^2.
\end{equation}
That means that the bulk spectrum includes all representations exactly once and primary operators have the same left and right label $j$, so we denote them as $\phi_{j,m,m'}(z,\bar z)$.

Classification of boundary theories is more complicated, so we will discuss these in the following subsection.

\subsection{Boundary theory}\label{sec:WZW:Boundary}
When it comes to boundary theory in \SUk WZW models, let us distinguish two types of boundary conditions:
\begin{itemize}
  \item Maximally symmetric boundary conditions. These boundary conditions preserve one half of the bulk current symmetry, that is one copy of the chiral algebra.
  \item Symmetry-breaking boundary conditions. They do not preserve the \SU current symmetry and, in general, they have only the Virasoro symmetry.
\end{itemize}

Boundary conditions of the first type are well understood and their boundary states are given by the Cardy solution (\ref{Cardy BS}). On the other hand, symmetry-breaking boundary states are not classified and only few examples are known. We will describe some of these in section \ref{sec:WZW:B-branes}.

In this section, we will focus on the maximally symmetric boundary conditions. They are characterized by gluing conditions of the form
\begin{equation}
J^a(z)=\Omega^a_{\ b}(g)\bar J^b(\zb)|_{z=\bar z}.
\end{equation}
Equivalently, the corresponding boundary states satisfy
\begin{equation}\label{Gluing J}
(J^a_n+\Omega^a_{\ b}(g)\bar J^b_{-n})\ww B\rra=0,
\end{equation}
where $g$ is an SU(2) element and $\Omega(g)$ is given by (\ref{Omega}). This matrix obeys
\begin{equation}
\Omega^T.K^{-1}.\Omega=K^{-1},
\end{equation}
which means that these gluing conditions automatically imply the gluing condition for the stress-energy tensor:
\begin{equation}
T(z)=\bar T(\zb)|_{z=\bar z}.
\end{equation}

The gluing conditions (\ref{Gluing J}) are solved by the \SU Ishibashi states $|j,g\rra$, which are built from bulk primaries. The Ishibashi state of spin $j$ is given by
\begin{equation}
|j,g\rra=\sum_{n_1,n_2}(-1)^{2j+|n_1|}G^{-1}_{n_1g(n_2)}|n_1\ra\overline{|n_2\ra},
\end{equation}
where the states $|n_{1,2}\ra$ form a chiral basis of the spin $j$ representation, $|n_1|$ is the eigenvalue of the number operator and $G^{-1}_{n_1g(n_2)}$ is the inverse of the Gram matrix of BPZ products $\la n_1|g(n_2)\ra$. See appendix \ref{app:Ishibashi} for more details and a derivation.

We are mostly interested in the lowest level components of Ishibashi states, which include only primary states. For the trivial gluing conditions $g=\Id$, we find
\begin{equation}
|j,\Id\rra=\sum_{m} (-1)^{j-m} |j,m,-m\ra+\dots
\end{equation}
and for generic gluing conditions
\begin{equation}\label{Ishibashi}
|j,g\rra=\sum_{m,n} (-1)^{j-m} D^j_{-mn}(g) |j,m,n\ra+\dots.
\end{equation}

Full boundary states are given by the Cardy solution. For each $g$, there is a set of boundary states
\begin{equation}\label{Cardy BS}
\ww J,g\rra=\sum_j B_J^{\ j}|j,g\rra=\sum_j\frac{S_J^{\ j}}{\sqrt{S_0^{\ j}}}|j,g\rra,
\end{equation}
where the label $J$ takes the same half-integer values as for the representations, $J=0,\frac{1}{2},\dots,\frac{k}{2}$, and the modular $S$-matrix of the \SUk WZW model reads
\begin{equation}\label{S matrix}
S_{i}^{\ j}=\sqrt{\frac{2}{k+2}}\sin \frac{(2i+1)(2j+1)\pi}{k+2}.
\end{equation}
These boundary states are formally very similar to boundary states in Virasoro minimal models. For a future reference, we provide boundary state components in the first few models in table \ref{tab:boundary states}.

\begin{table}[!]
\centering
\begin{tabular}{|l|lll|}
\mc{4}{c}{$k=2$}                                              \\\hline
$j$            & $0       $ & $\ps 1/2     $ & $\ps 1       $ \\\hline
$\ww 0   \rra$ & $0.707107$ & $\ps 0.840896$ & $\ps 0.707107$ \\
$\ww 1/2 \rra$ & $1.      $ & $\ps 0.      $ & $   -1.      $ \\
$\ww 1   \rra$ & $0.707107$ & $   -0.840896$ & $\ps 0.707107$ \\\hline
\end{tabular}\vspace{5mm}
\begin{tabular}{|l|llll|}
\mc{5}{c}{$k=3$}                                                               \\\hline
$j$            & $0       $ & $\ps 1/2     $ & $\ps 1       $ & $\ps 3/2     $ \\\hline
$\ww 0   \rra$ & $0.609711$ & $\ps 0.775565$ & $\ps 0.775565$ & $\ps 0.609711$ \\
$\ww 1/2 \rra$ & $0.986534$ & $\ps 0.479325$ & $   -0.479325$ & $   -0.986534$ \\
$\ww 1   \rra$ & $0.986534$ & $   -0.479325$ & $   -0.479325$ & $\ps 0.986534$ \\
$\ww 3/2 \rra$ & $0.609711$ & $   -0.775565$ & $\ps 0.775565$ & $   -0.609711$ \\\hline
\end{tabular}\vspace{5mm}
\begin{tabular}{|l|lllll|}
\mc{6}{c}{$k=4$}                                                                                \\\hline
$j$            & $0       $ & $\ps 1/2     $ & $\ps 1       $ & $\ps 3/2     $ & $\ps 2       $ \\\hline
$\ww 0   \rra$ & $0.537285$ & $\ps 0.707107$ & $\ps 0.759836$ & $\ps 0.707107$ & $\ps 0.537285$ \\
$\ww 1/2 \rra$ & $0.930605$ & $\ps 0.707107$ & $\ps 0.      $ & $   -0.707107$ & $   -0.930605$ \\
$\ww 1   \rra$ & $1.074570$ & $\ps 0.      $ & $   -0.759836$ & $\ps 0.      $ & $\ps 1.074570$ \\
$\ww 3/2 \rra$ & $0.930605$ & $   -0.707107$ & $\ps 0.      $ & $\ps 0.707107$ & $   -0.930605$ \\
$\ww 2   \rra$ & $0.537285$ & $   -0.707107$ & $\ps 0.759836$ & $   -0.707107$ & $\ps 0.537285$ \\\hline
\end{tabular}\vspace{5mm}
\begin{tabular}{|l|llllll|}
\mc{7}{c}{$k=5$}                                                                                                 \\\hline
$j$            & $0       $ & $\ps 1/2     $ & $\ps 1       $ & $\ps 3/2     $ & $\ps 2       $ & $\ps 5/2     $ \\\hline
$\ww 0   \rra$ & $0.481581$ & $\ps 0.646457$ & $\ps 0.721887$ & $\ps 0.721887$ & $\ps 0.646457$ & $\ps 0.481581$ \\
$\ww 1/2 \rra$ & $0.867780$ & $\ps 0.806119$ & $\ps 0.321270$ & $   -0.321270$ & $   -0.806119$ & $   -0.867780$ \\
$\ww 1   \rra$ & $1.082104$ & $\ps 0.358757$ & $   -0.578908$ & $   -0.578908$ & $\ps 0.358757$ & $\ps 1.082104$ \\
$\ww 3/2 \rra$ & $1.082104$ & $   -0.358757$ & $   -0.578908$ & $\ps 0.578908$ & $\ps 0.358757$ & $   -1.082104$ \\
$\ww 2   \rra$ & $0.867780$ & $   -0.806119$ & $\ps 0.321270$ & $\ps 0.321270$ & $   -0.806119$ & $\ps 0.867780$ \\
$\ww 3/2 \rra$ & $0.481581$ & $   -0.646457$ & $\ps 0.721887$ & $   -0.721887$ & $\ps 0.646457$ & $   -0.481581$ \\\hline
\end{tabular}\vspace{5mm}
\begin{tabular}{|l|lllllll|}
\mc{8}{c}{$k=6$}                                                                                                                  \\\hline
$j$            & $0       $ & $\ps 1/2     $ & $\ps 1       $ & $\ps 3/2     $ & $\ps 2       $ & $\ps 5/2     $ & $\ps 3       $ \\\hline
$\ww 0   \rra$ & $0.437426$ & $\ps 0.594604$ & $\ps 0.679662$ & $\ps 0.707107$ & $\ps 0.679662$ & $\ps 0.594604$ & $\ps 0.437426$ \\
$\ww 1/2 \rra$ & $0.808258$ & $\ps 0.840896$ & $\ps 0.520190$ & $\ps 0.      $ & $   -0.520190$ & $   -0.840896$ & $   -0.808258$ \\
$\ww 1   \rra$ & $1.056040$ & $\ps 0.594604$ & $   -0.281525$ & $   -0.707107$ & $   -0.281525$ & $\ps 0.594604$ & $\ps 1.056040$ \\
$\ww 3/2 \rra$ & $1.143050$ & $\ps 0.      $ & $   -0.735660$ & $\ps 0.      $ & $\ps 0.735660$ & $\ps 0.      $ & $   -1.143050$ \\
$\ww 2   \rra$ & $1.056040$ & $   -0.594604$ & $   -0.281525$ & $\ps 0.707107$ & $   -0.281525$ & $   -0.594604$ & $\ps 1.056040$ \\
$\ww 3/2 \rra$ & $0.808258$ & $   -0.840896$ & $\ps 0.520190$ & $\ps 0.      $ & $   -0.520190$ & $\ps 0.840896$ & $   -0.808258$ \\
$\ww 3   \rra$ & $0.437426$ & $   -0.594604$ & $\ps 0.679662$ & $   -0.707107$ & $\ps 0.679662$ & $   -0.594604$ & $\ps 0.437426$ \\\hline
\end{tabular}
\caption{Numerical values of components $B_J^{\ j}$ of \SU Cardy boundary states from $k=2$ to $k=6$.} \label{tab:boundary states}
\end{table}

The label $J$ goes up to $\frac{k}{2}$, but, in practice, we can consider only boundary states with $J\leq \frac{k}{4}$ because there is a relation between boundary states with labels $J$ and $\frac{k}{2}-J$. Boundary state coefficients from (\ref{Cardy BS}) satisfy
\begin{equation}
B_J^{\ j}=(-1)^{2j}B_{k/2-J}^{\ j},
\end{equation}
which means that these boundary states are related as
\begin{equation}\label{boundary relation}
\ww J,g\rra=\ww k/2-J, (-\Id)\circ g\rra.
\end{equation}

In this paper, we impose the following condition on the string field (see section \ref{sec:SFT:String field}):
\begin{equation}\label{J03 psi 2}
J_0^3|\Psi\ra=0,
\end{equation}
Using Ellwood invariants, it can be shown that this condition puts a restriction on boundary states that can be described by our OSFT solutions. For Cardy boundary states, it implies that they must preserve the gluing condition for the current $J^3$:
\begin{equation}\label{Gluing J3}
(J^3_n+\bar J^3_{-n})\ww B\rra=0.
\end{equation}
Group elements that are compatible with this condition have $\psi=0$ and they can be characterized just by the angle $\theta$, which now lies in the interval $\theta\in(-\pi,\pi)$. This leads to a great simplification of most formulas. Most importantly, irreducible representations become diagonal
\begin{equation}\label{D spec}
D^j(\theta)=\left(\begin{array}{ccccc}
e^{-2j \theta i} & 0 & & & \\
0 & e^{-(2j-1) \theta i} & & & \\
 & & \ddots & & \\
 & & & e^{(2j-1) \theta i} & 0 \\
 & & & 0 & e^{2j \theta i}
\end{array} \right),
\end{equation}
and the matrix $\Omega$ simplifies to
\begin{equation}\label{Omega spec}
\Omega(\theta)=\left(\begin{array}{ccc}
e^{-2 \theta i} & 0 & 0 \\
0 & 1 & 0 \\
0 & 0 & e^{2 \theta i}
\end{array} \right).
\end{equation}

Finally, we have to mention the boundary spectrum. In a boundary conformal field theory, there are two types of boundary operators. Normal boundary operators $\psi_{j,m}(x)$, which have the same boundary conditions on both sides, and boundary condition changing operators, which connect two different boundaries $a=(J_a,g_a)$ and $b=(J_b,g_b)$. We will focus on boundary condition changing operators which connect boundaries characterized by the same group element $g$. In this case, the boundary spectrum includes operators $\psi_{j,m}^{(ab)}$, where the allowed values of $j$ are given by the fusion rule of the boundary labels, $j\in J_a\times J_b$.

We are mainly interested in the spectrum of ordinary boundary operators $\psi_{j,m}(x)$. By specializing to the case $J_a= J_b=J$, we find out that their spin $j$ is always an integer, $j=0,1,\dots,\min(2J,k-2J)$, and $m=-j,\dots,j$ as usual.

\subsection{Structure constants}\label{sec:WZW:structure}
In this subsection, we will describe structure of the OPE and correlation functions in the \SUk WZW model. The basic correlators can be written in terms of structure constants, which we need as an input for OSFT calculations. In particular, we need boundary structure constants for the OSFT action and bulk-boundary structure constants for evaluation of Ellwood invariants. For simplicity, we consider only the Cardy boundary states with the basic boundary conditions given by the group element $g=\Id$, which we take as our initial OSFT setting. Some of the formulas presented here come from \cite{RecknagelSchomerus}, but we have not found a detailed discussion of the \SUk WZW model structure constants in the literature.

The OPE and correlators of primary fields must follow the generic structure which is imposed by the conformal symmetry, which determines their coordinate dependance. In addition to that, the \SU symmetry allows a simplification of structure constants. Consider a 3-point correlator of primary fields $\la \phi_{j_1,m_1}(z_1) \phi_{j_2,m_2}(z_2)\phi_{j_3,m_3}(z_3)\ra$. By acting with zero modes of the currents on the primaries (through contour deformations), we get relations between correlators with different $J_0^3$ eigenvalues $m_i$, which allow us to derive an analogue of the Wigner-Eckart theorem from quantum mechanics. Therefore the key part of the structure constants depends only on the spins $j_i$, while the $m$-dependance is captured by Clebsch-Gordan coefficients.

First, let us have a look at structure of the OPE of two chiral primary operators, which we can write as
\begin{equation}
\phi_{j_1,m_1}(z) \phi_{j_2,m_2}(w) \sim \sum_{j_3}\la j_1,m_1,j_2,m_2|j_3,m_3\ra\, C\indices{_{j_1}_{j_2}^{j_3}}\, \phi_{j_3,m_3}(w)\, (z-w)^{h_{j_3}-h_{j_1}-h_{j_2}}+\dots
\end{equation}
As mentioned above, that the $m$-dependence of the OPE is fully captured in $\la j_1,m_1,j_2,m_2|j_3,m_3\ra$, which are the usual \SU Clebsch-Gordan coefficients. Therefore we can define two types of structure constants, 'bare' structure constants $C\indices{_{j_1}_{j_2}^{j_3}}$, which depend only on $j_i$, and 'dressed' structure constants, which include the $m$-dependence:
\begin{equation}
C\indices{_{(j_1,m_1)}_{(j_2,m_2)}^{(j_3,m_3)}}=C\indices{_{j_1}_{j_2}^{j_3}}\la j_1,m_1,j_2,m_2|j_3,m_3\ra.
\end{equation}

Similarly, we define structure constants with lower indices, which appear in 3-point functions
\begin{equation}\label{three point dressed}
C_{(j_1,m_1)(j_2,m_2)(j_3,m_3)}=C\indices{_{j_1}_{j_2}^{j_3}}C\indices{_{j_3}_{j_3}^{0}}\la j_1,m_1,j_2,m_2|j_3,-m_3\ra\la j_3,-m_3,j_3,m_3|0,0\ra.
\end{equation}
These constants can be also expressed in terms of \SU 3-j symbols using the identity
\begin{equation}
\left(  \begin{array}{ccc}
j_1 & j_2 & j_3 \\
m_1 & m_2 & m_3
\end{array}\right)
=(-1)^{j_1-j_2+j_3}\la j_1,m_1,j_2,m_2|j_3,-m_3\ra\la j_3,-m_3,j_3,m_3|0,0\ra.
\end{equation}

Next, let us move to the full theory, where there are three types of OPEs: bulk OPE, boundary OPE and bulk-boundary OPE. Their structure (in this order) is
\begin{eqnarray}\label{OPE bulk}
\phi_{j_1,m_1,n_1}(z,\bar z) \phi_{j_2,m_2,n_2}(w,\bar w) &\sim& \sum_{j_3} C\indices{_{j_1}_{j_2}^{j_3}}\, \phi_{j_3,m_3,n_3}(w,\bar w)\, |z-w|^{2(h_{j_3}-h_{j_1}-h_{j_2})}\\
& &\times \la j_1,m_1,j_2,m_2|j_3,m_3\ra \la j_1,n_1,j_2,n_2|j_3,n_3\ra +\dots,\nonumber
\end{eqnarray}
\vspace{-4mm}
\begin{equation}\label{OPE bound}
\phi_{j_1,m_1}^{(ab)}(z) \phi_{j_2,m_2}^{(bc)}(w) \sim \sum_{j_3} \la j_1,m_1,j_2,m_2|j_3,m_3\ra\, C\indices*{^{(abc)j_3}_{j_1j_2}} \phi_{j_3,m_3}^{(ac)}(w)\, (z-w)^{h_{j_3}-h_{j_1}-h_{j_2}}+\dots,
\end{equation}
\begin{equation}\label{OPE bulk-bound}
\phi_{j_1,m_1,n_1}(z,\bar z) \sim  \sum_{j_2}\la j_1,m_1,j_1,n_1|j_2,m_2\ra
 \tensor[^{(a)}]{B}{_{j_1}^{ j_2}}\phi_{j_2,m_2}^{(aa)}(x) (2y)^{h_{j_2}-2h_{j_1}}+\dots.
\end{equation}
In the last equation, we write $z$ as $x+iy$. We have written the boundary OPE for a generic configuration of boundary condition changing operators, but we will actually need only a simpler case with $a=b=c$ in our calculations.

The CFT is therefore fully determined by the bare structure constants $\Cbulk ijk$, $\Cbound abc ijk$ and $\Cbb ail$. The solution for these structure constants can be obtained by solving sewing relations for the \SU model, see appendix \ref{app:sewing:sewing}. The solution is almost the same as in the Virasoro minimal models \cite{Runkel1}, there are just some additional signs:
\begin{eqnarray}
\Cbulk ijk &=& (-1)^{i+j-k}\left(\Fmatrix k0jiij \right)^{-1}, \label{SC bulk}\\
\Cbound abc ijk &=& \Fmatrix bkiacj, \label{SC bound} \\
\Cbb bij &=&  \frac{S\indices{_0^i}}{S\indices{_0^0}} \sum_m e^{i \pi (2h_m-2h_b-2h_i+\frac{1}{2}h_j)} \Fmatrix 0mbbii \Fmatrix mjbiib. \label{SC bulk-bound}
\end{eqnarray}
The solution is expressed in terms of the fusion matrix of the \SUk WZW model. We provide an explicit formula for the F-matrix in appendix \ref{app:sewing:fusion}. Bulk and boundary structure constants are always real, while bulk-boundary structure constants are either real (for even $j$) or purely imaginary (for odd $j$).

Next, we will discuss the normalization of correlators and the BPZ product. The solution for structure constants is not unique because we have a freedom in normalization of primary operators. It would be ideal to use this freedom to set all two-point functions to 1, but that is problematic because the related Clebsch-Gordan coefficient
\begin{equation}
\la j,m,j,-m|0,0\ra=(-1)^{j-m}\frac{1}{\sqrt{2j-1}}
\end{equation}
has an alternating sign. Therefore, we settle with setting them to $\pm 1$\footnote{This may look strange because the theory is unitary, but it is important to notice that we talk about the BPZ product. The Hermitian product, which is connected to unitarity, is positively definite.}. Two-point bulk structure constants based on (\ref{SC bulk}) read
\begin{equation}
\Cbulk jj0 = \frac{S\indices{_0^j}}{S\indices{_0^0}}.
\end{equation}
Since this expression is always positive, we can choose the normalization of bulk primaries to be
\begin{equation}\label{norm bulk}
N_j^{bulk}=\sqrt{2j-1}\frac{\sqrt{S\indices{_0^0}}}{\sqrt{S\indices{_0^j}}}
\end{equation}
so that the bulk BPZ product is
\begin{equation}\label{BPZ bulk}
\la j,m_1,n_1|j,m_2,n_2\ra=(-1)^{2j-m_1-n_1}\delta_{m_1,-m_2}\delta_{n_1,-n_2}.
\end{equation}

Boundary structure constants are sometimes negative, so we choose the normalization of boundary operators to be
\begin{equation}
N_j^{bound}=(2j-1)^{1/4}\frac{1}{\sqrt{\left|\Cbound aaa jj0\right|}}.
\end{equation}
The boundary BPZ product is then given by
\begin{equation}\label{BPZ bound}
\la j,m|j,n\ra={\rm sgn} (\Cbound aaa jj0)\,(-1)^{j-m}\delta_{m,-n}\, \la \Id \ra^{(a)},
\end{equation}
where $\la \Id \ra^{(a)}$ is the trivial boundary correlator, which is equal to the boundary state $g$-function $B_a^0$.

Finally, let us note that our result for boundary structure constants is different from \cite{Michishita} and we reached a conclusion that the corresponding expression in \cite{Michishita} is incorrect: compare the equation (A.1) from \cite{Michishita} with our expression for the F-matrix (\ref{F matrix}) with special entries given by (\ref{SC bound}) for $a=b=c$. The expression in \cite{Michishita} is missing the factors denoted by $\Lambda$, which means that it is based on an F-matrix in a wrong 'gauge'. We observe that $\Lambda$ approaches 1 as $k\rar\inf$, which explains why the accuracy of results in \cite{Michishita} improves with increasing $k$. We are essentially sure that our formulas are correct because we reproduce the expected boundary states with much better precision and because some of our solutions have duals in $m=2,3$ Virasoro minimal models.

\subsection{Symmetry-breaking boundary states}\label{sec:WZW:B-branes}
Construction of symmetry-breaking boundary states is generally quite difficult. The known examples do not describe completely generic boundary states. Instead, they rely on existence of some other symmetry, which is preserved even though the original symmetry is broken, or on orbifold constructions. A general description of these techniques is given in \cite{SymmetryBreakingBoundaries1}\cite{SymmetryBreakingBoundaries2}\cite{SymmetryBreakingBSD}\cite{BirkeWZWsymmetrybreaking}, more references can be found in \cite{SchomerusLectures}. In case of the \SUk WZW model, we have found two references that discuss symmetry-breaking boundary states  \cite{WZW B-branes}\cite{WZW Symmetry-breaking}. These constructions are based on decomposition of the \SUk WZW model into a parafermion theory and a free boson. So we will describe this decomposition first.

The SU(2) group includes a U(1) subgroup (conventionally generated by the $J^3$ current) and therefore we can decompose our model as ${\rm SU}(2)_k=\frac{{\rm SU}(2)_k}{{\rm U}(1)_k}\times {\rm U}(1)_k$. The U(1)$_k$ model describes a free boson theory on a circle of radius $R=\sqrt{k}$ and the coset $\frac{{\rm SU}(2)_k}{{\rm U}(1)_k}$ describes a parafermion theory. In special cases $k=2$ and $k=3$, the parafermion theory is equivalent to the Ising model and the Potts model respectively.

The free boson theory on a circle with radius of the form $R=\sqrt{k}$ is slightly different from a generic free boson theory because it has an extended symmetry, which is generated by additional generators $e^{\pm i 2 \sqrt{k} X}(z)$ with conformal dimension $k$. Representations of this theory are labeled by an integer $n$, which is defined modulo $2k$. These representations are built on states with momentum
\begin{equation}\label{free boson momentum}
p=\frac{n+2kl}{\sqrt{k}},\quad l\in \ZZ.
\end{equation}

This free boson theory includes two types of Ishibashi states which respect the symmetry: Dirichlet Ishibashi states
\begin{equation}
|r,r\rra_D=\exp\left(\sum_{n=1}^\inf \frac{1}{n}\alpha_{-n}\bar\alpha_{-n}\right)\sum_{l\in \ZZ} \left| \frac{r+2kl}{\sqrt{k}},\frac{r+2kl}{\sqrt{k}}\right\ra,\quad r=0,1,\dots,2k-1
\end{equation}
and Neumann Ishibashi states
\begin{equation}
|r,-r\rra_N=\exp\left(-\sum_{n=1}^\inf \frac{1}{n}\alpha_{-n}\bar\alpha_{-n}\right)\sum_{l\in \ZZ} \left| \frac{r+2kl}{\sqrt{k}},-\frac{r+2kl}{\sqrt{k}}\right\ra,\quad r=0,k.
\end{equation}

Using these Ishibashi states, we can construct Cardy boundary states. Dirichlet boundary states, which are denoted as A-branes using the terminology of \cite{WZW B-branes}, are
\begin{equation}
\ww D,n\rra=\frac{1}{(2k)^{1/4}}\sum_{m=0}^{2k-1} e^{-\frac{i\pi mn}{k}}|m,m\rra_D,\quad n=0,\dots,2k-1.
\end{equation}
These boundary states describe D0-branes that are positioned at one of $2k$ special points on the circle.
Neumann boundary states, denoted as B-branes in \cite{WZW B-branes}, are given by
\begin{equation}
\ww N,\eta\rra=\frac{k^{1/4}}{2^{1/4}}\left(|0,0\rra_N+\eta |k,-k\rra_N\right),\quad \eta=\pm1.
\end{equation}
They are interpreted as D1-branes with two special values of Wilson line. The theory of course admits D-branes with an arbitrary position or a Wilson line, but these boundary states do not respect the extended symmetry.

Next, let us have a look at the parafermion theory. Irreducible representations of this theory are labeled by pairs of numbers $(j,n)$, where $j$ is a half-integer in the range $0\leq j\leq \frac{k}{2}$ as in the \SUk WZW theory and $n$ is an integer defined modulo $2k$. The pairs are further restricted by a condition $2j+n=0\ {\rm mod}\ 2$ and by an equivalence relation $(j,n)\sim (\frac{k}{2}-j,k+n)$. If we choose $n$ from the interval $-2j\leq n\leq 2k-2j-2$, the weights of primaries are then given by
\begin{equation}
h_{j,n}=\begin{cases}
\frac{j(j+1)}{k+2}-\frac{n^2}{4k} &\ -j\leq \frac{n}{2} \leq j, \\
\frac{j(j+1)}{k+2}-\frac{n^2}{4k}+\frac{n-2j}{2} & \ j\leq \frac{n}{2} \leq k-j.
\end{cases}
\end{equation}

Cardy boundary states in the parafermion theory read
\begin{equation}
\ww J,n\rra=\sum_{(j,m)}\frac{S_{(J,n)}^{PF\,(j,m)}}{\sqrt{S_{(0,0)}^{PF\,(j,m)}}} |j,m\rra,
\end{equation}
where the parafermion $S$-matrix is related to the \SUk WZW model $S$-matrix as
\begin{equation}
S^{PF\,(j,m)}_{(J,n)}=\sqrt{\frac{2}{k}}e^{\frac{i\pi mn}{k}}S_J^{SU\,j}.
\end{equation}
These boundary states are denoted as A-branes in \cite{WZW B-branes}.
%This reference also introduces so-called B-branes in the parafermion theory, but we do not need explicit formulas for these boundary states.
B-branes in the parafermion theory are given by
\begin{equation}
\ww B,J\rra= \sqrt{k}\sum_{j} \frac{S_{(J,0)}^{PF\,(j,0)}}{\sqrt{S_{(j,0)}^{PF\,(j,m)}}} |B,j,0\rra,
\end{equation}
where B-type Ishibashi states are defined by equation (3.17) in \cite{WZW B-branes}. There are some subtleties regarding these boundary states, but we will not discuss them because they do not appear in the construction of B-branes in the \SUk WZW model.

\begin{figure}
   \centering
   \includegraphics[width=7cm]{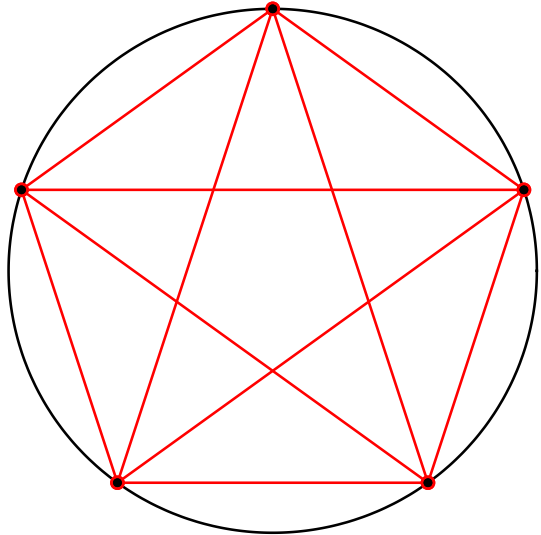}
   \caption{Graphical representations of D-branes in the parafermion theory for $k=5$. $J=0$ branes are located at 5 equidistant points on the circle, $J=\frac{1}{2}$ branes are represented by the shorter lines connecting the points and $J=1$ branes by the longer lines. The second parameter $n$ determines the angle of rotation of the branes.}
   \label{fig:Branes parafermion}
\end{figure}

Cardy boundary states (A-branes) in the parafermion theory have a nice geometrical interpretation. Consider a circle with $k$ regularly placed special points. Boundary states with $J=0$ or $J=k/2$ are D0-branes at these special points and states with other $J$ are represented by lines that connect two points separated by $2J$ segments. The other number $n$ determines rotation of these branes with respect to some reference position. See figure \ref{fig:Branes parafermion} for illustration.

Now, let us go back to the \SUk WZW model. The representations in this theory have a nontrivial decomposition into the free boson and parafermion representations. The character (\ref{character SU}) decomposes as
\begin{equation}\label{character decomposition}
\chi^{SU}_j(q,z)=\sum_{n=0}^{2k-1}\chi_{j,n}^{PF}(q)\chi^{U(1)}_n(q,z).
\end{equation}
The variable $z$ tells us that there is a relation between momenta in the free boson theory and eigenvalues of $J_0^3$ in the \SUk WZW model. Concretely, the quantum number $n$ from (\ref{free boson momentum}) corresponds to $2m$.

In \cite{WZW B-branes}, it was shown that Cardy boundary states in the \SUk WZW model can be written in terms of Cardy boundary states in the constituent models\footnote{The reference \cite{WZW B-branes} does not specify for what \SU gluing conditions it holds. It cannot be for the basic identity gluing conditions, which imply $(J_n^3+\bar J_{-n}^3)||B\rra=0$, while Dirichlet boundary state is annihilated by the combination of modes $\alpha_n-\bar\alpha_{-n}$. A more likely candidate for the gluing matrix is
\begin{equation}
\Omega=\left(\begin{array}{ccc}
0 & 0 & 1 \\
0 & -1 & 0 \\
1 & 0 & 0
\end{array} \right).\nn
\end{equation}}:
\begin{equation}
||J,g\rra^{SU} =\frac{1}{\sqrt{k}}\sum_{n=0}^{2k-1} \ww J,n\rra^{PF}\ww D,n\rra^{U(1)},
\end{equation}
where the sum over $n$ goes only over values allowed by the parafermion representations.

The so-called B-branes\footnote{B-brane boundary states were also constructed using Coulomb-gas representation of the \SUk WZW model in \cite{Kawai B-branes}.} are given by a very similar expression, one just needs to replace Dirichlet boundary states by Neumann boundary states in the U(1) theory
\begin{equation}\label{B-brane BS}
||B,J,\eta\rra^{SU} =\frac{1}{\sqrt{k}}\sum_{n=0}^{2k-1} \ww J,n\rra^{PF}\ww N,\eta\rra^{U(1)},\quad \eta=\pm 1.
\end{equation}
The $g$-function of a B-brane is given by $\sqrt{k}$ times the $g$-function of the usual Cardy boundary state with the same $J$. The formula above describes just two representatives of B-branes, which however appear in much larger continuous families. The spectrum of B-branes includes five dimension one operators and therefore they have five-dimensional moduli space.

In the reference \cite{WZW Symmetry-breaking}, the authors constructed another two types of symmetry-breaking boundary states. These are less relevant for our work, so we will mention them only briefly. First, they consider a case when the level is an integer squared, $k=\kappa^2$. In these models, they found boundary states parameterized by parafermion labels $(j,n)$ and group elements $g\in {\rm SU}(2)/\mathbb{Z}_\kappa$ or $g\in {\rm SU}(2)/\mathbb{Z}_{\kappa/2}$ for odd and even $k$ respectively. One can easily see how additional symmetry-breaking boundary states arise in the simplest nontrivial case $k=4$. In this model, primary operators $\phi_{2,m}$ have weight one; therefore there is an extended set of marginal operators and symmetry-breaking boundary states can be generated by marginal deformations of the $J=1$ boundary state. Marginal deformations are however not a focus of this article and we will encounter only one real OSFT solution that could potentially belong to this group.

The second class of boundary states in \cite{WZW Symmetry-breaking} is inspired by the free boson boundary states from \cite{Janik}. They are defined for an arbitrary $k$, but they suffer from the same problem as the boundary states in \cite{Janik}. When one tries to verify the Cardy condition, modular transformation of an overlap of two boundary states leads to an integral over characters instead of the usual sum. It is not clear whether such boundary states have any physical meaning. Another problem is that we do not know normalization of these boundary states, so we cannot do even a simple comparison of $g$-function with our results.

\subsection{Extension to the \SLk WZW model}\label{sec:WZW:SL2}
Although our original intention was to work with the \SUk WZW model, we found that some OSFT solutions actually describe \SLk WZW model boundary states. Therefore let us make few comments regarding the relation between the two models.

The \SL group can be viewed as a complexification of the SU(2) group. We can write a generic element of both groups using the same three generators,
\begin{equation}
g=\exp \sum_{a=1}^3 \lambda_a J^a,
\end{equation}
but the difference is that the parameters $\lambda_a$ are real in the SU(2) group and complex in the \SL group. To describe \SL elements, we can also use the formula (\ref{SU2 group element}) with complex angles, which generate hyperbolic functions.

Similarly, we can understand some \SL boundary states as complexification of SU(2) boundary states. In this paper, we focus on boundary states that preserve the $J^3$ gluing condition. To describe these states in the \SLk WZW model, can use the formulas (\ref{Ishibashi}) and (\ref{D spec}), where we make the replacement
\begin{equation}\label{rho def}
\theta \rar \theta-i \log \rho,
\end{equation}
so that
\begin{equation}
e^{i n \theta}\rar \rho^n e^{i n \theta},
\end{equation}
where the parameter $\rho$ is a positive real number. In this parameterization, boundary state components depend on $\rho$ through integer powers.

Obviously, \SL boundary states are not real with respect to our complex conjugation. That means that the corresponding OSFT solutions must be also complex. However, the \SL solutions that we discuss in this paper still have some reality properties, so we call them pseudo-real.

\section{String field theory implementation}\label{sec:SFT}
In this section, we will discuss several topics regarding how the \SUk WZW model is incorporated into OSFT. We take some inspiration from \cite{Michishita}, but our approach to many calculations is different. The general framework and numerical algorithms that we use are based on the thesis \cite{KudrnaThesis} and we refer to this work for most technical details. However, we have to make some adjustments in order to deal with the \SU currents, so we will focus on the description of the differences that come with the \SUk WZW model. First, we will discuss properties of the string field and then we will move to topics which concern the analysis and identification of solutions. Additionally, appendix \ref{app:Numerics} provides a description of some of our numerical algorithms for evaluating of the OSFT action and Ellwood invariants in this model.

\subsection{String field}\label{sec:SFT:String field}
The form of the string field in OSFT which describes the \SUk WZW model is obtained by tensoring the Hilbert space (\ref{state space SU}) with state spaces of the remaining part of the matter theory and the ghost theory (where we impose Siegel gauge and the SU(1,1) singlet condition \cite{ZwiebachSU11}\cite{GaiottoRastelli}\cite{KudrnaUniversal}). We expand the string field as
\begin{equation}
\Psi=\sum_{A,I,J,K,j,m} t_{AIJKjm} J_{-I}^A L'_{-J}L'^{gh}_{-K} c_1|j,m\ra,
\end{equation}
where we use multiindices defined in the usual way, $I=\{ i_1,i_2,\dots,i_n\}$, $L'$ are universal matter Virasoro generators with central charge $c'=26-c^{SU}$ and $L'^{gh}$ are 'twisted' ghost Virasoros \cite{GaiottoRastelliSU11}. This form of the string field automatically implements Siegel gauge and also the SU(1,1) singlet condition. We treat the universal matter and ghost parts of the string field in the same way as in \cite{KudrnaThesis}, so let us focus on the \SU part.

Even though we impose Siegel gauge, OSFT equations still have an unfixed symmetry. Similarly to free boson theories, which have U(1) symmetries that correspond to translation of solutions along compact dimensions, our theory must respect the SU(2) symmetry. That means that if $|\Psi\ra$ is a solution of OSFT equations, then $e^{i \lambda_a J^a_0}|\Psi\ra$ must be also a solution. Therefore, with the exception of SU(2) singlets, OSFT solutions form continuous families, which prevents us from finding them using Newton's method (which searches for isolated solutions). To deal with this problem, we fix this symmetry using the same prescription as in \cite{Michishita}. We require that that the string field obeys
\begin{equation}\label{J03 psi}
J_0^3|\Psi\ra=0.
\end{equation}
This condition sets $\lambda_{\pm}$ to zero and $\lambda_3$ becomes irrelevant, so the \SU symmetry is fully fixed. As we mentioned before, this restriction of the string field means that it can describe only certain boundary states. The Ellwood invariant conservation law for $J_0^3$ (which is analogous to momentum conservation in free boson theories and it can be derived following \cite{KMS}\cite{KudrnaThesis}) reads
\begin{equation}
\la E[\vv]|J_0^3 |\Psi\ra=-\la E[(J_0^3 +\bar J_0^3)\vv]|\Psi\ra
\end{equation}
and it follows that boundary states described by our solutions must obey
 \begin{equation}\label{J03 B}
(J_0^3 +\bar J_0^3)\ww B \rra=0.
\end{equation}
For Cardy boundary states, it automatically implies that they preserve the $J^3$ gluing condition\footnote{For generic symmetry-breaking boundary states, the condition (\ref{J03 B}) can be rewritten as
\begin{equation}
(L_n -\bar L_{-n})(J_{-n}^3 +\bar J_n^3)\ww B \rra=0, \nn
\end{equation}
which is however weaker than the $J^3$ gluing condition.}, which leaves us with a one-parametric family of boundary states labeled by $\theta$. The remaining symmetry is broken by the level truncation approximation, so the equations of motion for the restricted string field have a discrete set of solutions and we can search for them using Newton's method.

In order to implement the condition (\ref{J03 psi}), we decompose the \SUk WZW model Hilbert space according to eigenvalues of $J_0^3$: $\hh=\bigoplus_{m\in \mathds{Z}} \hh^{(m)}$. Based on this decomposition, we decompose the string field as
\begin{equation}
|\Psi\ra=\sum_{m\in \mathds{Z}}|\Psi^{(m)}\ra,
\end{equation}
where $|\Psi^{(m)}\ra$ satisfies $J_0^3|\Psi^{(m)}\ra=m|\Psi^{(m)}\ra$. The condition (\ref{J03 psi}) therefore selects $|\Psi^{(0)}\ra$, but we also need $|\Psi^{(\pm 1)}\ra$ as auxiliary objects, which help us  compute OSFT vertices and Ellwood invariants, see appendix \ref{app:Numerics}.
Since we work only with a part of the Hilbert space, the number of states at a given level is significantly reduced and imposing (\ref{J03 psi}) therefore greatly speeds up our calculations.

For construction of OSFT solutions, the most important states are the tachyon $c_1|0,0\ra$, relevant primary fields $c_1|j,0\ra$ and sometimes the marginal state $J_{-1}^3 c_1|0,0\ra$. These states are enough to find seeds of most of well-convergent solutions, which describe the fundamental boundary states, although we usually also solve equations for first few descendant fields when searching for seeds to have access to some more unusual solutions.

\subsection{Twist symmetry and reality conditions}\label{sec:SFT:twist}
Next, we will describe the twist symmetry and reality conditions of the string field in the \SUk WZW model, which are different from most other OSFT settings.

When we analyze boundary three-point functions in our model (which determine the basic cubic vertices in OSFT), we notice that bare structure constants are fully symmetric, but dressed structure constants satisfy
\begin{equation}
C_{(j_1,m_1)(j_2,m_2)(j_3,m_3)}=(-1)^{j_1+j_2+j_3}C_{(j_3,m_3)(j_2,m_2)(j_1,m_1)}
\end{equation}
when we reverse the order of the three labels. This equation follows from properties of the Clebsch-Gordan coefficients. In OSFT, the reversal of the order of entries in the cubic vertex involves the twist symmetry. When we consider ghost number one string fields, we have
\begin{equation}
\la \Psi_1,\Psi_2,\Psi_3\ra=\la \Omega\Psi_3,\Omega\Psi_2,\Omega\Psi_1\ra,
\end{equation}
where $\Omega$ is the twist operator. If we further restrict the string fields just to primary fields, $|\Psi_i\ra=c_1|j_i,m_i\ra$, the cubic vertex becomes
\begin{equation}
\la \Psi_1,\Psi_2,\Psi_3\ra=\left(\frac{3\sqrt{3}}{4}\right)^{3-h_1-h_2-h_3}C_{(j_1,m_1)(j_2,m_2)(j_3,m_3)}.
\end{equation}
By combining the equations above, we find out that the twist symmetry acts on primary operators in a nontrivial way\footnote{Alternatively, one can replace $(-1)^j$ by $(-1)^{j+m}$ because the factors $(-1)^{m_i}$ cancel each other, but that does not change the fact that some important states are twist odd.}
\begin{equation}
\Omega |j,m\ra=(-1)^j |j,m\ra.
\end{equation}
If we consider more generic states in a spin $j$ representation, the action of the twist symmetry generalizes to
\begin{equation}
\Omega |\Psi\ra=(-1)^{N+j} |\Psi\ra,
\end{equation}
where $N$ is the eigenvalue of the number operator.

Properties of the twist symmetry do not let us to set the twist odd part of the string field to zero, which is usually done in most OSFT calculations. The reason is that twist odd states play a crucial role in many important solutions and we would lose these solutions by removing twist odd states.
In particular, we note that the primary state $c_1|1,0\ra$ is twist odd and the marginal state $J_{-1}^3c_1|0,0\ra$ too. We considered some alternative definitions of twist symmetry (which combine the twist symmetry with some $Z_2$ symmetry), but none of them seems to be useful for our purposes. They either remove important states too or they are not compatible with our basis, which makes their implementation too complicated.

The inclusion of twist odd states increases the overall time requirements of our calculations, but not too much because most time is typically consumed by evaluation of cubic vertices, see appendix \ref{app:Numerics:vertices}.

Next, let us move to reality properties of the string field.
Complex conjugation (which is defined as a combination of the BPZ and Hermitian conjugations) acts on modes $J_n^a$ as
\begin{eqnarray}
(J_n^3)^\ast=(-1)^{n+1}J_{n}^3,\\
(J_n^\pm)^\ast=(-1)^{n+1}J_{n}^\mp.
\end{eqnarray}
and on boundary and bulk primary states as
\begin{eqnarray}
|j,m\ra^\ast &=&(-1)^m |j,-m\ra, \\
|j,m,n\ra^\ast &=&(-1)^{m-n} |j,-m,-n\ra.
\end{eqnarray}
See appendix \ref{app:complex} for a derivation of these formulas.

These rules for complex conjugation allow us to derive reality conditions for individual components of the string field (for example, we find  that coefficients of primary fields $c_1 |j,0\ra$ must be real for real solutions), but testing reality of the string field directly is not very practical. Since the complex conjugation switches $J^+$ with $J^-$ and we choose our basis asymmetrically, we often find that real solutions satisfy
\begin{equation}
\Psi=\Psi^\ast+\chi,
\end{equation}
where $\chi$ is a null state. In other words, a real string field is usually real only up to a null state. Checking this condition requires quite a lot of effort, so it is much easier to check reality of gauge invariant observables. A real solution must have a real energy and its invariants (which are defined in the next subsection) must satisfy
\begin{equation}\label{reality}
E_{j,m}=(-1)^{2j}E_{j,-m}^\ast
\end{equation}
and
\begin{equation}\label{reality2}
J_{+-}^\ast=J_{-+},\quad J_{33}^\ast=J_{33}.
\end{equation}

We also encounter many pseudo-real solutions, which have real action, but their string field and invariants are not real. As examples, see the \SL solutions in section \ref{sec:SL}. These solutions usually satisfy some alternative reality condition of the form
\begin{equation}
\Psi=S(\Psi^\ast)+\chi,
\end{equation}
where $S$ is some $Z_2$ symmetry and $\chi$ is a null state.

\subsection{Observables}\label{sec:SFT:Ellwood}
In order to identify solutions as boundary states, we need gauge invariants observables. The first quantity we consider is the energy, which is defined in the usual way using the OSFT action:
\begin{equation}
E[\Psi]=E_J-S[\Psi]=E_J+\frac{1}{g_o^2}\left(\frac{1}{2}\la\Psi,Q\Psi\ra+\frac{1}{3}\la\Psi,\Psi\ast\Psi\ra\right),
\end{equation}
where $E_J$ is the energy of the $J$-brane background. The normalization of the energy is chosen so that it reproduces boundary state $g$-functions, which means that $E_J=g_J=B_J^{\ 0}$. However, knowing the $g$-function is not enough to identify a boundary state because there are continuous families of boundary states with the same $g$-function. Therefore we need more gauge invariant observables to make unique identification of solutions.

It is conjectured that components of a boundary state corresponding to a solution are given by so-called Ellwood invariants \cite{EllwoodInvariants}, which are defined using on-shell bulk primary operators $\vv$:
\begin{equation}\label{Ellwood definition}
E[\vv]=2\pi i\la I |\vv(i,-i)|\Psi-\Psi_{TV}\ra,
\end{equation}
where $\Psi_{TV}$ is the tachyon vacuum solution. Using the normalization conventions of \cite{KudrnaThesis}, Ellwood invariants should reproduce components of the matter part of the boundary state corresponding to the solution $\Psi$, $E[\vv]^{exp}=\la V^m ||B_\Psi \rra$, where $V^m$ is the matter part of $\vv$.

When we decompose the bulk Hilbert space of the \SUk WZW theory into irreducible representations with respect to the stress-energy tensor, we find that there is an infinite tower of primary operators and therefore it is possible to define an unlimited number of Ellwood invariants. However, most of primary fields have high conformal weights, which means that the associated invariants would be not converge for most solutions (see \cite{KudrnaThesis}). Therefore we consider only a limited number of simple invariants with low weights. Furthermore, the condition (\ref{J03 psi}) implies that Ellwood invariants can be nonzero only for bulk operators which satisfy
\begin{equation}\label{J03 V}
(J_0^3+\bar J_0^3) \vv=0,
\end{equation}
so we consider only bulk operators which have the opposite left and right eigenvalues of $J_0^3$.

% (\ref{J03 V}) fixes spin
We define two types of invariants. The main invariants that we  work with are based on SU(2) primary operators:
\begin{equation}\label{Elw def E}
E_{j,m} = 2\pi i\la E[c\bar c \phi_{j,m,-m}V^{aux}]|\Psi-\Psi_{TV}\ra,
\end{equation}
where $V^{aux}$ is an auxiliary vertex operator, which sets the overall conformal weight to zero, see \cite{KMS} for more details. If we set $j=m=0$, we get the universal invariant $E_{0,0}$ which measures the $g$-function.
Additionally, we decided to test some invariants that include the \SU currents. Most of them have high conformal weights, so we define only the simplest possible invariants $J_{ab}$, which are analogous to $D_{1\mu\nu}$ invariants from the free boson theory \cite{KudrnaThesis}:
\begin{equation}\label{Elw def J}
J_{ab} = 2\pi i N_{ab}\la E[c\bar c J^a\bar{J^b}]|\Psi-\Psi_{TV}\ra.
\end{equation}
The condition (\ref{J03 V}) implies that there are only three nonzero invariants: $J_{+-}$, $J_{-+}$ and $J_{33}$. The vertex operators that define these invariants lie in the same representation as the identity, which means that these invariants should be related to $E_{0,0}$ for Cardy boundary states. We choose their normalization to be
\begin{eqnarray}
&&N_{+-}=N_{-+}=-\frac{1}{k},\\
&&N_{33}=-\frac{2}{k},
\end{eqnarray}
which guarantees that they are equal to $E_{0,0}$ for universal solutions.

Expectation values of our invariants for Cardy boundary states follow from the formulas in section \ref{sec:WZW:Boundary}. Their absolute values are given by the matrix $B_J^{\ j}$ and their phases follow from Ishibashi states for the given gluing conditions. Their expectation values for a boundary state describing a $J$-brane with angle $\theta$ are
\begin{eqnarray}\label{Elw inv exp}
E_{j,m}^{exp}&=&(-1)^{j-m}B_J^{\ j} e^{2 i m \theta}, \\
J_{\pm \mp}^{exp} &=& B_J^{\ 0} e^{\pm 2i \theta}, \\
%J_{+-}^{exp} &=& B_J^{\ 0} e^{2i \theta}, \\
%J_{-+}^{exp} &=& B_J^{\ 0} e^{-2i \theta}, \\
J_{33}^{exp} &=& B_J^{\ 0}.
\end{eqnarray}

In addition to gauge invariant observables, we also compute the first 'out-of-Siegel' equation $\Delta_S$ \cite{KudrnaUniversal}\cite{KudrnaThesis} using the prescription
\begin{equation}\label{DeltaS}
\Delta_S=-\la 0| c_{-1}c_0 b_2|Q\Psi+\Psi\ast\Psi\ra.
\end{equation}
This quantity serves as a consistency check whether solutions satisfy equations that were projected out during implementation of Siegel gauge and it should approach zero.

\subsection{Search for solutions}\label{sec:SFT:solutions}
Now, let us move to the topic of search for solutions and their analysis.

Our method to find OSFT solutions closely follows the algorithms described in  \cite{KudrnaThesis}, so we will summarize it only briefly. We first compute low level seeds (typically at level 2) using the homotopy continuation method which allows us to find all solutions at a given level. Out of these seeds, we select those that have promising properties and we improve them using Newton's method to the highest level we can reach using the available computer resources (which is 10 to 14 depending on the background).

We typically find many seed whose energy is of the same order as $g$-function of the initial D-brane.
Typically, only few seeds are real, more of them are pseudo-real solutions and the vast majority are generic complex solutions. Unfortunately, we cannot restrict the analysis to real solutions because pseudo-real solutions describe \SL branes and some complex seeds become real or pseudo-real at higher levels.
For low $k$ and $J$, the number of solutions is manageable, but it grows very quickly as we consider higher models. For $k=8$, we had to deal with over ten thousand of solutions.

The number of potentially useful seeds is too high to be analyzed by hand, so we had to come up with an automatic procedure to reduce their number to a manageable amount by discarding those with undesirable properties, like large imaginary part or large $\Delta_S$. We used several rounds of elimination, during which we gradually applied more and more strict criteria with increasing level. At the highest available level, we decided keep for a more detailed analysis by hand only solutions with $\rm{Re}[E]>0$, $|\rm{Im}[E]|\lesssim 0.05$ and $|\Delta_S|\lesssim 0.05$. In most settings, these solutions still included a significant amount (usually few dozens) of complex solutions. However, our analysis showed that only few complex solutions allow a clear identification (they mostly represent two \SL 0-branes). These complex solutions are usually not very interesting, so, in the end, we decided that the results presented in this paper will include only solutions which are real or pseudo-real, or became so at some achievable level.

The search for solutions in this model is more complicated than usual due to a partial numerical instability of some solutions. In case of these solutions, Newton's method does not work properly at (some or all) odd levels, it either does not converges at all or it leads to a result that is too different from the previous even level. In order to deal with this problem, we use only even levels as seeds for Newton's method. Some solutions with this issue have a clear interpretation, so this instability does not necessarily disqualify solutions, we just have to use only even level data for their analysis.

This instability is most likely connected to the marginal field $\lambda\, J_{-1}^3c_1|0,0\ra$, we checked that unstable solutions excite this field (the value of $\lambda$ is typically purely imaginary or complex, so the instability mainly concerns \SL solutions). In principle, we should be able to see continuous families of solutions that correspond to boundary states with different $\theta$, which are connected by marginal deformations. This symmetry is broken by the level truncation approximation, but the potential for the marginal field is still quite flat. And it seems that some of its minima either disappear as some levels or the move far between levels, which is probably the cause of the instabilities. However, it is not clear why the instabilities occur for imaginary values of the marginal field and why they happen only at odd levels.

\subsection{Identification of solutions}\label{sec:SFT:identification}
First, let us divide solutions in our model into three categories based on which types of D-branes they describe:
\begin{itemize}
\item {\bf \SU solutions:} These solutions correspond to Cardy boundary states of the \SU model described in section \ref{sec:WZW:Boundary}, which preserve the maximal possible amount of symmetry. These solutions must be real.
\item {\bf \SL solutions:} Next, there are solutions that go beyond the original SU(2) model and describe Cardy boundary states of the extended \SL model. \SL boundary states are not real with respect to our reality condition, but they are mostly pseudo-real.
\item {\bf Exotic solutions:} Finally, there are solutions that do not match any combination of Cardy boundary states. Therefore we think that they describe symmetry-breaking boundary conditions either in the \SU or the \SL model. They can be either real or pseudo-real, but we will focus on real solutions which describe boundary states in the \SU model.
\end{itemize}

Precise identification of solutions in our model and their assignment into one of these groups is more difficult than in the free boson theory, where we have tachyon and energy density profiles to guide us. The energy is usually consistent only with few D-branes configurations, but determining whether there is a combination of parameters $\theta$ (and possibly $\rho$) so that all invariants match the expected values can be quite difficult, especially if a solution describes more than one D-brane.

Therefore we introduce a quantity $R^2$ measuring the difference between invariants of a solution and their expected values for a D-brane configuration with given parameters. We define it as
\begin{equation}\label{R squared}
R^2(J,\theta,\rho)=\sum_{j,m} (E_{j,m}-E_{j,m}^{exp}(J,\theta,\rho))^2,
\end{equation}
where $E_{j,m}$ are infinite level extrapolations of Ellwood invariants (or values from the last available level) and $E_{j,m}^{exp}(J,\theta,\rho)$ are their expected values for given parameters $J$, $\theta$ and $\rho$.

To identify a solution, we select D-brane configurations allowed by the energy and then we numerically minimize this quantity for each configuration. Since there can be more local minima, we try several different starting points for $\theta$ and $\rho$. The minimum with the smallest value of $R^2$ is then chosen as the final identification. Cases of more D-branes configurations having similar $R^2$ are rare, so identification of well-behaved solutions is usually unique.
Good solutions typically have $R^2<0.001$. If $R^2\gtrsim 0.01$, it indicates low precision of a solution (which is often accompanied by other problems), and if $R^2\gtrsim 0.1$, it usually means that the identification failed and the given solution does not describe a configuration of Cardy boundary states.

The definition of $R^2$ allows various adjustments, some of which are more convenient for certain types of solutions and less for others. One can choose various ranges of $j$ and $m$. We decided to take only $m=-j,j$ because these invariants usually have the best precision and they are most sensitive to the parameters $\theta$ and $\rho$, while invariants with $m$ around zero usually have larger errors. Additionally, for \SL solutions at high $k$, we restrict $j$ by $j_{max}=2$ because these solutions have very large values of invariants with high $j$, which leads to problems during the numerical minimization of $R^2$. It would be also possible to make other changes, like weighting invariants by their errors.
Alternative definitions of $R^2$ typically do not affect the final value of $\theta$ because this angle is unambiguously fixed to few discrete values by reality of some invariants for most solutions. They lead to small changes in $\rho$ for \SL solutions, but the results seem to be affected more by precision of infinite level extrapolations than by the definition of $R^2$.

For infinite level extrapolations, we use the method described in \cite{KudrnaThesis}. Let us quickly review the key points here.

To extrapolate a given quantity, we fit the known data points by a function of the form of a polynomial in $1/L$:
\begin{equation}
f^{(M)}(L)=\sum_{n=0}^M a_n L^{-N},
\end{equation}
where $M$ is the order of the extrapolation. We usually use functions of the highest possible order, which actually interpolate the data points. The infinite level extrapolation is then given by the limit $\lim_{L\rar \inf}f^{(M)}(L)$.

However, OSFT data usually do not allow a straightforward extrapolation. They contain more or less visible oscillations with period of 2 (energy, $\Delta_S$, string field coefficients) or 4 (Ellwood invariants) levels. Therefore we divide data points into 2 or 4 groups and extrapolate each of them separately. Then we take the average of the 2 or 4 values as the final result and the standard deviation as the error estimate. This type of error estimate is somewhat problematic because it tends to under/over-estimate the actual error depending on properties of the given quantity, see the discussion in \cite{KudrnaThesis}, but we have no better option and it gives us at least a rough idea about precision of extrapolations.

\subsection{Visualization of SU(2) D-branes}\label{sec:SFT:visualization}
Boundary states in the \SUk WZW model have a somewhat different interpretation from usual D-branes in free boson theory. They do not form hyperplanes, but they have a geometrical meaning. Cardy boundary states are associated with conjugacy classes of the SU(2) group \cite{D-braneGeometry1}\cite{WZW B-branes}\cite{SchomerusLectures}. However, they are not localized exactly to conjugacy classes, but they are smeared objects around them. The localization is least definite for low $k$ and it gets sharper with increasing $k$.

\begin{figure}
   \centering
   \begin{subfigure}{0.45\textwidth}
   \includegraphics[width=\textwidth]{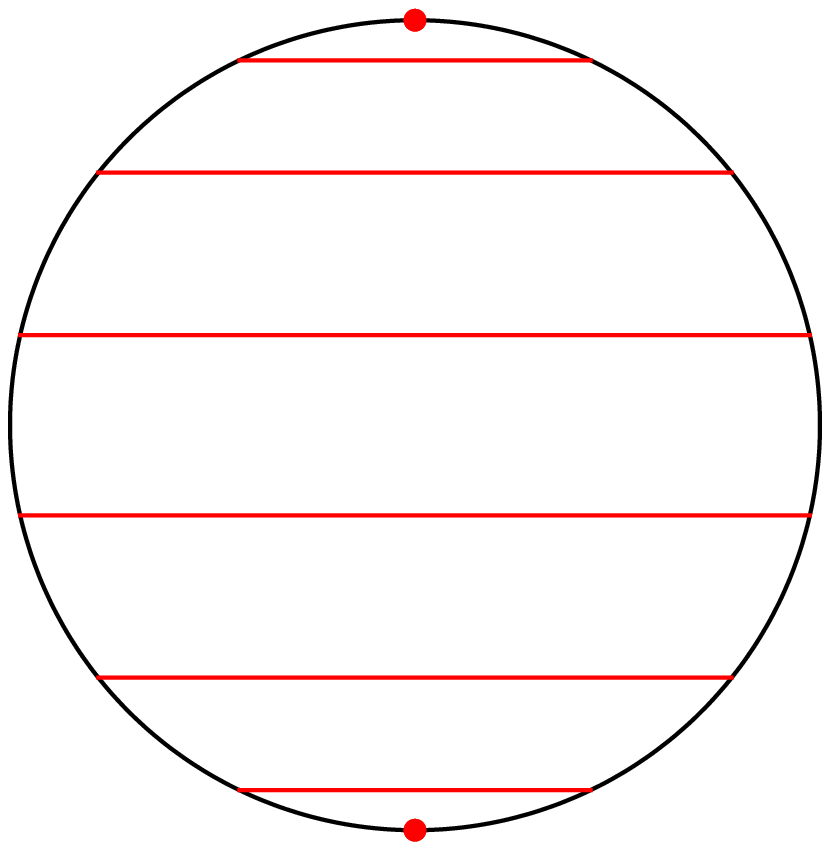}
   \end{subfigure}\qquad
   \begin{subfigure}{0.45\textwidth}
   \includegraphics[width=\textwidth]{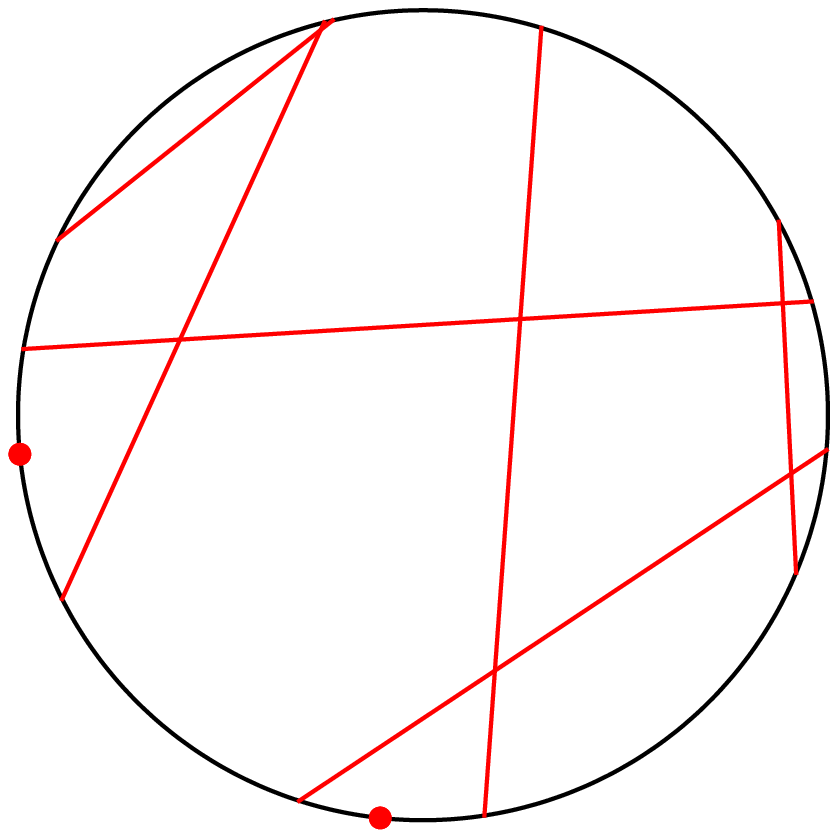}
   \end{subfigure}
   \caption{Visualizations of \SU Cardy branes for $k=7$. The figure on the left shows all D-branes with $\theta=0$, the right figure shows the same D-branes rotated by random angles $\theta$.}
   \label{fig:Branes k=7}
\end{figure}

The SU(2) group manifold is isomorphic to a 3-sphere and conjugacy classes form either points or 2-spheres on the 3-sphere. Points correspond to 0-branes or $\frac{k}{2}$-branes and 2-spheres to other D-branes. Visualization of the 3-sphere would be difficult, but the fact that our solutions preserve the gluing condition for the $J^3$ current fortunately makes the problem much simpler. These Cardy boundary states are characterized just by the angle $\theta$, so we can replace the 3-sphere just by a circle, which corresponds to a 'side view' of the 3-sphere. 0-branes are represented by points on the circle and other $J$-branes by lines connecting two points on the circle separated by a distance determined by $J$. The angle $\theta$ corresponds to rotation of D-branes with respect to the vertical axis. See figure \ref{fig:Branes k=7} for illustration, where we visualize some D-branes in $k=7$ model as an example.

Notice that visualization of Cardy boundary states preserving the $J^3$ gluing condition is very similar to visualization of parafermion D-branes mentioned in section \ref{sec:WZW:B-branes}. The only difference is that D-branes in the \SUk WZW model are not fixed to special points and they can be rotated by an arbitrary angle.

\section{\SU solutions}\label{sec:regular}
In this section, we will discuss OSFT solutions describing SU(2) Cardy boundary states, which are the basic results expected based on background independence of OSFT.
The $k=1$ model is not interesting for us because it is dual to free boson on the self-dual radius $R=1$\footnote{It is easy to check that the central charge for $k=1$ has the correct value $c=1$ and that the three WZW currents can be constructed using a single chiral free field:
\begin{equation*}
J^3 (z)=i\del X (z), \quad J^\pm (z)=e^{\pm2i X}(z).
\end{equation*}} \cite{DiFrancesco} and all fundamental boundary states are related by marginal deformations.
Therefore we will consider $k\geq 2$. We will discuss $k=2,3,4$ models in more detail and then we will show a summary of solutions that we have found up to $k=8$ and conjecture some generic properties of these solutions.

When it comes to the initial boundary conditions, we can restrict our attention just to $\frac{1}{2}\leq J\leq \frac{k}{4}$. Boundary conditions with $J>\frac{k}{4}$ are related to $\frac{k}{2}-J$ just by an \SU rotation (see (\ref{boundary relation})). We also skip all backgrounds with $J=0$ because they have trivial boundary spectrum and therefore all solutions on such backgrounds can be also found on an arbitrary background with the same $k$. The argument why it happens is quite simple. Consider a nontrivial boundary field $\phi_{j,m}$. One-point function of a such field is always zero, which means these fields always contribute to the action at least quadratically. The same holds for their descendants. Therefore it is possible to set all fields corresponding to representations with $j>0$ to zero and the remaining equations of motion for fields corresponding to descendants of the identity are the same as equations of motion for $J=0$. This shows that solutions based on the identity representation are shared by all backgrounds for given $k$.

\FloatBarrier
\subsection{$k=2$ solutions}\label{sec:regular:k=2}
Let us begin with the with the $k=2$ model, where we choose $J=\frac{1}{2}$ boundary conditions (with $\theta=0$) as the background.
By solving level 2 equations, we found approximately 1000 seed solutions. Out of them, there are only two meaningful real solutions\footnote{There are also some interesting pseudo-real and complex solutions, some of which will be discussed later.}\footnote{These solutions are analogous to some solutions from \cite{Michishita}, but this reference identifies them differently as 0-branes with $\theta=0,\pi$.}, which differ by a sign of the invariant $E_{1/2,1/2}$. We picked the solution with the positive sign as a representative and we improved it up to level 14 using Newton's method, see the data in table \ref{tab:sol 2 1/2 0-brane data}.

Gauge invariants of the solution have several symmetries and, because of that, only few of them are independent. Table \ref{tab:sol 2 1/2 0-brane data} includes only the independent observables and the remaining invariants can be obtained using relations
\begin{eqnarray}\label{symmetries 2}
J_{33}=-E_{1,1}=-E_{1,-1}&=&E_{0,0}, \nn \\
J_{+-}=J_{-+}&=&E_{1,0},  \\
E_{\dt1,-\dt1}&=&E_{\dt1,\dt1}.\nn
\end{eqnarray}
These symmetries can be also seen in table \ref{tab:sol 2 1/2 0-brane extrapolation}, which summarizes extrapolations of the observables and compares them to the expected values. Some of the symmetries are the same as the conditions (\ref{reality}) and (\ref{reality2}), which tell us that the solution is real, but there are also additional 'accidental' symmetries with uncertain origin. Similar symmetries also hold for many other real solutions and we will discuss them in more detail later.

\begin{table}[!]
\centering
\begin{tabular}{|l|lllll|}\hline
Level    & Energy   & $E_{0,0}$ & $E_{1/2,1/2}$ & $\ps E_{1,0}$  & $\ps \Delta _S$  \\\hline
2        & 0.749172 & 0.733703  & $0.739416 i$  & $   -0.893387$ & $\ps 0.0210426 $ \\
3        & 0.738953 & 0.725226  & $0.765890 i$  & $   -0.945626$ & $\ps 0.0063005 $ \\
4        & 0.726558 & 0.722133  & $0.778236 i$  & $   -0.487621$ & $\ps 0.0040500 $ \\
5        & 0.723823 & 0.719333  & $0.796483 i$  & $   -0.500237$ & $\ps 0.0026609 $ \\
6        & 0.719329 & 0.715848  & $0.801822 i$  & $   -0.721123$ & $\ps 0.0019628 $ \\
7        & 0.718253 & 0.714764  & $0.807270 i$  & $   -0.730309$ & $\ps 0.0014777 $ \\
8        & 0.715961 & 0.714011  & $0.810113 i$  & $   -0.629844$ & $\ps 0.0011938 $ \\
9        & 0.715423 & 0.713460  & $0.814922 i$  & $   -0.631591$ & $\ps 0.0009406 $ \\
10       & 0.714036 & 0.712159  & $0.816787 i$  & $   -0.704802$ & $\ps 0.0008065 $ \\
11       & 0.713724 & 0.711832  & $0.818892 i$  & $   -0.706614$ & $\ps 0.0006497 $ \\
12       & 0.712795 & 0.711535  & $0.820182 i$  & $   -0.664923$ & $\ps 0.0005809 $ \\
13       & 0.712596 & 0.711319  & $0.822338 i$  & $   -0.665365$ & $\ps 0.0004740 $ \\
14       & 0.711929 & 0.710651  & $0.823308 i$  & $   -0.701302$ & $\ps 0.0004371 $ \\\hline
%$\inf$   & 0.707097 & 0.7069    & $0.8396   i$    & $   -0.7072  $ & $   -0.000028  $ \\
%$\sigma$ & 0.000001 & 0.0002    & $0.0001   i$    & $\ps 0.0068  $ & $\ps 0.000003  $ \\\hline
%/Exp.     & 0.707107 & 0.707107  & $0.840896 i$    & $   -0.707107$ & $\ps 0         $ \\\hline
\end{tabular}
\caption{Independent observables of a solution describing 0-brane with $\theta=\frac{\pi}{2}$ in the $k=2$ model with $J=\frac{1}{2}$ boundary conditions. We provide data up to level 14, higher level data can be obtained through the duality to the Ising model. Extrapolations of these quantities and their expected values are shown in table \ref{tab:sol 2 1/2 0-brane extrapolation}.} \label{tab:sol 2 1/2 0-brane data}
\vspace{7mm}
\begin{tabular}{|c|lll|}\cline{1-4}
          & $\ps $Energy      & $\ps E_{0,0}     $ & $\ps \Delta_S    $ \\ \cline{1-4}
$\inf$    & $\ps 0.707097   $ & $\ps 0.7069      $ & $   -0.000028    $ \\
$\sigma$  & $\ps 0.000001   $ & $\ps 0.0002      $ & $\ps 0.000003    $ \\
Exp.      & $\ps 0.707107   $ & $\ps 0.707107    $ & $\ps 0           $ \\\cline{1-4}
          & $\ps J_{+-}     $ & $\ps J_{-+}      $ & $\ps J_{33}      $ \\\cline{1-4}
$\inf$    & $   -0.7072     $ & $   -0.7072      $ & $\ps 0.7069      $ \\
$\sigma$  & $\ps 0.0068     $ & $\ps 0.0068      $ & $\ps 0.0002      $ \\
Exp.      & $   -0.707107   $ & $   -0.707107    $ & $\ps 0.707107    $ \\\cline{1-4}
          & $\ps E_{1/2,1/2}$ & $\ps E_{1/2,-1/2}$ & \mc{1}{|c}{}       \\\cline{1-3}
$\inf$    & $\ps 0.8396   i $ & $\ps 0.8396   i  $ & \mc{1}{|c}{}       \\
$\sigma$  & $\ps 0.0001   i $ & $\ps 0.0001   i  $ & \mc{1}{|c}{}       \\
Exp.      & $\ps 0.840896 i $ & $\ps 0.840896 i  $ & \mc{1}{|c}{}       \\\cline{1-4}
          & $\ps E_{1,1}    $ & $\ps E_{1,0}     $ & $\ps E_{1,-1}    $ \\\cline{1-4}
$\inf$    & $   -0.7069     $ & $   -0.7072      $ & $   -0.7069      $ \\
$\sigma$  & $\ps 0.0002     $ & $\ps 0.0068      $ & $\ps 0.0002      $ \\
Exp.      & $   -0.707107   $ & $   -0.707107    $ & $   -0.707107    $ \\\cline{1-4}
\end{tabular}
\caption{Comparison of extrapolations of observables to their expected values for a solution describing $J=0$ boundary state with $\theta=\frac{\pi}{2}$ in the $k=2$ model with $J=\frac{1}{2}$ boundary conditions. For each observable, the first line shows the infinite level extrapolation, the second line the estimated error of the extrapolation and the final line the expected value based on $\theta=\frac{\pi}{2}$. The extrapolations are based on the level 14 solution which we found in the WZW model, but an improvement is possible using the dual Ising model solution, see table 8.4 in \cite{KudrnaThesis}. } \label{tab:sol 2 1/2 0-brane extrapolation}
\end{table}

The numbers in table \ref{tab:sol 2 1/2 0-brane data} are familiar to us because they match observables of the main Ising model solution \cite{Ising}\cite{KudrnaThesis} for $\Id$-brane. This is not surprising because the Ising model can be obtained as a coset of the SU(2)$_2$ WZW model (see section \ref{sec:WZW:B-branes}). The Ising model data can be used to predict values of observables of our solution up to level 22 and obtain better extrapolations, see \cite{KudrnaThesis}, but that will not be important for our analysis.

Extrapolations of observables and their errors are shown \ref{tab:sol 2 1/2 0-brane extrapolation}. The energy ($g$-function) is close to $1/\sqrt{2}$, which suggests that the solution can be identified as a 0-brane (see table \ref{tab:boundary states}, which summarizes boundary state coefficients for Cardy boundary states), but it remains to check whether its Ellwood invariants are also consistent with this identification.
First, let us consider the invariant $E_{\dt1,\dt1}$, which can be used to determine $\theta$. Using (\ref{Elw inv exp}), we find
\begin{equation}
\theta=\frac{\pi}{2}.
\end{equation}
The predictions of observables in table \ref{tab:sol 2 1/2 0-brane extrapolation} are based this angle. The other solution with the opposite sign of $E_{\dt1,\dt1}$ describes boundary state with $\theta=-\frac{\pi}{2}$. We can see that there is a good match between infinite level extrapolations and predictions, so we have no doubts that the solution really describes a 0-brane. The out-of-Siegel equation $\Delta_S$ (\ref{DeltaS}) is satisfied quite well, which suggests that the solution is consistent.

We notice that error estimates of some extrapolations do not match the actual errors very well. They are underestimated for well convergent quantities (energy, $E_{1/2,1/2}$, $\Delta_S$) and overestimated for $E_{1,0}$, which oscillates with level. This is the same type of behavior which was observed in \cite{KudrnaThesis} and we have to take it into account when analyzing error estimates made using this method.

It is interesting that we can determine the angle exactly, even though numerical results are never precise. That is possible because some invariants depend on $\theta$ through a complex phase and, since the invariants of this solution are either real or purely imaginary, $\theta=\pm\frac{\pi}{2}$ are the only possibilities.

The angles $\theta=\pm\frac{\pi}{2}$ are a somewhat unexpected result. In analogue with lump solutions in the free boson theory, we expected that the basic solutions would have $\theta=0,\pi$. However, that does not happen and, as we are going to see on more examples later, solutions in the \SUk WZW model follows different rules than in the free boson theory. There are solutions with $\theta=0$ on some backgrounds, but they are less common and they have worse properties.

The geometric interpretation of the two 0-brane solutions is shown in figure \ref{fig:Branes k=2,3} on the left. We observe that the 0-branes form around points which lie on the initial $\dt1$-brane. If we relaxed the condition $J_0^3 |\Psi\ra=0$, 0-brane solutions would form a 2-parametric family (which can be obtained by action of \SU group elements on this solution) and the 0-branes would lie somewhere on the 2-sphere that corresponds to the initial  $\dt1$-brane.

In addition to this solution, we also found a solution dual to the positive energy solution from the Ising model \cite{Ising}\cite{KudrnaThesis}. This solution has a complex seed, but its imaginary part decreases with level and it becomes real at level 14. Up to some signs and imaginary units, invariants of solutions in both models are again identical. Therefore we will not repeat the analysis of this solution here and we will just mention its physical interpretation. On the $\dt1$-brane, it can be identified as two 0-branes with $\theta_1=-\theta_2=\dt{\pi})$ (it is represented by both of the red dots in figure \ref{fig:Branes k=2,3} on the left). This solution can be also found when choosing the trivial $J=0$ boundary conditions. On this background, it represents $\dt1$-brane with $\theta=\dt{\pi}$.

\begin{figure}[t]
   \centering
   \begin{subfigure}{0.33\textwidth}
   \includegraphics[width=\textwidth]{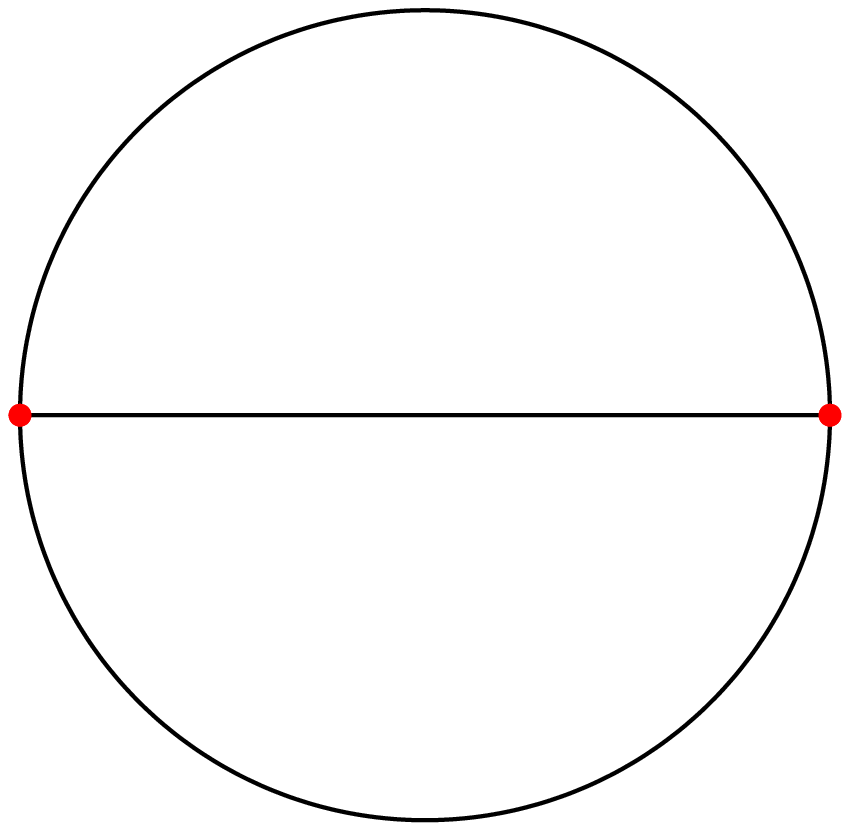}
   \end{subfigure}\qquad\qquad\qquad
   \begin{subfigure}{0.33\textwidth}
   \includegraphics[width=\textwidth]{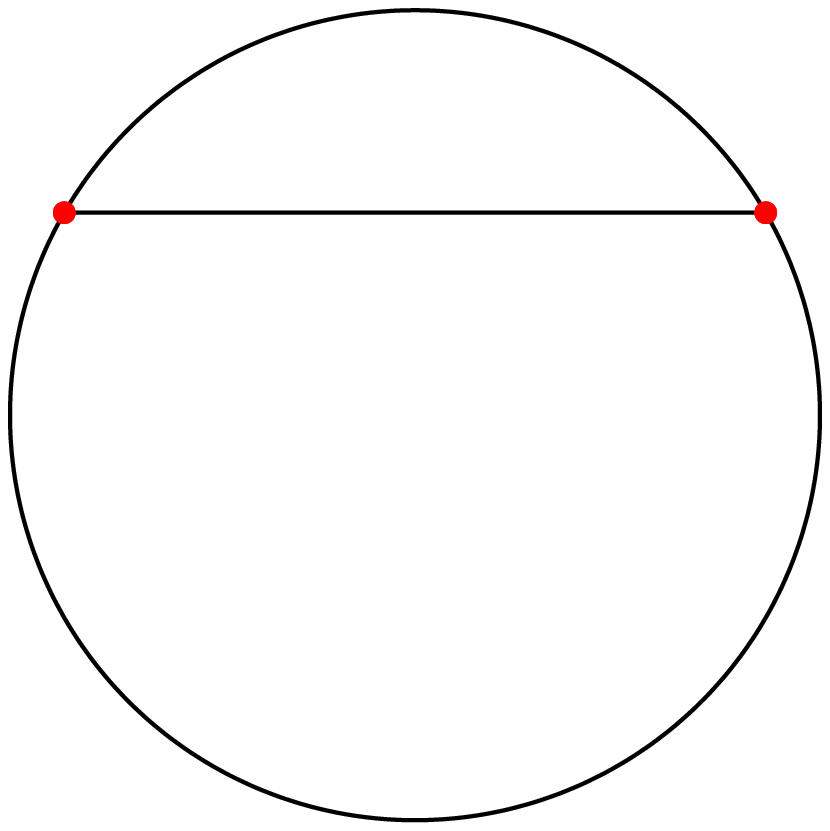}
   \end{subfigure}
   \caption{Graphical representations of 0-brane solutions for $k=2$ (on the left) and $k=3$ (on the right). The initial $\dt1$-brane is represented by the black line and the two solutions describing 0-branes by the red points which lie on the $\dt1$-brane. A two 0-brane solution would be represented by both of the red points.}
   \label{fig:Branes k=2,3}
\end{figure}

\FloatBarrier
\subsection{$k=3$ solutions}\label{sec:regular:k=3}
Next, we move to the $k=3$ model. The only interesting boundary conditions in this model are again $J=\frac{1}{2}$ because other boundary conditions do not offer any new solutions.

In this model, we find a pair of real solutions describing 0-branes, which are in many aspects similar to the solutions from the previous subsection. Properties of one of them are shown in table \ref{tab:sol 3 1/2 0-brane}. Its gauge invariants have a large amount of symmetry, so the table includes only the independent observables for simplicity.
Compared to $k=2$, there is one more independent invariant and some invariants now take generic complex values.
The remaining observables follow either from the reality conditions (\ref{reality}) and (\ref{reality2}) or from 'accidental' symmetries given by relations
\begin{eqnarray}\label{symmetries 3}
J_{33}=-E_{\dt3,\dt3}&=&E_{0,0}, \nn \\
-E_{1,1}&=&E_{\dt1,\dt1},  \\
J_{+-}&=&E_{\dt3,\dt1}. \nn
\end{eqnarray}
Notice that the equations (\ref{symmetries 3}) have a form similar to (\ref{symmetries 2}). Other solutions in this section also have similar symmetries, which suggests that the accidental symmetries follow a generic patter, which will be discussed in subsection \ref{sec:regular:other}.

\begin{table}[!]
\centering
{\scriptsize
\begin{tabular}{|l|llllll|}\hline
$L$      & Energy       & $E_{0,0}$ & $E_{1/2,1/2}$         & $\ps E_{1,0}$  & $\ps E_{3/2,1/2}$         & $\ps \Delta _S$ \\\hline
2        & 0.661072     & 0.652465  & $0.397689+0.574302 i$ & $   -0.842526$ & $   -0.506071           $ & $\ps 0.0257948$ \\
3        & 0.645043     & 0.638284  & $0.395050+0.610150 i$ & $   -0.901786$ & $   -0.566326-0.672324 i$ & $\ps 0.0092582$ \\
4        & 0.630980     & 0.632061  & $0.393469+0.618405 i$ & $   -0.656516$ & $   -0.097475-0.695914 i$ & $\ps 0.0054534$ \\
5        & 0.627511     & 0.628466  & $0.391996+0.633087 i$ & $   -0.663006$ & $   -0.103174-0.372796 i$ & $\ps 0.0040414$ \\
6        & 0.622583     & 0.623410  & $0.391858+0.636782 i$ & $   -0.768041$ & $   -0.361218-0.374128 i$ & $\ps 0.0027889$ \\
7        & 0.621308     & 0.622055  & $0.391288+0.642876 i$ & $   -0.772453$ & $   -0.369680-0.561248 i$ & $\ps 0.0023045$ \\
8        & 0.618848     & 0.620449  & $0.390765+0.644878 i$ & $   -0.732819$ & $   -0.229205-0.569512 i$ & $\ps 0.0017555$ \\
9        & 0.618235     & 0.619763  & $0.390448+0.648911 i$ & $   -0.733702$ & $   -0.229364-0.459290 i$ & $\ps 0.0015076$ \\
10       & 0.616768     & 0.617881  & $0.390184+0.650229 i$ & $   -0.765339$ & $   -0.327392-0.461067 i$ & $\ps 0.0012220$ \\
11       & 0.616423     & 0.617473  & $0.389993+0.652582 i$ & $   -0.766181$ & $   -0.329073-0.538361 i$ & $\ps 0.0010714$ \\
12       & 0.615451     & 0.616792  & $0.389775+0.653499 i$ & $   -0.750970$ & $   -0.264749-0.541515 i$ & $\ps 0.0009060$ \\\hline
%$\inf$   & 0.609695     & 0.6094    & $0.38775 +0.6701   i$ & $   -0.788   $ & $   -0.31    -0.54     i$ & $   -0.00006  $ \\
%$\sigma$ & 0.000009     & 0.0004    & $0.00053 +0.0010   i$ & $\ps 0.024   $ & $\ps 0.05    +0.05     i$ & $\ps 0.00004  $ \\\hline
%Exp.     & 0.609711     & 0.609711  & $0.387782+0.671659 i$ & $   -0.775565$ & $   -0.304856-0.528026 i$ & $\ps 0        $ \\\hline
\mc{7}{c}{} \\[-8pt]
\mc{7}{c}{Extension by minimal model data} \\\hline
13       & 0.615234     & 0.616523  &                       & $   -0.751208$ &                           & $\ps 0.0008049$ \\
14       & 0.614544     & 0.615552  &                       & $   -0.766236$ &                           & $\ps 0.0007016$ \\
15       & 0.614398     & 0.615362  &                       & $   -0.766493$ &                           & $\ps 0.0006292$ \\
16       & 0.613882     & 0.614994  &                       & $   -0.758508$ &                           & $\ps 0.0005613$ \\
17       & 0.613779     & 0.614853  &                       & $   -0.758590$ &                           & $\ps 0.0005069$ \\
18       & 0.613379     & 0.614262  &                       & $   -0.767355$ &                           & $\ps 0.0004603$ \\
19       & 0.613302     & 0.614152  &                       & $   -0.767452$ &                           & $\ps 0.0004180$ \\
20       & 0.612983     & 0.613925  &                       & $   -0.762549$ &                           & $\ps 0.0003851$ \\\hline
$\inf$   & 0.609706     & 0.60972   &                       & $   -0.7750  $ &                           & $   -0.000008 $ \\
$\sigma$ & $4\dexp{-7}$ & 0.00001   &                       & $\ps 0.0003  $ &                           & $\ps 0.000001  $ \\\hline
\end{tabular}}
\caption{Independent gauge invariants of a solution describing 0-brane with $\theta=\pi/3$ in the $k=3$ model with $J=\frac{1}{2}$ boundary conditions. The second part of the table includes higher level predictions based on $m=5$ minimal model data.} \label{tab:sol 3 1/2 0-brane}
\vspace{7mm}
\begin{tabular}{|c|llll|}\cline{1-4}
          & $\ps $Energy              & $\ps E_{0,0}            $ & $\ps \Delta_S           $ & \mc{1}{|c}{}       \\ \cline{1-4}
$\inf$    & $\ps 0.609695           $ & $\ps 0.6094             $ & $   -0.00006            $ & \mc{1}{|c}{}       \\
$\sigma$  & $\ps 0.000009           $ & $\ps 0.0004             $ & $\ps 0.00004            $ & \mc{1}{|c}{}       \\
Exp.      & $\ps 0.609711           $ & $\ps 0.609711           $ & $\ps 0                  $ & \mc{1}{|c}{}       \\\cline{1-4}
          & $\ps J_{+-}             $ & $\ps J_{-+}             $ & $\ps J_{33}             $ & \mc{1}{|c}{}       \\\cline{1-4}
$\inf$    & $   -0.31    +0.54     i$ & $   -0.31    -0.54     i$ & $\ps 0.6094             $ & \mc{1}{|c}{}       \\
$\sigma$  & $\ps 0.05    +0.05     i$ & $\ps 0.05    +0.05     i$ & $\ps 0.0004             $ & \mc{1}{|c}{}       \\
Exp.      & $   -0.304856+0.528026 i$ & $   -0.304856-0.528026 i$ & $\ps 0.609711           $ & \mc{1}{|c}{}       \\\cline{1-4}
          & $\ps E_{1/2,1/2}        $ & $\ps E_{1/2,-1/2}       $ & \mc{2}{|c}{}                                   \\\cline{1-3}
$\inf$    & $\ps 0.3877  +0.670    i$ & $   -0.3877  +0.670    i$ & \mc{2}{|c}{}                                   \\
$\sigma$  & $\ps 0.0005  +0.001    i$ & $\ps 0.0005  +0.001    i$ & \mc{2}{|c}{}                                   \\
Exp.      & $\ps 0.387782+0.671659 i$ & $   -0.387782+0.671659 i$ & \mc{2}{|c}{}                                   \\\cline{1-4}
          & $\ps E_{1,1}            $ & $\ps E_{1,0}            $ & $\ps E_{1,-1}           $ & \mc{1}{|c}{}       \\\cline{1-4}
$\inf$    & $   -0.3877  +0.670    i$ & $   -0.79               $ & $   -0.3877  -0.670    i$ & \mc{1}{|c}{}       \\
$\sigma$  & $\ps 0.0005  +0.001    i$ & $\ps 0.02               $ & $\ps 0.0005  +0.001    i$ & \mc{1}{|c}{}       \\
Exp.      & $   -0.387782+0.671659 i$ & $   -0.775565           $ & $   -0.387782-0.671659 i$ & \mc{1}{|c}{}       \\\cline{1-5}
          & $\ps E_{3/2,3/2}        $ & $\ps E_{3/2,1/2}$         & $\ps E_{3/2,-1/2}       $ & $\ps E_{3/2,-3/2}$ \\\cline{1-5}
$\inf$    & $   -0.6094             $ & $   -0.31    -0.54     i$ & $\ps 0.31    -0.54     i$ & $\ps 0.6094  $     \\
$\sigma$  & $\ps 0.0004             $ & $\ps 0.05    +0.05     i$ & $\ps 0.05    +0.05     i$ & $\ps 0.0004  $     \\
Exp.      & $   -0.609711           $ & $   -0.304856-0.528026 i$ & $\ps 0.304856-0.528026 i$ & $\ps 0.609711$     \\\cline{1-5}
\end{tabular}
\caption{Extrapolations of gauge invariants of a solution describing 0-brane with $\theta=\pi/3$ in the $k=3$ model with $J=\frac{1}{2}$ boundary conditions. } \label{tab:sol 3 1/2 0-brane extrapolation}
\end{table}

Similarly to $k=2$, this solution has a dual solution in minimal models. This time, the coset construction leads to $m=5$ minimal model\footnote{This model is called either the tetracritical Ising model or the Potts model based on the bulk partition function. Our calculations were done in the tetracritical Ising model with diagonal partition function, but the dual solution probably exists in the Potts model as well.}.
When working on \cite{KudrnaThesis}, we developed a code that allows us to do calculations in OSFT involving arbitrary Virasoro minimal model. Therefore we can compare results in the two models and we have found that the WZW model solution has a dual solution that describes $(3,3)$-brane going to $(1,3)$-brane in the minimal model. The solutions in both models are easiest to match using the tachyon coefficient, which is free from any normalization conventions. Matching observables is more difficult because both models have different normalization, but we managed to match energies and the invariants $E_{0,0}$ and $E_{1,0}$ (in the SU(2)$_3$ WZW model) with $E_{(1,1)}$ and $E_{(3,5)}$ (in the minimal model), all these quantities have proportionality coefficient $\frac{3^{1/4}}{2^{3/4}}$. We have not found any relations between other observables in the two models, which is probably because the decomposition of the SU(2)$_3$ WZW model is nontrivial and it mixes primaries in the two constituent models.

The duality can be used to predict behavior of some observables of the WZW model solution at higher levels using the minimal model data (which we have up to level 20), see the second part of table \ref{tab:sol 3 1/2 0-brane}. Comparison of extrapolations with table \ref{tab:sol 3 1/2 0-brane extrapolation} shows that the additional data significantly increase precision of extrapolations, but they are not critical for identification of the solution because the original data are good enough.

Table \ref{tab:sol 3 1/2 0-brane extrapolation} summarizes extrapolations of observables of the solution. By comparing the energy and the $E_{0,0}$ invariant with the numbers in table \ref{tab:boundary states}, we identified the solution as a 0-brane. Other Ellwood invariants can be used to compute the angle $\theta$ and we find
\begin{equation}
\theta=\frac{\pi}{3}.
\end{equation}
Unlike for $k=2$, some invariants have generic complex values, but the angle can be again determined exactly because the invariant $E_{3/2,3/2}$ is real. One can easily check that all invariants are consistent with expectation values based on this angle. The other solution differs by conjugation of complex invariants and it has $\theta=-\pi/3$.

A closer inspection of Ellwood invariants reveals one interesting property. In \cite{KudrnaThesis}, we observed that the rate of convergence of most of Ellwood invariants is related to their conformal weights. Invariants with large weights typically suffer from oscillations, which lead to large errors in infinite level extrapolations. However, the behavior of invariants of this solution is somewhat different. Sets of invariants with the same $j$ have the same weight, so one would expect that they should behave similarly, but there are sometimes big differences. For example, take invariants $E_{3/2,3/2}$ and $E_{3/2,1/2}$. We notice that $E_{3/2,3/2}$ has smaller oscillations than $E_{3/2,1/2}$ and its extrapolation is consequently more precise. Similarly, $E_{1,1}$ behaves better than $E_{1,0}$. Other examples of solutions in this section confirm this trend. We observe that behavior of invariants of this type of solutions does not depend only on the conformal weight (which follows from $j$), but also on $|m|$. Invariants with $m=0$ or $m=\pm \dt1$ usually have the worst behavior, which corresponds to their weights similarly as in \cite{KudrnaThesis}. As $m$ moves away from 0, the behavior improves and invariants with the maximal value of $|m|$, that is $E_{j,\pm j}$, usually converge very well regardless of their weight.
This property becomes more apparent with increasing $j$ (so it is more important for solutions in higher models) because the number of invariants grows and their conformal weight increases.
This unusual property of Ellwood invariants is actually quite helpful to us because there are more well-behaved invariants than we would expect just based on their conformal weights.

\FloatBarrier
\subsection{$k=4$, $J=1$ solutions}\label{sec:regular:k=4}
The final setting that we will discuss in detail is the $k=4$ model. In this model, there are two potentially interesting boundary conditions, $J=\dt1$ and $J=1$. We will focus on $J=1$ because there are more nontrivial solutions, while the $J=\dt1$ background offers only solutions similar to those we found for $k=2,3$.

When we consider the $J=1$ background, there are three interesting real solutions (disregarding multiplicities), which we evaluated up to level 11. One of them describes a $\dt1$-brane and the other two describe 0-branes. Two of the solutions are quite similar to the examples above, so we will go over them more quickly, and we will focus more on the last solution, which has somewhat different properties.

The results regarding the first solution are summarized in table \ref{tab:sol 4 1 1/2-brane extrapolation}. The number of invariants for this $k$ is already quite high, so showing all of them in detail would take a lot of space. Since the finite level data do not have any new interesting properties, we decided to show only the infinite level extrapolations. Once again, we notice that the solution has some 'accidental' symmetries, similar to other real solutions.

\begin{figure}
   \centering
   \includegraphics[width=7cm]{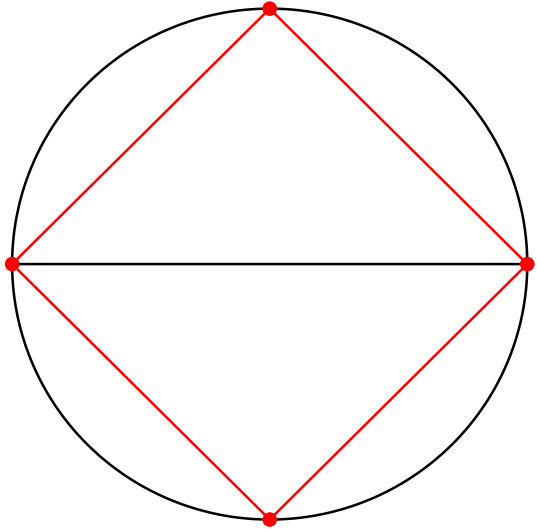}
   \caption{Graphical representations of solutions in $k=4$ model with $J=1$ boundary conditions. The initial $1$-brane divides the circle in two halves. The first solution in this subsection is represented by the four red lines, the second solution by the two red points that lie on the black line and the third solution by the two red points at the top and bottom of the circle.}
   \label{fig:Branes k=4}
\end{figure}

The energy indicates that the solution describes a $\dt 1$-brane. Some of its observables are complex, but the invariant $E_{2,2}$ is real, so we can again determine $\theta$ exactly as a nice multiple of $\pi$, it is equal to $\frac{\pi}{4}$. This solution has degeneracy 4, which is twice as much as for the previous 0-brane solutions. The three other solutions have $\theta=-\frac{\pi}{4}$ and $\theta=\pm\frac{3\pi}{4}$. See figure \ref{fig:Branes k=4}, which depicts the geometry of these solutions.

Precision of this solution is slightly lower than what we saw at lower $k$, but it is still quite good. We again observe the phenomenon that we noticed for the $k=3$ solution, precision of invariants with the same $j$ (which determines their conformal weight) varies depending on $|m|$ and invariants with the maximal value of $|m|$ are more precise than those with $m$ around 0.

At this point, we notice another interesting property of \SU Cardy solutions. It becomes apparent that the parameter $\theta$ follows a simple pattern. For $k=2$, we found a solution with $\theta=\frac{\pi}{2}$, for $k=3$, we got $\theta=\frac{\pi}{3}$ and, finally, now we have $\theta=\frac{\pi}{4}$. That suggests that the angles are given by multiples of $\frac{\pi}{k}$. In the next subsection, we will confirm this property and deduce further rules.

\begin{table}[!t]
\centering
%\begin{tabular}{|l|llllll|}\hline
%$L$      & Energy       & $E_{0,0}$ & $E_{1/2,1/2}$         & $\ps E_{1,1}$   & $\ps E_{2,0}$ & $\ps \Delta_S  $ \\\hline
%2        & 1.043867     & 0.971207  & $0.375988+0.267059 i$ & $   -0.0232451$ & $   -1.40290$ & $   -0.0014085 $ \\
%3        & 1.021183     & 0.935617  & $0.404428+0.300773 i$ & $\ps 0.0313188$ & $   -1.77470$ & $   -0.0100071 $ \\
%4        & 0.991918     & 0.936068  & $0.429662+0.369993 i$ & $   -0.0101171$ & $   -0.07632$ & $   -0.00678125$ \\
%5        & 0.984532     & 0.927668  & $0.432209+0.38517  i$ & $\ps 0.0032924$ & $   -0.13721$ & $   -0.00736707$ \\
%6        & 0.973161     & 0.927331  & $0.447784+0.408725 i$ & $   -0.0051378$ & $   -1.37295$ & $   -0.00607339$ \\
%7        & 0.969689     & 0.923733  & $0.448768+0.41502  i$ & $   -0.0017356$ & $   -1.40852$ & $   -0.00616536$ \\
%8        & 0.963546     & 0.925166  & $0.454443+0.427362 i$ & $   -0.0084851$ & $   -0.57728$ & $   -0.00538747$ \\
%9        & 0.961532     & 0.923203  & $0.454919+0.431962 i$ & $   -0.0042407$ & $   -0.58117$ & $   -0.00537846$ \\
%10       & 0.957641     & 0.923651  & $0.461159+0.439799 i$ & $   -0.0064717$ & $   -1.21426$ & $   -0.00483708$ \\
%11       & 0.956322     & 0.922428  & $0.461511+0.442391 i$ & $   -0.0050471$ & $   -1.22729$ & $   -0.00480085$ \\\hline
%\end{tabular}
%\caption{Selected gauge invariants of $\dt 1$-brane with $\theta=\pi/4$ in $k=4$ model and $J=1$ background.} \label{tab:sol 4 1 1/2-brane}
%\vspace{7mm}
\begin{tabular}{|c|lllll|}\cline{1-4}
          & $\ps $Energy        & $\ps E_{0,0}        $ & $\ps \Delta_S      $ & \mc{2}{|c}{}                            \\ \cline{1-4}
$\inf$    & $\ps 0.9320       $ & $\ps 0.919          $ & $   -0.0010        $ & \mc{2}{|c}{}                            \\
$\sigma$  & $\ps 0.0003       $ & $\ps 0.004          $ & $\ps 0.0001        $ & \mc{2}{|c}{}                            \\
Exp.      & $\ps 0.930605     $ & $\ps 0.930605       $ & $\ps 0             $ & \mc{2}{|c}{}                            \\\cline{1-4}
          & $\ps J_{+-}       $ & $\ps J_{-+}         $ & $\ps J_{33}        $ & \mc{2}{|c}{}                            \\\cline{1-4}
$\inf$    & $\ps 0.04+0.97   i$ & $\ps 0.04-0.97     i$ & $\ps 0.919         $ & \mc{2}{|c}{}                            \\
$\sigma$  & $\ps 0.08+0.07   i$ & $\ps 0.08+0.07     i$ & $\ps 0.004         $ & \mc{2}{|c}{}                            \\
Exp.      & $\ps 0   +0.9306 i$ & $\ps 0   -0.930605 i$ & $\ps 0.930605      $ & \mc{2}{|c}{}                            \\\cline{1-4}
          & $\ps E_{1/2,1/2}  $ & $\ps E_{1/2,-1/2}   $ & \mc{3}{|c}{}                                                   \\\cline{1-3}
$\inf$    & $\ps 0.482+0.487 i$ & $   -0.482+0.487   i$ & \mc{3}{|c}{}                                                   \\
$\sigma$  & $\ps 0.002+0.002 i$ & $\ps 0.002+0.002   i$ & \mc{3}{|c}{}                                                   \\
Exp.      & $\ps 0.5  +0.5   i$ & $   -0.5  +0.5     i$ & \mc{3}{|c}{}                                                   \\\cline{1-4}
          & $\ps E_{1,1}      $ & $\ps E_{1,0}        $ & $\ps E_{1,-1}      $ & \mc{2}{|c}{}                            \\\cline{1-4}
$\inf$    & $   -0.009        $ & $\ps 0.09           $ & $   -0.009         $ & \mc{2}{|c}{}                            \\
$\sigma$  & $\ps 0.004        $ & $\ps 0.03           $ & $\ps 0.004         $ & \mc{2}{|c}{}                            \\
Exp.      & $\ps 0            $ & $\ps 0              $ & $\ps 0             $ & \mc{2}{|c}{}                            \\\cline{1-5}
          & $\ps E_{3/2,3/2}  $ & $\ps E_{3/2,1/2}    $ & $\ps E_{3/2,-1/2}  $ & $\ps E_{3/2,-3/2}   $ & \mc{1}{|c}{}    \\\cline{1-5}
$\inf$    & $\ps 0.482-0.487 i$ & $\ps 0.59+0.54     i$ & $   -0.59+0.54    i$ & $   -0.482-0.487   i$ & \mc{1}{|c}{}    \\
$\sigma$  & $\ps 0.002+0.002 i$ & $\ps 0.12+0.05     i$ & $\ps 0.12+0.05    i$ & $\ps 0.002+0.002   i$ & \mc{1}{|c}{}    \\
Exp.      & $\ps 0.5  -0.5   i$ & $\ps 0.5 +0.5      i$ & $   -0.5 +0.5     i$ & $   -0.5  -0.5     i$ & \mc{1}{|c}{}    \\\cline{1-6}
          & $\ps E_{2,2}      $ & $\ps E_{2,1}        $ & $\ps E_{2,0}       $ & $\ps E_{2,-1}       $ & $\ps E_{2,-2} $ \\\cline{1-6}
$\inf$    & $\ps 0.919        $ & $   -0.04+0.97     i$ & $   -0.98          $ & $   -0.04-0.97     i$ & $\ps 0.919    $ \\
$\sigma$  & $\ps 0.004        $ & $\ps 0.08+0.07     i$ & $\ps 0.15          $ & $\ps 0.08+0.07     i$ & $\ps 0.004    $ \\
Exp.      & $\ps 0.930605     $ & $\ps 0.  +0.930605 i$ & $   -0.930605      $ & $\ps 0.  -0.930605 i$ & $\ps 0.930605 $ \\\cline{1-6}
\end{tabular}
\caption{Extrapolations of gauge invariants of a solution describing $J=\dt 1$ boundary state with $\theta=\frac{\pi}{4}$ in the $k=4$ model with $J=1$ boundary conditions.} \label{tab:sol 4 1 1/2-brane extrapolation}
\end{table}

\begin{table}[!t]
\centering
%\begin{tabular}{|l|lllllll|}\hline
%$L$      & Energy       & $E_{0,0}$ & $E_{1/2,1/2}$ & $\ps E_{1,1} $ & $\ps E_{3/2,1/2}$ & $\ps E_{2,0} $ & $\Delta_S $  \\\hline
%2        & 0.616264     & 0.614424  & $0.577380  i$ & $   -0.758380$ & $\ps 0.192460  i$ & $   -0.735362$ & $0.0270908$ \\
%3        & 0.583770     & 0.584006  & $0.633916  i$ & $   -0.758100$ & $   -0.890124  i$ & $   -0.960632$ & $0.0130627$ \\
%4        & 0.565541     & 0.571404  & $0.642395  i$ & $   -0.760568$ & $   -0.955495  i$ & $\ps 1.467816$ & $0.0051707$ \\
%5        & 0.560251     & 0.566677  & $0.659683  i$ & $   -0.764571$ & $   -0.518163  i$ & $\ps 1.610353$ & $0.0050812$ \\
%6        & 0.553888     & 0.558339  & $0.663036  i$ & $   -0.759139$ & $   -0.524961  i$ & $   -0.356280$ & $0.0024502$ \\
%7        & 0.552050     & 0.556819  & $0.671520  i$ & $   -0.758427$ & $   -0.744682  i$ & $   -0.419067$ & $0.0025954$ \\
%8        & 0.548923     & 0.553471  & $0.673289  i$ & $   -0.758697$ & $   -0.754043  i$ & $\ps 1.101988$ & $0.0013806$ \\
%9        & 0.548049     & 0.552794  & $0.678430  i$ & $   -0.760453$ & $   -0.628522  i$ & $\ps 1.151624$ & $0.0015109$ \\
%10       & 0.546211     & 0.549708  & $0.679565  i$ & $   -0.758437$ & $   -0.631506  i$ & $   -0.019181$ & $0.0008416$ \\
%11       & 0.545716     & 0.549351  & $0.682923  i$ & $   -0.758330$ & $   -0.716584  i$ & $   -0.041157$ & $0.0009461$ \\\hline
%\end{tabular}}
%\caption{Selected gauge invariants of $0$-brane with $\theta=\frac{\pi}{2}$ in $k=4$ model and $J=1$ background.} \label{tab:sol 4 1 0-brane 2}
%\vspace{7mm}
\begin{tabular}{|c|lllll|}\cline{1-4}
          & $\ps $Energy      & $\ps E_{0,0}     $ & $\ps \Delta_S    $ & \mc{2}{|c}{}                         \\ \cline{1-4}
$\inf$    & $\ps 0.537311   $ & $\ps 0.536       $ & $   -0.00009     $ & \mc{2}{|c}{}                         \\
$\sigma$  & $\ps 0.000008   $ & $\ps 0.001       $ & $\ps 0.00008     $ & \mc{2}{|c}{}                         \\
Exp.      & $\ps 0.537285   $ & $\ps 0.537285    $ & $\ps 0           $ & \mc{2}{|c}{}                         \\\cline{1-4}
          & $\ps J_{+-}     $ & $\ps J_{-+}      $ & $\ps J_{33}      $ & \mc{2}{|c}{}                         \\\cline{1-4}
$\inf$    & $   -0.55       $ & $   -0.55        $ & $\ps 0.536       $ & \mc{2}{|c}{}                         \\
$\sigma$  & $\ps 0.02       $ & $\ps 0.02        $ & $\ps 0.001       $ & \mc{2}{|c}{}                         \\
Exp.      & $   -0.537285   $ & $   -0.537285    $ & $\ps 0.537285    $ & \mc{2}{|c}{}                         \\\cline{1-4}
          & $\ps E_{1/2,1/2}$ & $\ps E_{1/2,-1/2}$ & \mc{3}{|c}{}                                              \\\cline{1-3}
$\inf$    & $\ps 0.704     i$ & $\ps 0.704      i$ & \mc{3}{|c}{}                                              \\
$\sigma$  & $\ps 0.001     i$ & $\ps 0.001      i$ & \mc{3}{|c}{}                                              \\
Exp.      & $\ps 0.707107  i$ & $\ps 0.707107   i$ & \mc{3}{|c}{}                                              \\\cline{1-4}
          & $\ps E_{1,1}    $ & $\ps E_{1,0}     $ & $\ps E_{1,-1}    $ & \mc{2}{|c}{}                         \\\cline{1-4}
$\inf$    & $   -0.757      $ & $   -0.76        $ & $   -0.757       $ & \mc{2}{|c}{}                         \\
$\sigma$  & $\ps 0.001      $ & $\ps 0.01        $ & $\ps 0.001       $ & \mc{2}{|c}{}                         \\
Exp.      & $   -0.759836   $ & $   -0.759836    $ & $   -0.759836    $ & \mc{2}{|c}{}                         \\\cline{1-5}
          & $\ps E_{3/2,3/2}$ & $\ps E_{3/2,1/2} $ & $\ps E_{3/2,-1/2}$ & $\ps E_{3/2,-3/2}$ & \mc{1}{|c}{}    \\\cline{1-5}
$\inf$    & $   -0.704     i$ & $   -0.69       i$ & $   -0.69       i$ & $   -0.704      i$ & \mc{1}{|c}{}    \\
$\sigma$  & $\ps 0.001     i$ & $\ps 0.10       i$ & $\ps 0.10       i$ & $\ps 0.001      i$ & \mc{1}{|c}{}    \\
Exp.      & $   -0.707107  i$ & $   -0.707107   i$ & $   -0.707107   i$ & $   -0.707107   i$ & \mc{1}{|c}{}    \\\cline{1-6}
          & $\ps E_{2,2}    $ & $\ps E_{2,1}     $ & $\ps E_{2,0}     $ & $\ps E_{2,-1}    $ & $\ps E_{2,-2} $ \\\cline{1-6}
$\inf$    & $\ps 0.536      $ & $\ps 0.55        $ & $\ps 0.7         $ & $\ps 0.55        $ & $\ps 0.536    $ \\
$\sigma$  & $\ps 0.001      $ & $\ps 0.02        $ & $\ps 0.1         $ & $\ps 0.02        $ & $\ps 0.001    $ \\
Exp.      & $\ps 0.537285   $ & $\ps 0.537285    $ & $\ps 0.537285    $ & $\ps 0.537285    $ & $\ps 0.537285 $ \\\cline{1-6}
\end{tabular}
\caption{Extrapolations of gauge invariants of a solution describing 0-brane with $\theta=\pi/2$ in the $k=4$ model with $J=1$ boundary conditions. } \label{tab:sol 4 1 0-brane 2 extrapolation}
\end{table}

Properties of the next solution solution are shown in table \ref{tab:sol 4 1 0-brane 2 extrapolation}. This solution is especially nice because all of its observables are either real or purely imaginary. A quick analysis shows that this solution represents a 0-brane, so we have found representatives of both types of D-branes with $g$-function lower than the background. Its invariants are consistent with the angle $\theta=\frac{\pi}{2}$ (as usual, there is also a solution with the opposite angle). This number is twice the basic value of $\frac{\pi}{k}$, but we will show later that it fits a generic pattern. Otherwise, this solution is very similar to the 0-brane solutions that we found at $k=2,3$.

Finally, we get to the third solution, which is different from the examples above. In table \ref{tab:sol 4 1 0-brane 0}, we show behavior of some of its invariants at finite levels and table \ref{tab:sol 4 1 0-brane 0 extrapolation} summarizes extrapolations of its observables.
The most apparent difference compared to the other two solutions is that this solution has much stronger level dependence, which, unfortunately, leads to a lower precision. For example, take a look at its energy. It starts approximately at $0.98$, which is close to $g$-function of a $\dt1$-brane, but it quickly decreases, its level 11 value is roughly $0.69$ and the infinite level extrapolation is around $0.58$. Therefore the most likely interpretation of this solution is a 0-brane.

In case of the first two solutions, their identification was unambiguous, but now we are not entirely sure. Out of Cardy boundary states, a 0-brane is the only option, but there is a small possibility that it is an exotic solution or a fake solution that appears as an artefact of the level truncation approximation. The 0-brane however seems to be the most likely option because all invariants show at least a rough agreement with the expected values and there are analogous solutions at higher $k$ that fit a certain pattern.

As a consistency check which helps us decide whether the solution has a physical meaning, we computed the out-of-Siegel equation $\Delta_S$ (\ref{DeltaS}). The extrapolated value of this quantity is approximately $|\Delta_S^{(\inf)}|\sim 0.01$. This is a much worse result than what is expected for a typical Siegel gauge solution at this level (compare with the other examples of solutions) and it is comparable, for example, to the positive energy Ising model solution \cite{KudrnaThesis}. This is an indication that this solution is problematic, but $\Delta_S$ is still low enough to accept the solution as physical.

As we mentioned above, the extrapolated value of its energy is around $0.58$, while it should be equal to $0.537285$. The error of the extrapolation is therefore at the second decimal place, which is a pretty bad precision. Surprisingly, the extrapolation of the invariant $E_{0,0}$ is closer to the expected value of the $g$-function than the energy, which does not happen very often because the energy tends to be the most precise invariant. When we go over other invariants, we observe that they also have errors at first or second decimal places. Their values are consistent with the expected results, but the precision is not nearly good enough to say that they converge towards the expected values with certainty.

When it comes to the angle $\theta$, we notice that all invariants are real, which restricts the possible values of $\theta$ to $0$ and $\pi$. This solution has $\theta=0$ and there is one more related solution with $\theta=\pi$. Therefore we have finally found the value of $\theta$ which we expected as the basic result and which eluded us at $k=2,3$. However, it appeared only for a quite problematic solution, which is not a representative solution in this model.

\begin{table}[!]
\centering
{%\scriptsize
\begin{tabular}{|l|lllllll|}\hline
$L$      & Energy       & $E_{0,0}$ & $E_{1/2,1/2}$ & $ E_{1,1}$ & $\ps E_{3/2,1/2} $ & $\ps E_{2,0}$ & $\ps \Delta_S $  \\\hline
2        & 0.983991     & 0.804398  & 0.545513      & 0.398939   & $\ps 2.363891    $ & $   -3.03723$ & $   -0.0190631$ \\
3        & 0.944328     & 0.754182  & 0.561370      & 0.483559   & $\ps 2.432603    $ & $   -3.45429$ & $   -0.0219647$ \\
4        & 0.824155     & 0.669008  & 0.619364      & 0.555732   & $   -0.503812    $ & $\ps 1.26504$ & $   -0.0238877$ \\
5        & 0.807669     & 0.654517  & 0.615335      & 0.585534   & $   -0.614354    $ & $\ps 1.35439$ & $   -0.0228702$ \\
6        & 0.757275     & 0.621488  & 0.636096      & 0.608182   & $\ps 0.745131    $ & $   -1.92255$ & $   -0.0223986$ \\
7        & 0.748563     & 0.614428  & 0.633373      & 0.618536   & $\ps 0.736658    $ & $   -1.95637$ & $   -0.0216490$ \\
8        & 0.720638     & 0.599660  & 0.639537      & 0.624000   & $   -0.452194    $ & $\ps 0.93105$ & $   -0.0210229$ \\
9        & 0.715108     & 0.595335  & 0.637546      & 0.634766   & $   -0.487965    $ & $\ps 0.99759$ & $   -0.0204840$ \\
10       & 0.697031     & 0.585025  & 0.644810      & 0.640832   & $\ps 0.268710    $ & $   -1.26871$ & $   -0.0198852$ \\
11       & 0.693133     & 0.582048  & 0.643394      & 0.645612   & $\ps 0.265116    $ & $   -1.29018$ & $   -0.0194766$ \\\hline
\end{tabular}}
\caption{Selected gauge invariants of a solution probably describing $0$-brane with $\theta=0$ in the $k=4$ model with $J=1$ boundary conditions.} \label{tab:sol 4 1 0-brane 0}
\vspace{7mm}
\begin{tabular}{|c|lllll|}\cline{1-4}
          & $\ps $Energy      & $\ps E_{0,0}     $ & $\ps \Delta_S    $ & \mc{2}{|c}{}                         \\ \cline{1-4}
$\inf$    & $\ps 0.580      $ & $\ps 0.528       $ & $   -0.0110      $ & \mc{2}{|c}{}                         \\
$\sigma$  & $\ps 0.003      $ & $\ps 0.005       $ & $\ps 0.0007      $ & \mc{2}{|c}{}                         \\
Exp.      & $\ps 0.537285   $ & $\ps 0.537285    $ & $\ps 0           $ & \mc{2}{|c}{}                         \\\cline{1-4}
          & $\ps J_{+-}     $ & $\ps J_{-+}      $ & $\ps J_{33}      $ & \mc{2}{|c}{}                         \\\cline{1-4}
$\inf$    & $\ps 0.47       $ & $\ps 0.47        $ & $\ps 0.528       $ & \mc{2}{|c}{}                         \\
$\sigma$  & $\ps 0.42       $ & $\ps 0.42        $ & $\ps 0.005       $ & \mc{2}{|c}{}                         \\
Exp.      & $\ps 0.537285   $ & $\ps 0.537285    $ & $\ps 0.537285    $ & \mc{2}{|c}{}                         \\\cline{1-4}
          & $\ps E_{1/2,1/2}$ & $\ps E_{1/2,-1/2}$ & \mc{3}{|c}{}                                              \\\cline{1-3}
$\inf$    & $\ps 0.657      $ & $   -0.657       $ & \mc{3}{|c}{}                                              \\
$\sigma$  & $\ps 0.007      $ & $\ps 0.007       $ & \mc{3}{|c}{}                                              \\
Exp.      & $\ps 0.707107   $ & $   -0.707107    $ & \mc{3}{|c}{}                                              \\\cline{1-4}
          & $\ps E_{1,1}    $ & $\ps E_{1,0}     $ & $\ps E_{1,-1}    $ & \mc{2}{|c}{}                         \\\cline{1-4}
$\inf$    & $\ps 0.689      $ & $   -0.56        $ & $\ps 0.689       $ & \mc{2}{|c}{}                         \\
$\sigma$  & $\ps 0.006      $ & $\ps 0.06        $ & $\ps 0.006       $ & \mc{2}{|c}{}                         \\
Exp.      & $\ps 0.759836   $ & $   -0.759836    $ & $\ps 0.759836    $ & \mc{2}{|c}{}                         \\\cline{1-5}
          & $\ps E_{3/2,3/2}$ & $\ps E_{3/2,1/2} $ & $\ps E_{3/2,-1/2}$ & $\ps E_{3/2,-3/2}$ & \mc{1}{|c}{}    \\\cline{1-5}
$\inf$    & $\ps 0.657      $ & $   -0.5         $ & $\ps 0.5         $ & $   -0.657       $ & \mc{1}{|c}{}    \\
$\sigma$  & $\ps 0.007      $ & $\ps 0.1         $ & $\ps 0.1         $ & $\ps 0.007       $ & \mc{1}{|c}{}    \\
Exp.      & $\ps 0.707107   $ & $   -0.707107    $ & $\ps 0.707107    $ & $   -0.707107    $ & \mc{1}{|c}{}    \\\cline{1-6}
          & $\ps E_{2,2}    $ & $\ps E_{2,1}     $ & $\ps E_{2,0}     $ & $\ps E_{2,-1}    $ & $\ps E_{2,-2} $ \\\cline{1-6}
$\inf$    & $\ps 0.528      $ & $   -0.47        $ & $\ps 0.3         $ & $   -0.47        $ & $\ps 0.528    $ \\
$\sigma$  & $\ps 0.005      $ & $\ps 0.42        $ & $\ps 0.3         $ & $\ps 0.42        $ & $\ps 0.005    $ \\
Exp.      & $\ps 0.537285   $ & $   -0.537285    $ & $\ps 0.537285    $ & $   -0.537285    $ & $\ps 0.537285 $ \\\cline{1-6}
\end{tabular}
\caption{Extrapolations of gauge invariants of a solution probably describing $0$-brane with $\theta=0$ in the $k=4$ model with $J=1$ boundary conditions and their expected values. } \label{tab:sol 4 1 0-brane 0 extrapolation}
\end{table}

Geometry of all solutions that we have found on this background is depicted in figure \ref{fig:Branes k=4}. Unlike in figure \ref{fig:Branes k=2,3}, which is rather trivial, the branes now form a lattice of four points and lines connecting them. We will discuss the rules governing positions of Cardy branes in the next subsection.

\FloatBarrier
\subsection{Generic properties of \SU solutions}\label{sec:regular:other}
In the previous subsections, we analyzed several examples of solutions describing SU(2) Cardy boundary states. However, we have found several dozens of such solutions and we do not have enough space to discuss all of them in such detail. It would not make much sense anyway because their properties are mostly similar to the examples above. Therefore we will present only a summary of these solutions and we will focus on deducing generic rules governing their properties.

A survey of basic properties of the solutions we have found up to $k=8$ is given in tables \ref{tab:solutions regular 1} and \ref{tab:solutions regular 2}. The tables provide information about which boundary states are described by the solutions, comparison of their energies with the expected $g$-functions and with energy of the background and some additional notes when necessary.

First, let us discuss which types of boundary states can be found. When we go over the list of solutions, we observe that all of them describe either one or two D-branes. 0-branes are the most common, the number of $\dt1$-branes is smaller and there are only few 1-brane solutions. Two D-brane solutions almost exclusively describe two 0-branes (an example of a such solution is given in appendix \ref{app:Data}), there are just two solutions describing a 0-brane and a $\dt1$-brane (and both have some numerical problems).
Some backgrounds at higher $k$ have enough energy to allow three D-brane solutions, but we have found none. There are some solutions with energy similar to what we would expect from three 0-branes (see table \ref{tab:solutions exotic 2}), but these solutions are pseudo-real (so they can describe at best \SL branes anyway) and we have not found any parameters $\theta$ and $\rho$ that would be consistent with their Ellwood invariants. Therefore we concluded that they are exotic solutions describing some symmetry-breaking boundary states. However, it is possible that solutions describing three or more D-branes will appear if one considers backgrounds with even higher $k$ and $J$.

As usual for Siegel gauge solutions, the energy of most solutions is lower than the energy of the reference D-brane, which means that the final $J$ is lower than its initial value. Therefore there are less opportunities to see branes with high $J$ and it explains why there are so many 0-brane solutions. Nevertheless, OSFT allows us to find solutions with positive energy. Two 0-brane solutions on backgrounds with $J=\dt1$ can serve as examples. Some of these solutions are complex at low levels, but many of them have real seeds at level 2. The seeds are however usually quite far away from the expected results and the solutions evolve rapidly with level, so the precision of their extrapolations is usually worse than for other solutions.

\begin{table}[!]
\centering
\begin{tabular}{|l|l|l|l|l|l|l|}\hline
$k$           & $J$               & Identification       & $E^{(\inf)}$ & $E^{(exp)}$ & $E_J$               & Notes                         \\\hline
\mr{2}{*}{2}  & \mr{2}{*}{$\dt1$} & $(0,\pm 1)$          & 0.707097     & 0.707107    & \mr{2}{*}{1.000000} & appears in Ising model        \\
              &                   & $(0,1)+(0,-1)$       & 1.56367$^*$  & 1.41421     &                     & R(14), appears in Ising model \\\hline\hline
\mr{1}{*}{3}  & \mr{1}{*}{$\dt1$} & $(0,\pm 1)$          & 0.609695     & 0.609711    & \mr{1}{*}{0.986534} & appears in Potts model        \\\hline\hline
\mr{5}{*}{4}  & \mr{2}{*}{$\dt1$} & $(0,\pm 1)$          & 0.537275     & 0.537285    & \mr{2}{*}{0.930605} &                               \\
              &                   & $(0,1)+(0,-1)$       & 1.093        & 1.07457     &                     & slow convergence              \\\cline{2-7}
              & \mr{3}{*}{1}      & $(\dt1,\pm 1(\pm3))$ & 0.9320       & 0.930605    & \mr{3}{*}{1.074570} &                               \\
              &                   & $(0,0(4))$           & 0.580        & 0.537285    &                     & slow convergence              \\
              &                   & $(0,\pm 2)$          & 0.53731      & 0.537285    &                     &                               \\\hline\hline
\mr{6}{*}{5}  & \mr{2}{*}{$\dt1$} & $(0,\pm 1)$          & 0.481562     & 0.481581    & \mr{2}{*}{0.867780} &                               \\
              &                   & $(0,1)+(0,-1)$       & 0.958        & 0.963163    &                     & slow convergence              \\\cline{2-7}
              & \mr{4}{*}{1}      & $(0,2)+(0,-2)$       & 0.9646       & 0.963163    & \mr{4}{*}{1.082104} &                               \\
              &                   & $(\dt1,\pm 1)$       & 0.8688       & 0.867780    &                     &                               \\
              &                   & $(0,0)$              & 0.510        & 0.481581    &                     & slow convergence              \\
              &                   & $(0,\pm2)$           & 0.48175      & 0.481581    &                     &                               \\\hline\hline
\mr{10}{*}{6} & \mr{2}{*}{$\dt1$} & $(0,\pm 1)$          & 0.43740      & 0.437426    & \mr{2}{*}{0.808258} &                               \\
              &                   & $(0,1)+(0,-1)$       & 0.869        & 0.874852    &                     & slow convergence              \\\cline{2-7}
              & \mr{4}{*}{1}      & $(0,2)+(0,-2)$       & 0.8755       & 0.874852    & \mr{4}{*}{1.056040} &                               \\
              &                   & $(\dt1,\pm 1)$       & 0.8088       & 0.808258    &                     &                               \\
              &                   & $(0,0)$              & 0.454        & 0.437426    &                     & slow convergence              \\
              &                   & $(0,\pm2)$           & 0.43756      & 0.437426    &                     &                               \\\cline{2-7}
              & \mr{4}{*}{$\dt3$} & $(\dt1,0(6))$        & 0.845        & 0.808258    & \mr{4}{*}{1.143050} & slow convergence              \\
              &                   & $(0,3)+(0,-3)$       & 0.8755       & 0.874852    &                     &                               \\
              &                   & $(0,\pm 3)$          & 0.43763      & 0.437426    &                     &                               \\
              &                   & $(\dt1,\pm2(\pm4))$  & 0.816        & 0.808258    &                     & R(3)                          \\\hline
\end{tabular}
\caption{List of solutions in SU(2)$_k$ WZW models with $k\leq 8$ describing \SU Cardy boundary states, part 1. The result are grouped according to the level $k$ and then by the boundary condition $J$. We show the most likely identification of solutions, comparison of extrapolated energies $E^{(\inf)}$ with their expected values $E^{(exp)}$ and with energy of the background $E_J$ and sometimes additional notes. Boundary states are described by their half-integer label $J$ and by $\theta$ in multiples of $\frac{\pi}{k}$. For solutions with higher degeneracy, we show all values of $\theta$. The table includes only solutions which are real or become real at some accessible level. The note R($L$) means that a solution has a complex seed and it becomes real at level $L$. If a solution becomes real at a too high level to do a meaningful extrapolation, we show its energy from the highest available level. Such energies are denoted by the symbol $^*$.} \label{tab:solutions regular 1}
\end{table}

\begin{table}[!]
\centering
\begin{tabular}{|l|l|l|l|l|l|l|}\hline
$k$           & $J$               & Identification    & $E^{(\inf)}$ & $E^{(exp)}$ & $E_J$               & Notes                 \\\hline
\mr{13}{*}{7} & \mr{2}{*}{$\dt1$} & $(0,\pm 1)$       & 0.401510     & 0.401534    & \mr{2}{*}{0.754638} &                       \\
              &                   & $(0,1)+(0,-1)$    & 0.796        & 0.803069    &                     & slow convergence      \\\cline{2-7}
              & \mr{4}{*}{1}      & $(0,2)+(0,-2)$    & 0.8034       & 0.803069    & \mr{4}{*}{1.016721} &                       \\
              &                   & $(\dt1,\pm 1)$    & 0.75496      & 0.754638    &                     &                       \\
              &                   & $(0,0)$           & 0.412        & 0.401534    &                     & slow convergence      \\
              &                   & $(0,\pm2)$        & 0.40163      & 0.401534    &                     &                       \\\cline{2-7}
              & \mr{7}{*}{$\dt3$} & $(\dt1,0)$        & 0.778        & 0.754638    & \mr{7}{*}{1.156172} & slow convergence      \\
              &                   & $(\dt1,\pm2)$     & 0.7601       & 0.754638    &                     &                       \\
              &                   & $(0,3)+(0,-3)$    & 0.80356      & 0.803069    &                     &                       \\
              &                   & $(0,\pm 1)$       & 0.453        & 0.401534    &                     & slow convergence      \\
              &                   & $(0,\pm 3)$       & 0.40175      & 0.401534    &                     &                       \\
              &                   & $(1,\pm 1)$       & 1.021        & 1.01672     &                     & R(4)                  \\
              &                   & $(0,-3)+(\dt1,4)$ & 1.26722$^*$  & 1.15617     &                     & R(8)                  \\\hline\hline
\mr{15}{*}{8} & \mr{2}{*}{$\dt1$} & $(0,\pm 1)$       & 0.371724     & 0.371748    & \mr{2}{*}{0.707107} &                       \\
              &                   & $(0,1)+(0,-1)$    & 0.736        & 0.743496    &                     & slow convergence      \\\cline{2-7}
              & \mr{4}{*}{1}      & $(0,-2)+(0,2)$    & 0.74365      & 0.743496    & \mr{4}{*}{0.973249} &                       \\
              &                   & $(\dt1,\pm 1)$    & 0.7073       & 0.707107    &                     &                       \\
              &                   & $(0,0)$           & 0.3780       & 0.371748    &                     &                       \\
              &                   & $(0,\pm 2)$       & 0.37182      & 0.371748    &                     &                       \\\cline{2-7}
              & \mr{6}{*}{$\dt3$} & $(\dt1,0)$        & 0.723        & 0.707107    & \mr{6}{*}{1.144123} & slow convergence      \\
              &                   & $(\dt1,\pm 2)$    & 0.7111       & 0.707107    &                     &                       \\
              &                   & $(0,-3)+(0,3)$    & 0.74385      & 0.743496    &                     &                       \\
              &                   & $(0,\pm 1)$       & 0.408        & 0.371748    &                     & slow convergence      \\
              &                   & $(0,\pm 3)$       & 0.37195      & 0.371748    &                     &                       \\
              &                   & $(1,\pm 1)$       & 0.9750       & 0.973249    &                     & R(4)                  \\\cline{2-7}
              & \mr{3}{*}{2}      & $(0,-4)+(0,4)$    & 0.7440       & 0.743496    & \mr{3}{*}{1.203002} &                       \\
              &                   & $(0,\pm 4)$       & 0.3720       & 0.371748    &                     &                       \\
              &                   & $(0,4)+(\dt1,-5)$ & 1.094        & 1.07885     &                     & low level instability \\\hline
\end{tabular}
\caption{List of solutions in the \SUk WZW model with $k\leq 8$ describing \SU Cardy boundary states, part 2. The table has the same format as table \ref{tab:solutions regular 1}.} \label{tab:solutions regular 2}
\end{table}

Next, let us focus on geometry of these solutions.
First of all, the data in tables \ref{tab:solutions regular 1} and \ref{tab:solutions regular 2} confirm the previous observation that the angle $\theta$ is always an integer multiple of $\frac{\pi}{k}$. As for the examples above, the parameter $\theta$ of real solutions can be determined exactly due to reality of the invariant $E_{k/2,k/2}$, so there is no ambiguity or error.

As we analyze solutions on the individual backgrounds, we notice recurring groups of similar solutions. Concretely, when we fix the boundary condition $J$ and vary the level $k$, we often find the same number solutions which describe D-branes with the same parameters (with $\theta$ expressed in multiples of $\pik$)\footnote{The reason for this repeated structure is that the boundary condition $J$ determines the spectrum of boundary primaries. Therefore OSFT equations for fixed $J$ have a similar structure and the same types of solutions.}.
For example, if we choose $J=\frac{1}{2}$, we typically find a 0-brane solution with $\theta=\pm \frac{\pi}{k}$ and a two 0-brane solution with $\theta_1=-\theta_2=\frac{\pi}{k}$. For $J=1$, we find a 0-brane solution with $\theta=\pm 2\frac{\pi}{k}$ and a $\frac{1}{2}$-brane solution with $\theta=\pm \frac{\pi}{k}$, etc.

\begin{figure}
   \centering
   \begin{subfigure}{0.45\textwidth}
   \includegraphics[width=\textwidth]{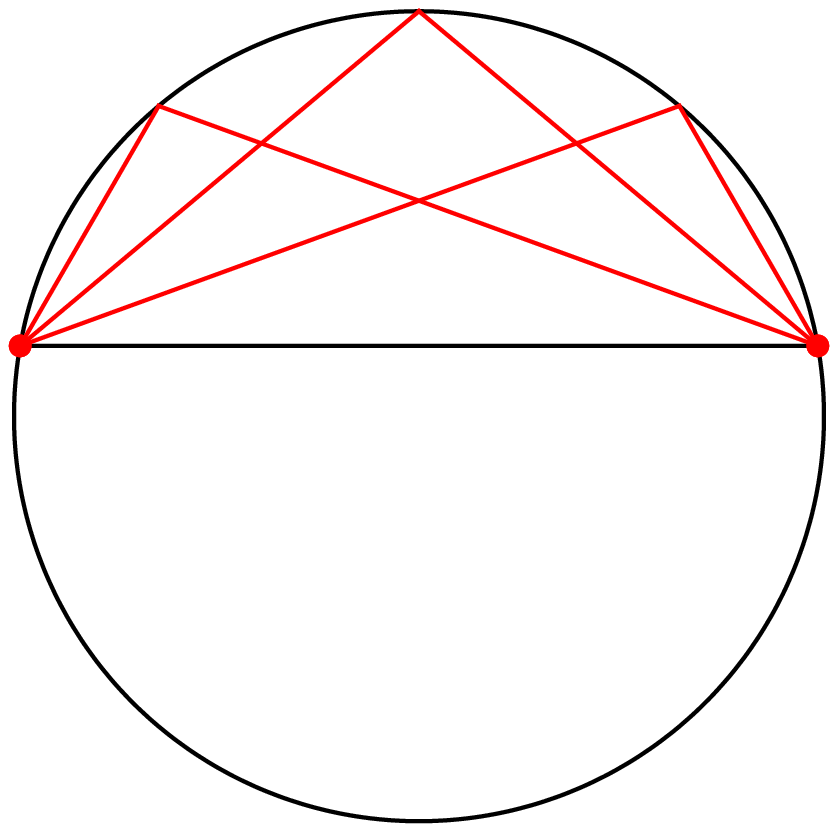}
   \end{subfigure}\qquad
   \begin{subfigure}{0.45\textwidth}
   \includegraphics[width=\textwidth]{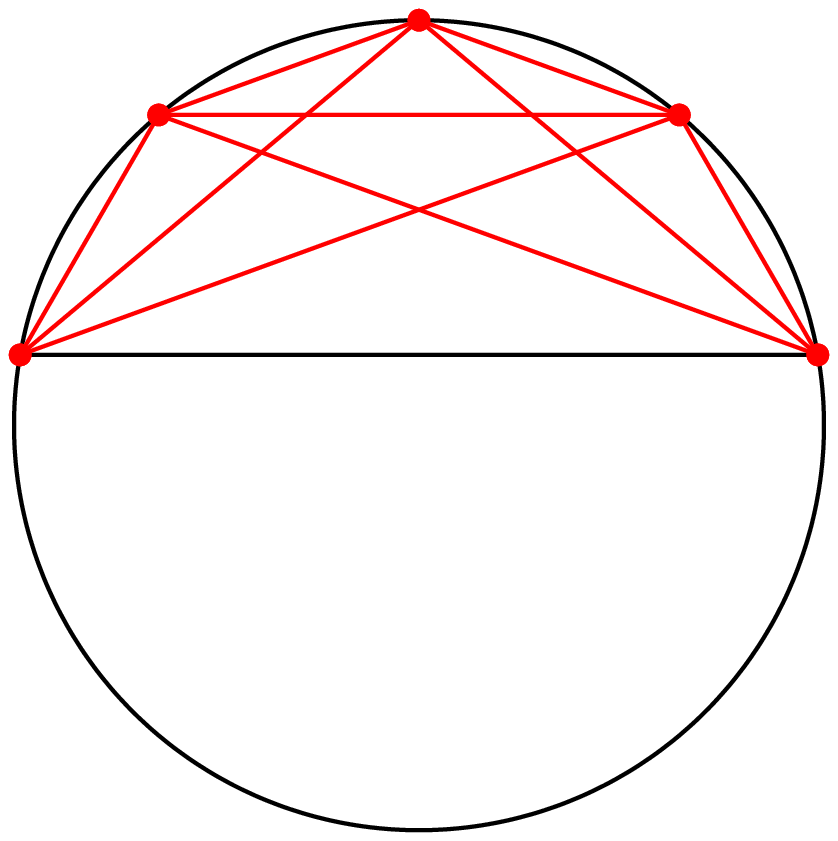}
   \end{subfigure}
   \caption{Visualization of the conjectured rules governing solutions describing \SU Cardy branes for $k=9$ and $J_i=2$ background. The initial D-brane is denoted by the black line and the possible configurations of final D-branes by red lines. The left figure illustrates the relation (\ref{theta 1}). In this figure, all red lines touch the black one at one point. On the right, there is an illustration of the more generic relation (\ref{theta 2}), where the points and lines form a net based on 5 points regularly positioned in the segment.}
   \label{fig:Branes k=9}
\end{figure}

If we consider 0-brane solutions and analyze what $\theta$ they have on different backgrounds, we discover another pattern. For $J=\dt1$, they always have $\theta=\pm \frac{\pi}{k}$, for $J=1$, they usually have $\theta=\pm 2\frac{\pi}{k}$, next is $\theta=\pm 3\frac{\pi}{k}$ for $J=\dt3$, etc. These observations allow us to guess a relation between $\theta$, the initial boundary condition $J_i$ and the final boundary condition $J_f$.
First, let us focus only on well convergent solutions. Their parameters satisfy the following relation:
\begin{equation}\label{theta 1}
\theta=\pm(J_f-J_i)\frac{2\pi}{k}.
\end{equation}
Geometrically, this relation means that D-branes described by these solutions touch the initial D-brane at one point. Furthermore, unless $J=k/4$, the final D-branes lay in the circular segment given by the initial D-brane. See figure \ref{fig:Branes k=9} on the left for illustration. If $J=k/4$, solutions describe D-branes in both halves of the circle and $\theta$ can be shifted by an additional factor of $\pi$. This case is illustrated in figure \ref{fig:Branes k=4} for $k=4$.

The equation (\ref{theta 1}) also holds for multiple brane solutions. For two 0-brane solutions, we additionally find that they are always symmetric around the origin, which means $\theta_1=-\theta_2$.

What about more slowly converging solutions, like the 0-brane solution with $\theta=0$ from table \ref{tab:sol 4 1 0-brane 0 extrapolation}? To describe geometry of these solutions, we need to modify the relation (\ref{theta 1}). The difference is that they have $\theta$ shifted by an additional multiple of $\frac{2\pi}{k}$.
Therefore, we propose the following generalization of the equation (\ref{theta 1})
\begin{equation}\label{theta 2}
\theta=\pm(J_f-J_i+n)\frac{2\pi}{k},\quad n\in \mathbb{Z}.
\end{equation}
Similarly as before, unless $J=k/4$, the sign and the integer $n$ are such that the final D-branes lie in the circular segment given by the initial D-brane.
So far, we have seen only solutions with $n=\pm 1$, but we predict that $n$ can also take larger values. The only analyzed setting where we could possibly find solutions with $n=2$ is $k=8$ and $J=2$, but, for unknown reason, we have found only few real solutions on this background and this one is missing.

This more generic relation is illustrated in figure \ref{fig:Branes k=9} on the right. We observe that the D-branes given by relation (\ref{theta 2}) form a net of points and lines, which lie on or connect points which are regularly positioned in the circular segment.
The geometry of branes described by (\ref{theta 2}) is in fact the same as geometry of D-branes in parafermion models, which we mentioned in section \ref{sec:WZW:B-branes}. This supports the possibility that all \SU Cardy solutions have dual solutions in parafermion theories.

Next, we will discuss properties of Ellwood invariants of \SU Cardy solutions. On the examples in the previous subsections, we observed that solutions of this type have many symmetries. These symmetries hold for all \SU Cardy solutions and also for some real exotic solutions (for example, for the B-branes from section \ref{sec:exotic:B-branes}). Some of their symmetries follow from the reality conditions, but there are also 'accidental' symmetries without an obvious explanation. It turns out that these symmetries follow a relatively simple pattern. The equations (\ref{symmetries 2}) and (\ref{symmetries 3}) suggest that the invariant $J_{33}$ is given by
\begin{equation}\label{symmetries su 1}
J_{33}=E_{0,0}.
\end{equation}
Similarly, the invariant $J_{+-}$ is also related to one of the $E_{j,m}$ invariants. We get
\begin{equation}\label{symmetries su 2}
J_{+-}=(-1)^{2J+1}E_{k/2,k/2-1}^\ast. %corrected sign
\end{equation}
Therefore none of the $J_{ab}$ invariants are independent for this class of solutions.
The remaining symmetries involve $E_{j,m}$ invariants with the maximal value of $m$. These invariants are related as
\begin{equation}\label{symmetries su 3}
E_{j,j}=(-1)^{2J} E_{k/2-j,k/2-j}^\ast.
\end{equation}
Interestingly, the signs in these equation is given purely by the initial boundary condition $J$ and they does not depend on properties of the solutions themselves.

The exact mechanism behind these 'accidental' symmetries is not entirely clear to us. We noticed that these symmetries are related to absence of the marginal field $J_{-1}^3 c_1|0\ra$. We checked that \SU solutions do not excite the marginal field. On the other hand, \SL solutions in the next section, which include the marginal field, do not have any such symmetries. Another possibility how to excite the marginal field is to look for marginal solutions using the standard approach of Sen and Zwiebach \cite{MarginalSen}. Such solutions are real, but the lack any accidental symmetries, which supports the conjecture that accidental symmetries are allowed only when the marginal field is absent\footnote{However, it is not a sufficient condition. There are some real solutions (although clearly unphysical) without the marginal field that do not have these symmetries.}.

Another related issue may be the duality to parafermion theories. At $k=2,3$, we observed dualities between \SUk WZW model solutions and minimal model solutions. We expect that similar dualities extend to higher $k$ and some (maybe all) \SU Cardy solutions are dual to parafermion solutions. We are however not able to confirm this explicitly because we have not constructed OSFT for parafermion theories. \SUk WZW model solutions which have duals in parafermion theories cannot excite states that correspond to the free boson primaries in the decomposition of the theory (which include the marginal field), which could be the origin the 'accidental' symmetries.

Numerical precision \SU Cardy solutions is quite diverse. There are many solutions which converge well and extrapolations of their invariants are close to the expected values despite the fact that we cannot reach levels as high as in simpler OSFT models. The best solutions describe 0-branes with $\theta$ following (\ref{theta 1}), solutions describing branes with higher $J$ tend to have slightly worse precision.
But there are also solutions with quite poor precision. Some of them are D-branes with $\theta$ described by (\ref{theta 2}) with $n\neq 0$, for example, the last solution in subsection \ref{sec:regular:k=4}. These solutions always have worse precision than solutions describing the same D-brane with $n=0$. Low precision is also typical for positive energy solutions or solutions which are complex at low levels, which is however a generic property of all OSFT models \cite{KudrnaThesis}.

When it comes to relative precision of Ellwood invariants for a given solution, we noticed on the examples in previous subsections that the precision depends not only on conformal weights of invariants (which is determined only by the label $j$), but also on the absolute value of $m$. Invariants with high $|m|$ have smaller oscillations and better precision than invariants with $|m|$ around zero.
This property of Ellwood invariants does not seem to be fully universal. It holds for real solutions, both for solutions in this section and for real exotic solutions from section \ref{sec:exotic}, but it is less apparent for pseudo-real solutions from section \ref{sec:SL}, where it holds only in some cases. These solutions are usually highly asymmetric and there are large differences between absolute and relative errors, so it also depends what kind error one chooses to compare.

A partial explanation of this behavior of Ellwood invariant comes from the symmetry (\ref{symmetries su 3}), which relates invariants $E_{j,j}$. Invariants with $j>\frac{k}{4}$ are related to invariants with lower conformal weights and therefore they behave better than expected. However, this relation involves only few invariants.
A more fundamental explanation may be that string fields describing \SU Cardy solutions are in some sense partially universal. Consider one of the setting discussed in \cite{KudrnaThesis}, free boson on a two-dimensional torus. There are solutions (for example some D1-branes) that depend trivially (only through the stress-energy tensor) on one of the free boson coordinates, say $X$. Therefore invariants constructed only using the $X$ field are proportional to the $E_{00}$ invariant and they behave better that invariants which include the other field $Y$. A similar mechanism may work here. If the expected duality to parafermion theories holds, it means that our solutions do not excite primaries from the free boson part of the theory, which could lead to the special properties of Ellwood invariants.

\FloatBarrier
\section{SL(2,$\CC$) solutions}\label{sec:SL}
The original purpose of this project was to study the \SUk WZW theory, but our OSFT setup naturally allows solutions that describe \SL boundary states. Analogously as the \SL group can be viewed as complexification of the \SU group, some \SL boundary states can be obtained by complexification of \SU boundary states and they are described by complex solutions in the \SUk WZW model. The \SL solutions presented in this section are not generic complex solutions, but pseudo-real solutions. Their energy and some invariants are real, but they do not satisfy the reality condition (\ref{reality}). The energy of these solutions can be real because a part of their string field (including the marginal field) is purely imaginary and because the imaginary part contributes to the action quadratically. If one attempts to do marginal deformations following the approach of \cite{MarginalSen} using a generic complex value of the marginal parameter \cite{KudrnaThesis}, the result is a complex solution without any special properties, with a small imaginary part of the energy.

In this section, we will briefly discuss solutions corresponding to \SL Cardy boundary states. We will illustrate their properties on a $k=2$ example and then we will provide a survey of solutions at higher $k$ and describe their generic properties. Finally, we will mention one universal marginal solution that belongs to this group. Few more examples of \SL solutions can be found in appendix \ref{app:Data}.

\subsection{$k=2$ solutions}\label{sec:SL:k=2}
Let us once again consider the $k=2$ model with $J=\dt1$ boundary conditions. In subsection \ref{sec:regular:k=2}, we discussed a pair of real solutions describing 0-branes with $\theta=\pm \frac{\pi}{2}$, but this model admits one more interesting group of four solutions. Gauge invariants of one of these solutions are shown in table \ref{tab:sol 2 1/2 SL 0-brane} and their extrapolations in table \ref{tab:sol 2 1/2 SL 0-brane extrapolation}. Unlike the previous examples, this solution does not have any symmetries, so there are more independent invariants than before. The lack of any symmetries also means that it does not satisfy the reality conditions. However, its energy is real and Ellwood invariants are either real or purely imaginary, so this solution is pseudo-real. Similarly, the string field combines real and purely imaginary states; for example, the tachyon field is real and the marginal field is purely imaginary. The out-of-Siegel equation $\Delta_S$, which we use as a consistency check, is satisfied quite well, which confirms that the solution can have a physical meaning despite not being real.

The energy suggests that the solution describes a 0-brane. Its precision is similar as for the real \SU solution, but Ellwood invariants tend to oscillate more, which leads to larger errors of extrapolations.
Determining parameters of the boundary state is more difficult than for real solutions. It cannot be described by a single real angle, so we have to consider the parameter $\rho$ introduced in (\ref{rho def}). By minimizing the quantity $R^2$ defined in (\ref{R squared}), we get
\begin{eqnarray}
  \theta &=& \frac{\pi}{2}, \\
  \rho &\approx & 2.07.
\end{eqnarray}
The angle $\theta$ can be once again determined exactly due to reality of some observables and it satisfies the relation (\ref{theta 1}), but we get only an approximate result for $\rho$, which seems to be a generic real number. It is not easy so say what is the precision of $\rho$. The relevant invariants used to compute $\rho$ have estimated errors of order of 1\%, but we have seen that these error estimates are not always reliable. Based on behavior of the invariants, the errors are probably somewhat overestimated, so it is reasonable to assume that the relative error $\rho$ is around 1\% or less.

As for \SU solutions, there are several solutions that share the same energy and which differ only by simple transformations. This time, there are 4 solutions that are related by complex conjugation or by exchange of invariants $E_{j,m}\leftrightarrow E_{j,-m}$. The parameters of corresponding boundary states are related by transformations $\theta\rar -\theta$ or $\rho\rar 1/\rho$.

When we take a closer look at values of the invariants, we notice that there is a disproportion between them. In case of real solutions from the previous section, absolute values of invariants with the same $j$ were roughly the same, while now there are large differences, for example, $|E_{1,1}|$ is much larger than $|E_{1,-1}|$. This illustrates that the $m$-dependence of $E_{j,m}$ invariants of \SL solutions is different from \SU solutions. Invariants of \SU solutions behave as $E_{j,m}\sim e^{2im\theta}$, so they differ just by a complex phase, while for \SL solutions, we have
\begin{equation}
E_{j,m}\sim \rho^{2m} e^{2im\theta}.
\end{equation}
We can see that absolute values of these invariants are proportional to $\rho^{2m}$, which can make them either very small or very large. The ratio $|E_{j,m}|/|E_{j,-m}|$ is expected to be $\rho^{4m}$, so for this solution, which has $\rho$ higher than 2, $|E_{1,1}|$ is more than 16 times higher than $|E_{1,-1}|$ and the invariants $J_{+-}$ and $J_{-+}$ differ by a similar ratio. The differences are not so crucial in models with low $k$, where all invariants still take moderate values, but they become problematic for larger $k$. If we consider invariants with the highest $m$ that we encountered in this project, $E_{4,\pm 4}$ in $k=8$ model, their ratio is $\rho^{16}$, which can be of order of tens of thousands for similar values of $\rho$. So these invariants are spread across many orders, which can cause problems in some numerical calculations.

\begin{table}[!]
\centering
\begin{tabular}{|l|llllll|}\hline
Level    & Energy   & $E_{0,0}$ & $\ps J_{+-}$  & $\ps J_{-+}$   & $\ps J_{33}  $ & $\ps \Delta _S$ \\\hline
2        & 0.732856 & 0.756306  & $\ps 0.25674$ & $   -1.728354$ & $\ps 1.836561$ & $\ps 0.0337496$ \\
3        & 0.723239 & 0.749342  & $   -2.28109$ & $\ps 1.641892$ & $\ps 2.843920$ & $\ps 0.0167758$ \\
4        & 0.720008 & 0.735311  & $   -3.10667$ & $   -0.077581$ & $   -0.178054$ & $\ps 0.0095397$ \\
5        & 0.717007 & 0.733258  & $   -3.32678$ & $   -0.666911$ & $   -0.437828$ & $\ps 0.0077107$ \\
6        & 0.715071 & 0.725776  & $   -2.20887$ & $   -0.040855$ & $\ps 1.419582$ & $\ps 0.0057583$ \\
7        & 0.713881 & 0.724882  & $   -2.56862$ & $\ps 0.093424$ & $\ps 1.531108$ & $\ps 0.0049433$ \\
8        & 0.712798 & 0.721990  & $   -2.76483$ & $   -0.298796$ & $\ps 0.292591$ & $\ps 0.0041006$ \\
9        & 0.712189 & 0.721490  & $   -3.04900$ & $   -0.246259$ & $\ps 0.249433$ & $\ps 0.0036194$ \\
10       & 0.711516 & 0.718856  & $   -2.65424$ & $   -0.068778$ & $\ps 1.063174$ & $\ps 0.0031645$ \\
11       & 0.711152 & 0.718528  & $   -2.78922$ & $   -0.097844$ & $\ps 1.089831$ & $\ps 0.0028473$ \\
12       & 0.710698 & 0.717312  & $   -2.79057$ & $   -0.240629$ & $\ps 0.497605$ & $\ps 0.0025669$ \\
13       & 0.710458 & 0.717081  & $   -2.99280$ & $   -0.179035$ & $\ps 0.485697$ & $\ps 0.0023430$ \\
14       & 0.710132 & 0.715741  & $   -2.78204$ & $   -0.112116$ & $\ps 0.912584$ & $\ps 0.0021543$ \\\hline
%$\inf$   & 0.707093 & 0.7076    & $   -3.05   $ & $   -0.16    $ & $\ps 0.69    $ & $   -0.000017 $ \\
%$\sigma$ & 0.000003 & 0.0001    & $\ps 0.19   $ & $\ps 0.16    $ & $\ps 0.35    $ & $\ps 0.000002 $ \\\hline
%Exp.     & 0.707107 & 0.707107  & $   -3.02982$ & $   -0.165026$ & $\ps 0.707107$ & 0               \\\hline
\end{tabular}\vspace{3mm}
\begin{tabular}{|l|lllll|}\hline
Level    & $E_{1/2,1/2}$ & $E_{1/2,-1/2}$ & $\ps E_{1,1}$ & $\ps E_{1,0} $ & $\ps E_{1,-1}$ \\\hline
2        & $1.49750i$    & $0.181299i$    & $   -2.28898$ & $   -1.275935$ & $   -0.303887$ \\
3        & $2.01348i$    & $0.295306i$    & $   -3.54596$ & $   -1.366885$ & $   -0.047299$ \\
4        & $1.63766i$    & $0.397805i$    & $   -2.65845$ & $   -0.548245$ & $   -0.175615$ \\
5        & $1.78121i$    & $0.345793i$    & $   -3.16269$ & $   -0.558834$ & $   -0.173484$ \\
6        & $1.65939i$    & $0.372642i$    & $   -2.75013$ & $   -0.806901$ & $   -0.194656$ \\
7        & $1.77487i$    & $0.371102i$    & $   -3.08903$ & $   -0.815452$ & $   -0.155406$ \\
8        & $1.68733i$    & $0.395409i$    & $   -2.82897$ & $   -0.679071$ & $   -0.179102$ \\
9        & $1.75285i$    & $0.376313i$    & $   -3.05834$ & $   -0.680402$ & $   -0.169132$ \\
10       & $1.69375i$    & $0.390029i$    & $   -2.85837$ & $   -0.751813$ & $   -0.181081$ \\
11       & $1.75443i$    & $0.384421i$    & $   -3.04825$ & $   -0.752593$ & $   -0.163572$ \\
12       & $1.70358i$    & $0.397976i$    & $   -2.88787$ & $   -0.701759$ & $   -0.175908$ \\
13       & $1.74718i$    & $0.386308i$    & $   -3.03991$ & $   -0.701910$ & $   -0.168075$ \\
14       & $1.70655i$    & $0.395863i$    & $   -2.90209$ & $   -0.734085$ & $   -0.176403$ \\\hline
%$\inf$   & $   -1.744  i$    & $   -0.405   i$    & $   -3.029  $ & $   -0.714   $ & $   -0.163   $ \\
%$\sigma$ & $\ps 0.015  i$    & $\ps 0.003   i$    & $\ps 0.030  $ & $\ps 0.014   $ & $\ps 0.005   $ \\\hline
%Exp.     & $   -1.74064i$    & $   -0.406234i$    & $   -3.02982$ & $   -0.707107$ & $   -0.165026$ \\\hline
\end{tabular}
\caption{Gauge invariants of an \SL solution describing 0-brane with $\theta=\frac{\pi}{2}$ and $\rho\approx 2.07$ in the $k=2$ model with $J=\frac{1}{2}$ boundary conditions.} \label{tab:sol 2 1/2 SL 0-brane}
\end{table}

\begin{table}[!]
\centering
\begin{tabular}{|c|lll|}\cline{1-4}
          & $\ps $Energy      & $\ps E_{0,0}     $ & $\ps \Delta_S    $ \\ \cline{1-4}
$\inf$    & $\ps  0.707093  $ & $\ps 0.7076      $ & $   -0.000016    $ \\
$\sigma$  & $\ps  0.000003  $ & $\ps 0.0001      $ & $\ps 0.000002    $ \\
Exp.      & $\ps  0.707107  $ & $\ps 0.707107    $ & $\ps 0           $ \\\cline{1-4}
          & $\ps J_{+-}     $ & $\ps J_{-+}      $ & $\ps J_{33}      $ \\\cline{1-4}
$\inf$    & $   -3.05       $ & $   -0.16        $ & $\ps 0.69        $ \\
$\sigma$  & $\ps 0.19       $ & $\ps 0.16        $ & $\ps 0.36        $ \\
Exp.      & $   -3.03       $ & $   -0.17        $ & $\ps 0.707107    $ \\\cline{1-4}
          & $\ps E_{1/2,1/2}$ & $\ps E_{1/2,-1/2}$ & \mc{1}{|c}{}       \\\cline{1-3}
$\inf$    & $\ps 1.744    i $ & $\ps 0.405       $ & \mc{1}{|c}{}       \\
$\sigma$  & $\ps 0.015    i $ & $\ps 0.003       $ & \mc{1}{|c}{}       \\
Exp.      & $\ps 1.741    i $ & $\ps 0.406       $ & \mc{1}{|c}{}       \\\cline{1-4}
          & $\ps E_{1,1}    $ & $\ps E_{1,0}     $ & $\ps E_{1,-1}    $ \\\cline{1-4}
$\inf$    & $   -3.029      $ & $   -0.714       $ & $   -0.163       $ \\
$\sigma$  & $\ps 0.030      $ & $\ps 0.014       $ & $\ps 0.005       $ \\
Exp.      & $   -3.030      $ & $   -0.707107    $ & $   -0.165       $ \\\cline{1-4}
\end{tabular}
\caption{Extrapolations of gauge invariants of an \SL solution describing 0-brane with $\theta=\frac{\pi}{2}$ and $\rho\approx 2.07$ in the $k=2$ model with $J=\frac{1}{2}$ boundary conditions. Expected values of invariants that depend on $\rho$ are not exact because they are based on its estimated value $\rho=2.07$.} \label{tab:sol 2 1/2 SL 0-brane extrapolation}
\end{table}

\FloatBarrier
\subsection{Generic properties of solutions}\label{sec:SL:other}
Properties of \SL solutions describing Cardy boundary states are more uniform that properties of other solutions, so after showing the $k=2$ example, we will go straight to discussion of their properties in general. Up to $k=8$, we have found several dozens of such solutions and we show their survey in tables \ref{tab:solutions SL 1} and \ref{tab:solutions SL 2} in a similar way as in subsection \ref{sec:regular:other}. The tables provide identification of solutions, their energies and some additional notes when needed.

The two tables are dominated mostly by 0-brane solutions, other than that, there are just few $\dt1$-branes. We have not found any \SL solutions with $J\geq 1$. Some of the solutions describe a single brane, most solutions correspond to two 0-branes and we have found just one combination of a 0-brane and a $\dt1$-brane.

In addition to $J$, \SL solutions are described by parameters $\theta$ and $\rho$. The parameter $\theta$ behaves very predictably and it always follows the relation (\ref{theta 1}), both for solutions describing one and two D-branes. We have not seen any solutions where the angle would be shifted by a multiple of $\frac{2\pi}{k}$ as for some \SU solutions. For two 0-brane solutions, their parameters additionally satisfy $\theta_1=-\theta_2$.

On the other hand, the parameter $\rho$ seems to take generic values. It typically has moderate values which are neither too high nor too low, say $0.5\lesssim\rho\lesssim 2$. We also notice that it does not take values close to 1, except for few solutions which probably describe combinations of an \SU brane and an \SL brane.
Despite not being able to identify exact values of $\rho$, we notice that behavior of this parameter is not completely random and there are traces of some rules. For example, if we select 0-brane solutions on $J=\dt1$ backgrounds, the parameter $\rho$ always decreases as we increase the level $k$. But the sequence does not fit any easily recognizable formula.

As we mentioned above, the parameter $\rho$ different from 1 leads to large differences between absolute values of invariants. Invariants with small absolute values, i.e. those with negative $m$ for $\rho>1$ and with positive $m$ for $\rho<1$, often have large relative errors (sometimes over 100\%, see table \ref{tab:sol 5 1 SL1 extrapolation} as an example). Therefore the value of $\rho$ is, in practice, determined mainly from invariants with large absolute values.

When it comes to solutions describing two 0-branes, we notice that there are rules relating their parameters $\rho_1$ and $\rho_2$, which sometimes lead to additional symmetries. The parameters of our solutions always follow one of three different patterns:
\begin{itemize}
  \item $\rho_1=\rho_2$. These solutions have real invariants, but the solutions themselves do not satisfy the reality condition (\ref{reality}) because there is no exact relation between $E_{j,m}$ and $E_{j,-m}$.
  \item $\rho_1=1/\rho_2$. Invariants of these solutions are given by generic complex numbers, but they have an additional symmetry $E_{j,m}=(-1)^{2j}E_{j-,m}$.
  \item $\rho_1\approx 1$ or $\rho_2\approx 1$. These solutions probably describe combinations of an \SU brane and an \SL brane. Invariants of these solutions do not have any special properties.
\end{itemize}
Examples of these three types of solutions are given in appendix \ref{app:Data} for $k=5$.

As we described in section \ref{sec:SFT:solutions}, a common trait of many \SL solutions is a numerical instability at odd levels, which is probably connected to the fact that these solutions excite the marginal field. This instability means that Newton's method, which we use to find solutions, either does not converge within a reasonable number of iterations or it converges to a point that is too far from nearby even level data. This problem can manifest either only at some levels (typically 3 and 5) or at all odd levels. We can still identify these solutions based on even level data, but having lesser number of data points (whose number is low to begin with) obviously leads to lower precision of results. We denote solutions with numerical instabilities in tables \ref{tab:solutions SL 1} and \ref{tab:solutions SL 2} by the symbol I in the notes.

\begin{table}[!]
\centering
\begin{tabular}{|l|l|l|l|l|l|l|l|}\hline
$k$           & $J$               & Identification           & $E^{(\inf)}$ & $E^{(exp)}$ & $E_J$               & Notes            \\\hline
\mr{1}{*}{2}  & \mr{1}{*}{$\dt1$} & $(0,1,2.07)$             & 0.707093     & 0.707107    & \mr{1}{*}{1.000000} &                  \\\hline\hline
\mr{2}{*}{3}  & \mr{2}{*}{$\dt1$} & $(0,1,1.80)$             & 0.60969      & 0.609711    & \mr{2}{*}{0.986534} &                  \\
              &                   & $(0,1,1.63)+(0,-1,0.82)$ & 1.221        & 1.21942     &                     & I, low precision \\\hline\hline
\mr{3}{*}{4}  & \mr{1}{*}{$\dt1$} & $(0,1,1.61)$             & 0.53721      & 0.537285    & \mr{1}{*}{0.930605} &                  \\\cline{2-7}
              & \mr{2}{*}{1}      & $(0,2,1.79)$             & 0.537295     & 0.537285    & \mr{2}{*}{1.074570} & I(3,5)           \\
              &                   & $(0,2,0.48)+(0,-2,2.10)$ & 1.0755       & 1.07457     &                     & I(3)             \\\hline\hline
\mr{7}{*}{5}  & \mr{2}{*}{$\dt1$} & $(0,1,1.48)$             & 0.481507     & 0.481581    & \mr{2}{*}{0.867780} &                  \\
              &                   & $(0,1,0.63)+(0,-1,0.99)$ & 0.9642       & 0.963163    &                     & I, \SU+\SL       \\\cline{2-7}
              & \mr{5}{*}{1}      & $(0,2,1.69)$             & 0.481544     & 0.481581    & \mr{5}{*}{1.082104} & I(3)             \\
              &                   & $(\dt1,1,2.05)$          & 0.869        & 0.86778     &                     & low precision    \\
              &                   & $(0,2,0.50)+(0,-2,0.50)$ & 0.9636       & 0.963163    &                     &                  \\
              &                   & $(0,2,0.99)+(0,-2,1.82)$ & 0.9640       & 0.963163    &                     & I(3,5), \SU+\SL  \\
              &                   & $(0,2,0.49)+(0,-2,2.05)$ & 0.9619       & 0.963163    &                     & I                \\\hline\hline
\mr{18}{*}{6} & \mr{3}{*}{$\dt1$} & $(0,1,1.39)$             & 0.43736      & 0.437426    & \mr{3}{*}{0.808258} &                  \\
              &                   & $(0,1,1.53)+(0,-1,0.98)$ & 0.8756       & 0.874852    &                     & I, \SU+\SL       \\
              &                   & $(0,1,1.27)+(0,-1,0.79)$ & 0.8724       & 0.874852    &                     &                  \\\cline{2-7}
              & \mr{8}{*}{1}      & $(0,2,1.60)$             & 0.43736      & 0.437426    & \mr{8}{*}{1.056040} & I(3)             \\
              &                   & $(0,2,0.57)+(0,-2,1.74)$ & 0.874881     & 0.874852    &                     & I                \\
              &                   & $(0,2,0.56)+(0,-2,0.56)$ & 0.8750       & 0.874852    &                     &                  \\
              &                   & $(\dt1,1,1.79)$          & 0.8086       & 0.808258    &                     & low precision    \\
              &                   & $(0,2,1.70)+(0,-2,0.98)$ & 0.8752       & 0.874852    &                     & I(3,5), \SU+\SL  \\
              &                   & $(0,2,0.47)+(0,-2,2.11)$ & 0.8738       & 0.874852    &                     & I                \\
              &                   & $(0,2,2.11)+(0,-2,0.47)$ & 0.8737       & 0.874852    &                     & I                \\
              &                   & $(0,2,2.52)+(0,-2,0.40)$ & 0.870        & 0.874852    &                     & I                \\\cline{2-7}
              & \mr{7}{*}{$\dt3$} & $(0,3,1.66)$             & 0.437451     & 0.437426    & \mr{7}{*}{1.143050} & I(3,5)           \\
              &                   & $(0,3,0.56)+(0,-3,1.78)$ & 0.8745       & 0.874852    &                     & I                \\
              &                   & $(0,3,0.55)+(0,-3,0.55)$ & 0.8752       & 0.874852    &                     & I(3,5)           \\
              &                   & $(0,3,1.02)+(0,-3,0.58)$ & 0.8753       & 0.874852    &                     & I(3,5), \SU+\SL  \\
              &                   & $(0,3,0.41)+(0,-3,2.47)$ & 0.8730       & 0.874852    &                     & I                \\
              &                   & $(0,3,2.50)+(0,-3,0.40)$ & 0.867        & 0.874852    &                     & I                \\
              &                   & $(0,3,3.11)+(0,-3,0.32)$ & 0.89         & 0.874852    &                     & I                \\\hline
\end{tabular}
\caption{List of \SL Cardy solutions in models with $k\leq 8$, part 1. The table has the same format as table \ref{tab:solutions regular 1}. Identification of solutions includes the parameter $J$, the angle $\theta$ in multiples of $\frac{\pi}{k}$ and the parameter $\rho$. In case of solutions with higher degeneracy, we show only one example. These solutions are pseudo-real, so the note R means that the given solution changes from generically complex to pseudo-real. The note I denotes instability at odd levels. If there are numbers in brackets, the instability appears only at these levels, otherwise all odd levels are unstable. The notes also mark solutions that probably describe combinations of \SU and \SL branes.} \label{tab:solutions SL 1}
\end{table}

\begin{table}[!]
\centering
\begin{tabular}{|l|l|l|l|l|l|l|}\hline
$k$           & $J$               & Identification              & $E^{(\inf)}$ & $E^{(exp)}$ & $E_J$               & Notes            \\\hline
\mr{16}{*}{7} & \mr{3}{*}{$\dt1$} & $(0,1,1.33)$                & 0.40147      & 0.401534    & \mr{3}{*}{0.754638} &                  \\
              &                   & $(0,1,0.57)+(0,-1,1.75)$    & 0.808009$^*$ & 0.803069    &                     & R(10)            \\
              &                   & $(0,1,0.79)+(0,-1,1.27)$    & 0.8022       & 0.803069    &                     &                  \\\cline{2-7}
              & \mr{8}{*}{1}      & $(0,2,1.52)$                & 0.40144      & 0.401534    & \mr{8}{*}{1.016721} & I(3)             \\
              &                   & $(0,2,1.62)+(0,-2,0.62)$    & 0.8065       & 0.803069    &                     & I                \\
              &                   & $(0,2,0.60)+(0,-2,0.60)$    & 0.80305      & 0.803069    &                     &                  \\
              &                   & $(\dt1,1,1.65)$             & 0.7548       & 0.754638    &                     & low precision    \\
              &                   & $(0,2,0.99)+(0,-2,1.60)$    & 0.80324      & 0.803069    &                     & I(3), \SU+\SL    \\
              &                   & $(0,2,2.16)+(0,-2,0.46)$    & 0.8023       & 0.803069    &                     & I                \\
              &                   & $(0,2,2.20)+(0,-2,0.45)$    & 0.799        & 0.803069    &                     & I                \\
              &                   & $(0,2,0.36)+(0,-2,2.79)$    & 0.808        & 0.803069    &                     & I                \\\cline{2-7}
              & \mr{5}{*}{$\dt3$} & $(0,3,1.61)$                & 0.40150      & 0.401534    & \mr{5}{*}{1.156172} & I(3,5)           \\
              &                   & $(0,3,1.68)+(0,-3,0.60)$    & 0.8024       & 0.803069    &                     & I                \\
              &                   & $(0,3,1.69)+(0,-3,1.69)$    & 0.8033       & 0.803069    &                     & I(3,5)           \\
              &                   & $(0,3,1.01)+(0,-3,0.60)$    & 0.8033       & 0.803069    &                     & I(3,5), \SU+\SL  \\
              &                   & $(0,-3,1.91)+(\dt1,4,1.10)$ & 1.25491$^*$  & 1.156172    &                     & R(10)            \\\hline\hline
\mr{18}{*}{8} & \mr{4}{*}{$\dt1$} & $(0,1,1.29)$                & 0.37169      & 0.371748    & \mr{4}{*}{0.707107} &                  \\
              &                   & $(0,1,1.63)+(0,-1,0.62)$    & 0.746884$^*$ & 0.743496    &                     & R(8)             \\
              &                   & $(0,1,1.42)+(0,-1,0.96)$    & 0.7440       & 0.743496    &                     & I                \\
              &                   & $(0,1,0.80)+(0,-1,1.26)$    & 0.74315      & 0.743496    &                     &                  \\\cline{2-7}
              & \mr{7}{*}{1}      & $(0,2,1.45)$                & 0.37165      & 0.371748    & \mr{7}{*}{0.973249} & I(3)             \\
              &                   & $(0,2,1.56)+(0,-2,1.56)$    & 0.74339      & 0.743496    &                     &                  \\
              &                   & $(0,2,1.53)+(0,-2,0.65)$    & 0.74343      & 0.743496    &                     & I                \\
              &                   & $(\dt1,1,1.54)$             & 0.70721      & 0.707107    &                     & low precision    \\
              &                   & $(0,2,0.99)+(0,-2,1.52)$    & 0.74356      & 0.743496    &                     & I(3), \SU+\SL    \\
              &                   & $(0,2,2.17)+(0,-2,0.46)$    & 0.7430       & 0.743496    &                     & I                \\
              &                   & $(0,2,2.75)+(0,-2,2.75)$    & 0.752        & 0.743496    &                     & I, low precision \\\cline{2-7}
              & \mr{4}{*}{$\dt3$} & $(0,3,1.55)$                & 0.37169      & 0.371748    & \mr{4}{*}{1.144123} & I(3)             \\
              &                   & $(0,3,0.69)+(0,-3,1.59)$    & 0.7426       & 0.743496    &                     & I                \\
              &                   & $(0,3,0.62)+(0,-3,0.62)$    & 0.74363      & 0.743496    &                     & I(3,5)           \\
              &                   & $(0,3,0.99)+(0,-3,1.58)$    & 0.74363      & 0.743496    &                     & I(3,5), \SU+\SL  \\\cline{2-7}
              & \mr{3}{*}{2}      & $(0,4,1.58)$                & 0.37165      & 0.371748    & \mr{3}{*}{1.203002} & I(3)             \\
              &                   & $(0,4,1.62)+(0,-4,1.62)$    & 0.7437       & 0.743496    &                     & I(3,5)           \\
              &                   & $(0,4,1.01)+(0,-4,0.62)$    & 0.7437       & 0.743496    &                     & I(3), \SU+\SL    \\\hline
\end{tabular}
\caption{List of \SL Cardy solutions in models with $k\leq 8$, part 2.} \label{tab:solutions SL 2}
\end{table}

\FloatBarrier
\subsection{A complex marginal deformation solution}\label{sec:SL:marginal}
Finally, we will take a look at one curious \SL solution that appears on all backgrounds regardless of the chosen level $k$ or the boundary condition $J$. The reason is that it does not excite any \SU primaries, but only states in the Verma module of the identity field. The solution describes a specific marginal deformation by the field $J_{-1}^3 c_1 |0,0\ra$ with purely imaginary value of the marginal parameter.

In table \ref{tab:sol 2 1/2 mar data}, we show its observables for $k=2$ and $J=\dt1$, where we computed the solution to the highest level. The table shows only even levels because the solution suffers from the odd level instability that we described before. The solution changes quickly with increasing level, which together with the instability means that extrapolations of invariants are not very precise, see table \ref{tab:sol 2 1/2 mar extrapolation}.

The energy of the solution approaches 1 and the only boundary state with this $g$-function is a $\frac{1}{2}$-brane. This suggests that the solution describes marginal deformations with imaginary value of the marginal parameter. However, it is not a standard marginal solution because the marginal parameter quickly changes with level, see table \ref{tab:sol 2 1/2 mar data}. Ellwood invariants are consistent with a $\frac{1}{2}$-brane and they tell us that $\theta=0$, which agrees with (\ref{theta 1}) for $J_i=J_f$. The second parameter $\rho$ can be determined only using the invariant $E_{1,1}$, which is the only well-behaved invariant that depends on this parameter, and we get $\rho\approx 5.9$. However, this number has a large error because $E_{1,1}$ has poor precision. The parameter $\rho$ changes if we consider a different level $k$, see the values in table \ref{tab:sol mar rho}. It decreases as we go to higher $k$, but we have not managed to fit it by any simple function, possibly due to errors of the results. On the other hand, the parameter $\rho$ does not seem to change when we fix $k$ and vary the boundary condition $J$, its values obtained for different $J$ are very close to each other.

Interestingly, the solution has 'accidental' symmetries similar to \SU Cardy solutions described in subsection \ref{sec:regular:other}, but the symmetries themselves are different:
\begin{eqnarray}
&&E_{\dt1,\dt1} = E_{\dt1,-\dt1}=0, \\
&&E_{1,0} = E_{0,0}, \\
&&E_{1,1}=-J_{+-},  \\
&&E_{1,-1}=-J_{-+},
\end{eqnarray}
The traditional marginal solution \cite{MarginalSen}\cite{MarginalKMOSY}\cite{MarginalTachyonKM} is also a solution in the \SUk WZW model OSFT, but it does not have any accidental symmetries for generic value of the marginal parameter, so our solution is clearly exceptional. It most likely represents a special point in the family of marginal solutions (which changes with level) where the equation of motion for the marginal field, which is omitted in the traditional marginal approach, is satisfied. That probably leads to the accidental symmetries, although the exact mechanism is not clear to us.

\begin{table}[t]
\centering
\begin{tabular}{|l|lllllll|}\hline
Level    & Energy  & $E_{0,0}$ & $\ps E_{1,1}$ & $\ps E_{1,-1}$ & $\ps J_{33}$  & $\ps \Delta _S$ & $\ps \lambda  $ \\\hline
2        & 1.01246 & 0.997915  & $   -1.08383$ & $   -5.238689$ & $\ps 5.32460$ & $   -0.0169001$ & $   -0.330633i$ \\
4        & 1.00931 & 0.993375  & $   -8.09888$ & $   -0.142909$ & $   -1.53685$ & $   -0.0118150$ & $   -0.404850i$ \\
6        & 1.00905 & 0.988201  & $   -10.2092$ & $   -0.134406$ & $\ps 2.51075$ & $   -0.0095241$ & $   -0.485320i$ \\
8        & 1.00905 & 0.986204  & $   -12.8326$ & $   -0.156754$ & $\ps 1.35414$ & $   -0.0084797$ & $   -0.560089i$ \\
10       & 1.00896 & 0.984836  & $   -15.2304$ & $   -0.079354$ & $\ps 0.70920$ & $   -0.0078490$ & $   -0.625607i$ \\
12       & 1.00876 & 0.984682  & $   -17.4956$ & $   -0.113267$ & $\ps 2.11236$ & $   -0.0073784$ & $   -0.681621i$ \\
14       & 1.00848 & 0.984716  & $   -19.5194$ & $   -0.068453$ & $\ps 0.52919$ & $   -0.0069804$ & $   -0.729196i$ \\\hline
\end{tabular}
\caption{Independent observables and the marginal parameter $\lambda$ of the special complex marginal solution in the $k=2$ model with $J=\frac{1}{2}$ boundary conditions. We show only even level data because of the odd level instability.} \label{tab:sol 2 1/2 mar data}
\vspace{7mm}
\begin{tabular}{|c|llll|}\cline{1-5}
          & $\ps $Energy      & $\ps E_{0,0}     $ & $\ps \Delta_S    $ & $\ps \lambda$ \\ \cline{1-5}
$\inf$    & $\ps 0.9997     $ & $\ps 0.986       $ & $   -0.0005      $ & $   -1.21   $ \\
$\sigma$  & $\ps -          $ & $\ps 0.004       $ & $\ps -           $ & $\ps -      $ \\
Exp.      & $\ps 1          $ & $\ps 1           $ & $\ps 0           $ & $\ps -      $ \\\cline{1-5}
          & $\ps J_{+-}     $ & $\ps J_{-+}      $ & $\ps J_{33}      $ & \mc{1}{|c}{}  \\\cline{1-4}
$\inf$    & $\ps 34.5       $ & $\ps 0.02        $ & $\ps 2.8         $ & \mc{1}{|c}{}  \\
$\sigma$  & $\ps 4.4        $ & $\ps 0.09        $ & $\ps 0.7         $ & \mc{1}{|c}{}  \\
Exp.      & $\ps 34.5       $ & $\ps 0.03        $ & $\ps 1           $ & \mc{1}{|c}{}  \\\cline{1-4}
          & $\ps E_{1/2,1/2}$ & $\ps E_{1/2,-1/2}$ & \mc{2}{|c}{}                       \\\cline{1-3}
$\inf$    & $\ps 0          $ & $\ps 0           $ & \mc{2}{|c}{}                       \\
$\sigma$  & $\ps -          $ & $\ps -           $ & \mc{2}{|c}{}                       \\
Exp.      & $\ps 0          $ & $\ps 0           $ & \mc{2}{|c}{}                       \\\cline{1-4}
          & $\ps E_{1,1}    $ & $\ps E_{1,0}     $ & $\ps E_{1,-1}    $ & \mc{1}{|c}{}  \\\cline{1-4}
$\inf$    & $   -34.5       $ & $\ps 0.986       $ & $   -0.02        $ & \mc{1}{|c}{}  \\
$\sigma$  & $\ps 4.4        $ & $\ps 0.004       $ & $\ps 0.09        $ & \mc{1}{|c}{}  \\
Exp.      & $   -34.5       $ & $\ps 1           $ & $   -0.03        $ & \mc{1}{|c}{}  \\\cline{1-4}
\end{tabular}
\caption{Extrapolations of observables and the marginal parameter of the special complex marginal solution in the $k=2$ model with $J=\frac{1}{2}$ boundary conditions. The expected values are based on $\theta=0$ and $\rho\approx 5.9$. We do not have good error estimates for some observables.} \label{tab:sol 2 1/2 mar extrapolation}
\end{table}

\begin{table}[!b]
\centering
\begin{tabular}{|l|ccccccc|}\hline
$k$    & 2    & 3    & 4    & 5    & 6    & 7    & 8   \\\hline
$\rho$ & 5.88 & 3.83 & 2.87 & 2.59 & 2.40 & 2.27 & 2.18\\\hline
\end{tabular}
\caption{Values of the parameter $\rho$ of the special complex marginal solution for different $k$.} \label{tab:sol mar rho}
\end{table}

\FloatBarrier

\section{Exotic solutions}\label{sec:exotic}
In the last section dedicated to solutions, we will discuss solutions which seem to be physical, but which we failed to identify as \SU or \SL Cardy boundary states. We think that most of them describe symmetry-breaking boundary states, but it is possible that our identification failed in some cases because of low precision or numerical instabilities. Among the solutions, there is one group of real solutions with similar properties, which describe the B-brane boundary states from \cite{WZW B-branes}. We will discuss this group of solutions first, then we will discuss the $k=4$ background with $J=1$ boundary conditions, which has some special properties, then we will present two unidentified real exotic solutions for $k=8$ and finally we will provide a survey of all exotic solutions that we have found.

\subsection{B-brane solutions}\label{sec:exotic:B-branes}
During our analysis of exotic solutions, we noticed that there is one group of real solutions with similar properties, which we later identified as the so-called B-branes, whose properties are reviewed in section \ref{sec:WZW:B-branes}.
In order to compare our numerical results with B-brane boundary states (\ref{B-brane BS}), we need to know what are the expected values of Ellwood invariants. We have not worked out the exact map between the \SUk WZW model Hilbert space and the parafermion times free boson Hilbert space, but we can fortunately extract the information we need just based on conformal weights and free boson momentum.

We have found only solutions representing B-branes with $J=0$ because others have too high $g$-function to be seen in the numerical approach, so we will restrict our attention to this case.
The $g$-function of B-branes is multiplied by $\sqrt{k}$ compared to the usual Cardy branes with the same label $J$ and therefore we expect that a $J=0$ B-brane solution will have
\begin{equation}\label{B-brane pred 1}
E_{0,0}^{exp}=\sqrt{k}B_0^{\ 0}.
\end{equation}
Next, we notice that B-brane boundary states involve only $n=0,k$ free boson representations, which means that most invariants are equal to zero because they correspond to other momenta. We find that the trivial invariants are
\begin{equation}
E_{j,m}^{exp}=0,\quad m\neq 0,\pm\frac{k}{2}.
\end{equation}
The exceptions are invariants with $m=0$, which are equal to
\begin{equation}\label{B-brane pred 3}
E_{j,0}^{exp}=\sqrt{k}B_0^{\ j},
\end{equation}
and invariants with $j=\pm m=\frac{k}{2}$:
\begin{equation}
|E_{k/2,\pm k/2}^{exp}|=\sqrt{k}B_0^{\ 0}.
\end{equation}
This equation is written with an absolute value because determining the sign (which is related to $\eta$) would require knowledge of the exact correspondence between the two Hilbert spaces.
Finally, $J_{ab}$ invariants have expected values
\begin{eqnarray}
J_{+-}^{exp}&=&J_{-+}^{exp}=0,\\
J_{33}^{exp}&=&\sqrt{k}B_0^{\ 0}.\label{B-brane pred 6}
\end{eqnarray}
Overall, invariants of B-branes are much simpler than for \SU Cardy branes. Most invariants vanish and the rest are real numbers proportional to elements of the matrix of boundary state coefficients.

Now that we know what results to expect, we can move to actual solutions. We have found B-brane solutions in models with $k\geq 3$ with the exception of $k=4$. Furthermore, the background needs to have high enough $g$-function, so most of B-brane solutions appear on backgrounds with the highest $g$-function for the given $k$. Boundary states (\ref{B-brane BS}) should be described by real solutions, but some of them have complex seeds and they become real at a higher level. We also find some pseudo-real solutions with B-brane properties. We will discuss examples of B-brane solutions for $k=3$ and $k=6$ in more detail.

\begin{table}[!]
\centering
\begin{tabular}{|l|llllll|}
\mc{7}{c}{\SUk WZW model data} \\\hline
Level    & $\ps $Energy    & $\ps E_{0,0}  $ & $\ps E_{1/2,1/2}$ & $\ps E_{1,0}  $ & $\ps E_{3/2,1/2}$ & $\ps \Delta_S   $ \\\hline
2        & $\ps 1.16839  $ & $\ps 0.98818  $ & $   -0.276857   $ & $\ps 2.88659  $ & $   -2.674373   $ & $   -0.0319044  $ \\
         & $   -0.05599 i$ & $   +0.11488 i$ & $   +0.306363 i $ & $   -0.40927 i$ & $   +1.895532 i $ & $   +0.0217899 i$ \\
4        & $\ps 1.09180  $ & $\ps 1.00282  $ & $   -0.137331   $ & $\ps 1.32541  $ & $\ps 0.217832   $ & $   -0.0136018  $ \\
6        & $\ps 1.07642  $ & $\ps 1.02430  $ & $   -0.0759386  $ & $\ps 1.58556  $ & $   -0.676351   $ & $   -0.0075041  $ \\
8        & $\ps 1.07037  $ & $\ps 1.03273  $ & $   -0.049461   $ & $\ps 1.40505  $ & $\ps 0.017113   $ & $   -0.0052371  $ \\
10       & $\ps 1.06710  $ & $\ps 1.03719  $ & $   -0.0386589  $ & $\ps 1.46741  $ & $   -0.329571   $ & $   -0.0040362  $ \\
12       & $\ps 1.06506  $ & $\ps 1.04022  $ & $   -0.0300022  $ & $\ps 1.39810  $ & $   -0.024171   $ & $   -0.0032897  $ \\\hline
\mc{7}{c}{ }                                                                                                               \\[-13pt]
\mc{7}{c}{Extension by minimal model data}                                                                                      \\\hline
14       & $\ps 1.06366  $ & $\ps 1.04227  $ & $   -0.0256691  $ & $\ps 1.42623  $ & $   -0.207926   $ & $   -0.0027795  $ \\
16       & $\ps 1.06264  $ & $\ps 1.04389  $ & $   -0.0214132  $ & $\ps 1.38967  $ & $   -0.035069   $ & $   -0.0024084  $ \\
18       & $\ps 1.06187  $ & $\ps 1.04509  $ & $   -0.0191116  $ & $\ps 1.40559  $ & $   -0.148645   $ & $   -0.0021259  $ \\
20       & $\ps 1.06126  $ & $\ps 1.04610  $ & $   -0.0165916  $ & $\ps 1.38302  $ & $   -0.037352   $ & $   -0.0019035  $ \\\hline
$\inf$   & $\ps 1.05606  $ & $\ps 1.05546  $ & $\ps 0.0005     $ & $\ps 1.3442   $ & $\ps 0.002      $ & $   -0.000007   $ \\
$\sigma$ & $\ps -        $ & $\ps 0.00004  $ & $\ps 0.0004     $ & $\ps 0.0007   $ & $\ps 0.006      $ & $\ps -          $ \\\hline
Exp.     & $\ps 1.05605  $ & $\ps 1.05605  $ & $\ps 0          $ & $\ps 1.34332  $ & $\ps 0          $ & $\ps 0          $ \\\hline
\end{tabular}
\caption{Selected invariants of the B-brane solution in the $k=3$ model with $J=\frac{1}{2}$ boundary conditions. The first part of the table includes data found in the \SUk WZW model, while the second part shows higher level predictions based on a dual solution in the $m=5$ minimal model. We show only data from even levels because odd level data are the same.}
\label{tab:sol 3 1/2 exotic}
\vspace{5mm}
\begin{tabular}{|c|llll|}\cline{1-4}
          & $\ps $Energy      & $\ps E_{0,0}     $ & $\ps \Delta_S    $ & \mc{1}{|c}{}       \\ \cline{1-4}
$\inf$    & $\ps 1.05624    $ & $\ps 1.054       $ & $   -0.00014     $ & \mc{1}{|c}{}       \\
$\sigma$  & $\ps -          $ & $\ps 0.004       $ & $\ps -           $ & \mc{1}{|c}{}       \\
Exp.      & $\ps 1.05605    $ & $\ps 1.05605     $ & $\ps 0           $ & \mc{1}{|c}{}       \\\cline{1-4}
          & $\ps J_{+-}     $ & $\ps J_{-+}      $ & $\ps J_{33}      $ & \mc{1}{|c}{}       \\\cline{1-4}
$\inf$    & $\ps 0.06       $ & $\ps 0.06        $ & $\ps 1.054       $ & \mc{1}{|c}{}       \\
$\sigma$  & $\ps 0.18       $ & $\ps 0.18        $ & $\ps 0.004       $ & \mc{1}{|c}{}       \\
Exp.      & $\ps 0          $ & $\ps 0           $ & $\ps 1.05605     $ & \mc{1}{|c}{}       \\\cline{1-4}
          & $\ps E_{1/2,1/2}$ & $\ps E_{1/2,-1/2}$ & \mc{2}{|c}{}                            \\\cline{1-3}
$\inf$    & $\ps 0.006      $ & $   -0.006       $ & \mc{2}{|c}{}                            \\
$\sigma$  & $\ps 0.016      $ & $\ps 0.016       $ & \mc{2}{|c}{}                            \\
Exp.      & $\ps 0          $ & $\ps 0           $ & \mc{2}{|c}{}                            \\\cline{1-4}
          & $\ps E_{1,1}    $ & $\ps E_{1,0}     $ & $\ps E_{1,-1}    $ & \mc{1}{|c}{}       \\\cline{1-4}
$\inf$    & $   -0.006      $ & $\ps 1.31        $ & $   -0.006       $ & \mc{1}{|c}{}       \\
$\sigma$  & $\ps 0.016      $ & $\ps 0.03        $ & $\ps 0.016       $ & \mc{1}{|c}{}       \\
Exp.      & $\ps 0          $ & $\ps 1.34332     $ & $\ps 0           $ & \mc{1}{|c}{}       \\\cline{1-5}
          & $\ps E_{3/2,3/2}$ & $\ps E_{3/2,1/2} $ & $\ps E_{3/2,-1/2}$ & $\ps E_{3/2,-3/2}$ \\\cline{1-5}
$\inf$    & $   -1.054      $ & $\ps 0.06        $ & $   -0.06        $ & $\ps 1.054       $ \\
$\sigma$  & $\ps 0.004      $ & $\ps 0.18        $ & $\ps 0.18        $ & $\ps 0.004       $ \\
Exp.      & $   -1.05605    $ & $\ps 0           $ & $\ps 0           $ & $\ps 1.05605     $ \\\cline{1-5}
\end{tabular}
\caption{Extrapolations (based on levels 4 to 12) of the B-brane solution in the $k=3$ model with $J=\frac{1}{2}$ boundary conditions.} \label{tab:sol 3 1/2 exotic extrapolation}
\end{table}

The first B-brane solution can be found on the $k=3$ background with $J=\dt 1$ boundary conditions. Interestingly, the solution has degeneracy just one, so we have found only one of the two B-brane boundary states described by equation (\ref{B-brane BS}). Its data up to level 12 and infinite level extrapolations are given in tables \ref{tab:sol 3 1/2 exotic} and \ref{tab:sol 3 1/2 exotic extrapolation}. The basic identification can be done using the energy, which is approximately 1.056 and which does not matches any combination of \SU Cardy boundary states given in table \ref{tab:boundary states}. The energy of the B-brane solution is higher than the $g$-function of the initial $\dt1$-brane, so it is not surprising that the seed solution at level 2 is complex. This aspect of the solution reminds us of other positive energy solutions (see, for example, positive energy lumps or one of the Ising model solutions in \cite{KudrnaThesis}), but our solution fortunately becomes real much more quickly, the imaginary part disappears already at level 4. Therefore we have enough real data points to do a decent analysis.

In many aspects, the solution is similar to the 0-brane solution on the same background. The data in table \ref{tab:sol 3 1/2 exotic extrapolation} show that its invariants satisfy the reality conditions and that they have the same symmetries (\ref{symmetries su 1})-(\ref{symmetries su 3}).
When we take a closer look at their extrapolations, we observe that there is a good agreement with the predictions (\ref{B-brane pred 1})-(\ref{B-brane pred 6}), which makes the identification of the solution definitive. In particular, notice that a high number of invariants is close to zero, which is a characteristic of B-brane solutions.
The precision of extrapolations is somewhat lesser than expected (because we have only 5 data points from levels 4 to 12), but it can be improved if we realize that the solution has a dual in the $m=5$ minimal model like the 0-brane solution. This dual solution describes (3,5)-brane going to (1,2)-brane\footnote{It would be interesting to see the minimal model solution from the  perspective of the parafermion theory. Based on its energy, it seems likely that it describes a B-brane in the parafermion model.}.
Using the dual solution, we can predict behavior of its invariants up to level 20, see the second part of table \ref{tab:sol 3 1/2 exotic}. The table also includes extrapolations using these extra data and we observe that the precision improves approximately by one order compared to table \ref{tab:sol 3 1/2 exotic extrapolation}.

Next, we move to the $k=6$ background with $J=3/2$ boundary conditions. In this case, we find not one, but two different B-brane solutions (solutions number 18 and 3061 from table \ref{tab:solutions exotic 1}). Due to large amount of data, we decided to show only extrapolations of their invariants in table \ref{tab:sol 6 3/2 B-brane extrapolation}.

The first solution is not much different from the one at $k=3$. It is somewhat nicer because it has lower energy than the background and therefore it is real at all levels. That allows us to extrapolate its energy with better precision, even though we evaluated the solution only up to level 10. When it comes to Ellwood invariants, they have again the 'accidental' symmetries (\ref{symmetries su 1})-(\ref{symmetries su 3}), which are typical for Cardy solutions. Some higher invariants have large errors due to their conformal weights, but most of them agree very well with the expected values, so the solution is consistent with our predictions. We notice that the percentage of invariants going to zero is higher than for the solution above and all invariants with half-integer labels vanish identically, so these invariants were removed from the table to reduce its size.

The second solution is different. Although most of its invariants have similar values as for the first solution, we notice that they are no longer symmetric and therefore the solution is not real, but only pseudo-real. Most of the invariants go to zero and invariants with $m=0$ match the predictions (\ref{B-brane pred 3}), so the key difference lies in the invariants $E_{3,3}\approx -9$ and $E_{3,-3}\approx -0.3$. These invariants represent the the main asymmetry of the solution and they are the only ones that do not match the predictions. However, we think that they should satisfy
\begin{equation}
E_{3,3}=\rho^6 E_{0,0},\quad E_{3,-3}=\rho^{-6} E_{0,0}
\end{equation}
for some constant $\rho$, although exact verification of this relation is not possible due to large relative error of $E_{3,-3}$. This relation holds for \SL boundary states and therefore we think that the two B-brane solutions are related in the same way as ordinary \SU and \SL solutions. B-brane boundary states have several moduli, so the pseudo-real solution should be possible to obtain by marginal deformations of the real solution using a purely imaginary value of the marginal field related to $J_{-1}^3$. This solution has degeneracy two and the two solutions differ  by exchange $E_{j,m}\leftrightarrow E_{j,-m}$.

The remaining B-brane solutions that we found appear in the survey of exotic solutions in tables \ref{tab:solutions exotic 1} and \ref{tab:solutions exotic 2}. They have analogous properties as the examples above, so there is no need to discuss them in detail.

\begin{table}
\begin{tabular}{|c|lllllll|}
\mc{4}{c}{Real solution}  & \mc{4}{c}{}                                                                                       \\\cline{1-4}
          & $\ps $Energy  & $\ps E_{0,0}$ & $\ps \Delta_S$ & \mc{4}{|c}{}                                                     \\\cline{1-4}
$\inf$    & $\ps 1.07149$ & $\ps 1.0713 $ & $   -0.000036$ & \mc{4}{|c}{}                                                     \\
$\sigma$  & $\ps 0.00002$ & $\ps 0.0011 $ & $\ps 0.000003$ & \mc{4}{|c}{}                                                     \\
Exp.      & $\ps 1.07147$ & $\ps 1.07147$ & $\ps 0       $ & \mc{4}{|c}{}                                                     \\\cline{1-4}
          & $\ps J_{+-} $ & $\ps J_{-+} $ & $\ps J_{33}  $ & \mc{4}{|c}{}                                                     \\\cline{1-4}
$\inf$    & $\ps 0.00   $ & $\ps 0.00   $ & $\ps 1.0713  $ & \mc{4}{|c}{}                                                     \\
$\sigma$  & $\ps 0.05   $ & $\ps 0.05   $ & $\ps 0.0011  $ & \mc{4}{|c}{}                                                     \\
Exp.      & $\ps 0      $ & $\ps 0      $ & $\ps 1.07147 $ & \mc{4}{|c}{}                                                     \\\cline{1-4}
          & $\ps E_{1,1}$ & $\ps E_{1,0}$ & $\ps E_{1,-1}$ & \mc{4}{|c}{}                                                     \\\cline{1-4}
$\inf$    & $\ps 0.0003 $ & $\ps 1.667  $ & $\ps 0.0003  $ & \mc{4}{|c}{}                                                     \\
$\sigma$  & $\ps 0.0004 $ & $\ps 0.005  $ & $\ps 0.0004  $ & \mc{4}{|c}{}                                                     \\
Exp.      & $\ps 0      $ & $\ps 1.66482$ & $\ps 0       $ & \mc{4}{|c}{}                                                     \\\cline{1-6}
          & $\ps E_{2,2}$ & $\ps E_{2,1}$ & $\ps E_{2,0} $ & $\ps E_{2,-1}$ & $\ps E_{2,-2}$ & \mc{2}{|c}{}                   \\\cline{1-6}
$\inf$    & $   -0.0003 $ & $   -0.02   $ & $\ps 1.66    $ & $   -0.02    $ & $   -0.0003  $ & \mc{2}{|c}{}                   \\
$\sigma$  & $\ps 0.0004 $ & $\ps 0.06   $ & $\ps 0.19    $ & $\ps 0.06    $ & $\ps 0.0004  $ & \mc{2}{|c}{}                   \\
Exp.      & $\ps 0      $ & $\ps 0      $ & $\ps 1.66482 $ & $\ps 0       $ & $\ps 0       $ & \mc{2}{|c}{}                   \\\cline{1-8}
          & $\ps E_{3,3}$ & $\ps E_{3,2}$ & $\ps E_{3,1} $ & $\ps E_{3,0} $ & $\ps E_{3,-1}$ & $\ps E_{3,-2}$ & $\ps E_{3,-3}$\\\cline{1-8}
$\inf$    & $   -1.0713 $ & $\ps 0.00   $ & $\ps 0.0     $ & $\ps 0.9     $ & $\ps 0.0     $ & $\ps 0.00    $ & $   -1.0713  $\\
$\sigma$  & $\ps 0.0011 $ & $\ps 0.05   $ & $\ps 0.7     $ & $\ps 0.8     $ & $\ps 0.7     $ & $\ps 0.05    $ & $\ps 0.0011  $\\
Exp.      & $   -1.07147$ & $\ps 0      $ & $\ps 0       $ & $\ps 1.07147 $ & $\ps 0       $ & $\ps 0       $ & $   -1.07147 $\\\cline{1-8}
\mc{8}{c}{ }                                                                                                                  \\[-9pt]
\mc{4}{c}{Pseudo-real solution} & \mc{4}{c}{}                                                                                 \\\cline{1-4}
          & $\ps $Energy  & $\ps E_{0,0}$ & $\ps \Delta_S$ & \mc{4}{|c}{}                                                     \\\cline{1-4}
$\inf$    & $\ps 1.07149$ & $\ps 1.0718 $ & $\ps 0.000024$ & \mc{4}{|c}{}                                                     \\
$\sigma$  & $\ps 0.00006$ & $\ps 0.0003 $ & $\ps 0.000001$ & \mc{4}{|c}{}                                                     \\
Exp.      & $\ps 1.07147$ & $\ps 1.07147$ & $\ps 0       $ & \mc{4}{|c}{}                                                     \\\cline{1-4}
          & $\ps J_{+-} $ & $\ps J_{-+} $ & $\ps J_{33}  $ & \mc{4}{|c}{}                                                     \\\cline{1-4}
$\inf$    & $\ps 0.04   $ & $   -0.02   $ & $\ps 1.09    $ & \mc{4}{|c}{}                                                     \\
$\sigma$  & $\ps 0.11   $ & $\ps 0.07   $ & $\ps 0.52    $ & \mc{4}{|c}{}                                                     \\
Exp.      & $\ps 0      $ & $\ps 0      $ & $\ps 1.07147 $ & \mc{4}{|c}{}                                                     \\\cline{1-4}
          & $\ps E_{1,1}$ & $\ps E_{1,0}$ & $\ps E_{1,-1}$ & \mc{4}{|c}{}                                                     \\\cline{1-4}
$\inf$    & $   -0.01   $ & $\ps 1.665  $ & $   -0.001   $ & \mc{4}{|c}{}                                                     \\
$\sigma$  & $\ps 0.02   $ & $\ps 0.009  $ & $\ps 0.014   $ & \mc{4}{|c}{}                                                     \\
Exp.      & $\ps 0      $ & $\ps 1.66482$ & $\ps 0       $ & \mc{4}{|c}{}                                                     \\\cline{1-6}
          & $\ps E_{2,2}$ & $\ps E_{2,1}$ & $\ps E_{2,0} $ & $\ps E_{2,-1}$ & $\ps E_{2,-2}$ & \mc{2}{|c}{}                   \\\cline{1-6}
$\inf$    & $\ps 0.01   $ & $   -0.05   $ & $\ps 1.6     $ & $   -0.06    $ & $\ps 0.05    $ & \mc{2}{|c}{}                   \\
$\sigma$  & $\ps 0.05   $ & $\ps 0.09   $ & $\ps 0.3     $ & $\ps 0.16    $ & $\ps 0.08    $ & \mc{2}{|c}{}                   \\
Exp.      & $\ps 0      $ & $\ps 0      $ & $\ps 1.66482 $ & $\ps 0       $ & $\ps 0       $ & \mc{2}{|c}{}                   \\\cline{1-8}
          & $\ps E_{3,3}$ & $\ps E_{3,2}$ & $\ps E_{3,1} $ & $\ps E_{3,0} $ & $\ps E_{3,-1}$ & $\ps E_{3,-2}$ & $\ps E_{3,-3}$\\\cline{1-8}
$\inf$    & $   -9.2    $ & $\ps 0.04   $ & $\ps 0.5     $ & $\ps 1.0     $ & $   -0.2     $ & $   -0.1     $ & $   -0.3     $\\
$\sigma$  & $\ps 0.3    $ & $\ps 0.20   $ & $\ps 1.2     $ & $\ps 1.2     $ & $\ps 1.0     $ & $\ps 0.3     $ & $\ps 0.3     $\\
Exp.      & $\ps ?      $ & $\ps 0      $ & $\ps 0       $ & $\ps 1.07147 $ & $\ps 0       $ & $\ps 0       $ & $\ps ?       $\\\cline{1-8}
\end{tabular}
\caption{Extrapolations of observables of two B-brane solutions in the $k=6$ model with $J=\frac{3}{2}$ boundary conditions. All invariants $E_{j,m}$ with half-integer $j$ are identically equal to zero.}
\label{tab:sol 6 3/2 B-brane extrapolation}
\end{table}

\FloatBarrier
\subsection{$k=4$ solutions}\label{sec:exotic:k=4}
Next, we consider one special setting, the $k=4$ model with $J=1$. In addition to the \SU currents, the boundary spectrum of this model includes additional marginal fields given by $\phi_{2,m}$, which have weight 1 for this $k$. This setting therefore allows more types of marginal deformations, which lead to an 8-parametric family of boundary states, which presumably includes the first type of symmetry-breaking boundary states from \cite{WZW Symmetry-breaking}. In this paper, we impose the condition (\ref{J03 psi}), so only one of the new marginal fields survives and our string field includes two marginal states $J_{-1}^3 c_1 |0,0\ra$ and $c_1 |2,0\ra$. By solving the OSFT equations, we found few solutions (one real and few pseudo-real) with energy close to 1-brane $g$-function. The real solution is more interesting and we will discuss it first (it is solution number 745 from table \ref{tab:solutions exotic 1}).

Extrapolations of its invariants are given in table \ref{tab:sol 4 1 exotic extrapolation} and we observe that its properties (including 'accidental' symmetries) are similar to \SU solutions. However, its identification is difficult. Since it has energy similar to a 1-brane, we will first compare the extrapolations to expected values for 1-brane with $\theta=0$, which seems to be the closest match. We observe that some invariants match the 1-brane ($E_{0,0}$, $E_{2,2}$, $E_{2,0}$ and invariants with half-integer labels), but some are clearly different ($E_{1,1}$, $E_{1,0}$, $E_{2,1}$). Therefore the solution does not represent a 1-brane but some symmetry-breaking boundary state.
The coefficient of $c_1 |2,0\ra$ is nonzero while the coefficient of $J_{-1}^3 c_1 |0,0\ra$ disappears, which suggests that the solution represents a marginal deformation of the initial 1-brane by the additional marginal operator.

Let us check whether it belongs to the first group of boundary states from \cite{WZW Symmetry-breaking}. We do not have an exact map between the two Hilbert spaces, so we cannot do a full comparison, but some operators can be uniquely matched just based on their weights and the decomposition (\ref{character decomposition}), so we will at least compare the related invariants. Let us have a look at the boundary state (3.4) from \cite{WZW Symmetry-breaking} for $\kappa=2$. The $g$-function of a $J_\alpha=0$ boundary state\footnote{We had to add a factor $\sqrt{2}$ to the normalization to match the claim that boundary states for certain parameters match superpositions of $\kappa$ Cardy branes.} is $\mathcal{N} B_{(0,0)}^{PF(0,0)} D^0_{0,0}(g)=1.07457$, which matches the energy of our solution. The other label $n_\alpha$ affects only some signs, so we take $n_\alpha=0$ for simplicity. Next, we notice that the boundary state includes only integer $J$ primaries, so all invariants with half-integer labels should disappear. That also agrees with our results. However, the invariant $E_{1,0}$ should be $\mathcal{N} B_{(0,0)}^{PF(1,0)}D^0_{0,0}(g)=1.51967$, which does not match our results. Table \ref{tab:sol 4 1 exotic extrapolation} includes expected values for few more invariants whose values can be easily extracted from the boundary state. They depend on the parameter $a$ from the parameterization of group elements (\ref{SU2 group element abcd}), but there is no value of the parameter that would lead to agreement with our results. So we conclude that this solution does not represent a boundary state from the family described in \cite{WZW Symmetry-breaking}.

Next, let us briefly go back to other solutions in this model. We notice that we can make a complex linear combination of the two marginal operators so that the resulting operator has trivial OPE with itself (like $\del X^+$). Deformations by a such operator are exactly marginal and we have indeed found several solutions that describe exactly marginal deformations (table \ref{tab:solutions exotic 1} includes one example, solution number 697). They can be easily recognized because their OSFT action exactly disappears. However, they have big problems with numerical stability, which are worse than for \SL solutions and which seem to be caused by the trivial (constant) potential for the marginal field. Therefore it would be necessary to investigate them using the traditional marginal approach \cite{MarginalSen}\cite{MarginalKMOSY}\cite{MarginalTachyonKM}\cite{KudrnaThesis}.

\begin{table}[!t]
\centering
\begin{tabular}{|c|lllll|}\cline{1-4}
          & $\ps $Energy       & $\ps E_{0,0}      $ & $\ps \Delta_S     $ & \mc{2}{|c}{}                               \\ \cline{1-4}
$\inf$    & $\ps 1.074573    $ & $\ps 1.0748       $ & $   -0.000007     $ & \mc{2}{|c}{}                               \\
$\sigma$  & $\ps 0.000003    $ & $\ps 0.0006       $ & $\ps 0.000007     $ & \mc{2}{|c}{}                               \\
Exp. 1    & $\ps 1.074570    $ & $\ps 1.07457      $ & $\ps 0            $ & \mc{2}{|c}{}                               \\
Exp. 2    & $\ps 1.074570    $ & $\ps 1.07457      $ & $\ps 0            $ & \mc{2}{|c}{}                               \\\cline{1-4}
          & $\ps J_{+-}      $ & $\ps J_{-+}       $ & $\ps J_{33}       $ & \mc{2}{|c}{}                               \\\cline{1-4}
$\inf$    & $\ps 0.711       $ & $\ps 0.711        $ & $\ps 1.0748       $ & \mc{2}{|c}{}                               \\
$\sigma$  & $\ps 0.005       $ & $\ps 0.005        $ & $\ps 0.0006       $ & \mc{2}{|c}{}                               \\
Exp. 1    & $\ps 1.07457     $ & $\ps 1.07457      $ & $\ps 1.07457      $ & \mc{2}{|c}{}                               \\
Exp. 2    & $\ps ?           $ & $\ps ?            $ & $\ps 1.07457      $ & \mc{2}{|c}{}                               \\\cline{1-4}
          & $\ps E_{1/2,1/2} $ & $\ps E_{1/2,-1/2} $ & \mc{3}{|c}{}                                                     \\\cline{1-3}
$\inf$    & $   -0.013       $ & $\ps 0.013        $ & \mc{3}{|c}{}                                                     \\
$\sigma$  & $\ps 0.001       $ & $\ps 0.001        $ & \mc{3}{|c}{}                                                     \\
Exp. 1    & $\ps 0           $ & $\ps 0            $ & \mc{3}{|c}{}                                                     \\
Exp. 2    & $\ps 0           $ & $\ps 0            $ & \mc{3}{|c}{}                                                     \\\cline{1-4}
          & $\ps E_{1,1}     $ & $\ps E_{1,0}      $ & $\ps E_{1,-1}     $ & \mc{2}{|c}{}                               \\\cline{1-4}
$\inf$    & $   -0.367       $ & $\ps 1.342        $ & $   -0.367        $ & \mc{2}{|c}{}                               \\
$\sigma$  & $\ps 0.004       $ & $\ps 0.004        $ & $\ps 0.004        $ & \mc{2}{|c}{}                               \\
Exp. 1    & $   -0.759836    $ & $\ps 0.759836     $ & $   -0.759836     $ & \mc{2}{|c}{}                               \\
Exp. 2    & $\ps 1.51967 a   $ & $\ps 1.51967      $ & $\ps 1.51967 a^*  $ & \mc{2}{|c}{}                               \\\cline{1-5}
          & $\ps E_{3/2,3/2} $ & $\ps E_{3/2,1/2}  $ & $\ps E_{3/2,-1/2} $ & $\ps E_{3/2,-3/2} $ & \mc{1}{|c}{}         \\\cline{1-5}
$\inf$    & $   -0.013       $ & $   -0.047        $ & $\ps 0.047        $ & $\ps 0.013        $ & \mc{1}{|c}{}         \\
$\sigma$  & $\ps 0.001       $ & $\ps 0.006        $ & $\ps 0.006        $ & $\ps 0.001        $ & \mc{1}{|c}{}         \\
Exp. 1    & $\ps 0           $ & $\ps 0            $ & $\ps 0            $ & $\ps 0            $ & \mc{1}{|c}{}         \\
Exp. 2    & $\ps 0           $ & $\ps 0            $ & $\ps 0            $ & $\ps 0            $ & \mc{1}{|c}{}         \\\cline{1-6}
          & $\ps E_{2,2}     $ & $\ps E_{2,1}      $ & $\ps E_{2,0}      $ & $\ps E_{2,-1}     $ & $\ps E_{2,-2} $      \\\cline{1-6}
$\inf$    & $\ps 1.0748      $ & $   -0.711        $ & $\ps 1.07         $ & $   -0.711        $ & $\ps 1.0748   $      \\
$\sigma$  & $\ps 0.0006      $ & $\ps 0.005        $ & $\ps 0.19         $ & $\ps 0.005        $ & $\ps 0.0006   $      \\
Exp. 1    & $\ps 1.07457     $ & $   -1.07457      $ & $\ps 1.07457      $ & $   -1.07457      $ & $\ps 1.07457  $      \\
Exp. 2    & $\ps 1.07457 a^2 $ & $\ps 1.07457 a    $ & $\ps ?            $ & $\ps 1.07457 a^*  $ & $\ps 1.07457(a^*)^2$ \\\cline{1-6}
\end{tabular}
\caption{Extrapolations of a real exotic solution in the $k=4$ model with $J=1$ boundary conditions. There are two sets of expected values, the first one is for 1-brane with $\theta=0$ and the second one for the boundary state (3.4) from \cite{WZW Symmetry-breaking} with $J_\alpha=0$ and $n_\alpha=0$. The solution does not seem to match either of them.} \label{tab:sol 4 1 exotic extrapolation}
\end{table}

\FloatBarrier
\subsection{$k=8$ solutions}\label{sec:exotic:k=8}
In this subsection, we will describe two real exotic solutions that we found in the $k=8$ model, which is the highest model considered in this paper. The first solution is solution number 4669 from table \ref{tab:solutions exotic 2} for $J=3/2$ boundary conditions and the second solution has number 6534 and it was found for $J=2$ boundary conditions. These solutions are our best candidates for new symmetry-breaking boundary states. We provide some of their data for comparison in case that they are rediscovered by some other numerical or analytic method.

Although the two solutions were found on different backgrounds, their properties are quite similar. Both solutions have complex seeds at level 2, but they become real already at level 4 like one of the B-brane solutions. We managed to evaluate these solutions up to level 10, so we have enough data points to do extrapolations, although the precision is obviously not great. Fortunately, the solutions do not evolve too much with increasing level, which helps a bit. Both solutions satisfy the out-of-Siegel equation $\Delta_S$ quite well, so they most likely have an  interpretation as boundary states.

In table \ref{tab:sol 8 3/2 exotic data}, we provide values of several invariants of the first solution at levels 4 to 10 and table \ref{tab:sol 8 3/2 exotic extrapolation} shows extrapolations of the observables. The number of invariants in this model is too high, so we show $E_{j,m}$ invariants only up to $j=2$. The invariants again satisfy the 'accidental' symmetries (\ref{symmetries su 1})-(\ref{symmetries su 3}) and these relations provide values of few more invariants. The extrapolated energy of the solution $E\approx 1.1580$ is slightly higher than the $g$-function of the reference $\frac{3}{2}$-brane, which is $1.14412$. However, the solution clearly does not represent a $\frac{3}{2}$-brane because other invariants do not agree with the expected values. For example, $E_{1,0}$ is around 0 instead of 0.437016, $E_{2,2}$ is approximately 0.37 instead of 0, etc. Additionally, we checked that invariants of the solution do not match well enough any combination of \SU Cardy boundary states or B-branes and therefore we believe that it describes a yet unknown symmetry-breaking boundary state in the $k=8$ model.

The $g$-function of the new boundary state either matches the $g$-function of the $\frac{3}{2}$-brane or it is slightly above. The later option is more likely because the difference is over $30\sigma$, which is quite a lot, even though error estimates tend to be underestimated for the energy.

Tables \ref{tab:sol 8 2 exotic data} and \ref{tab:sol 8 2 exotic extrapolation} show data regarding the second exotic solution, the first table provides examples of finite level data and the second one infinite level extrapolations. This solution was found on $J=2$ background, but its properties are often analogous to the first solution. It has the same symmetries and its energy is also slightly higher than for the reference D-brane (1.239 compared to 1.203). Once again, we cannot rule out the possibility that the energy matches the 2-brane $g$-function, but it is more likely that it does not.

Since the two solutions have the same symmetries as real solutions describing Cardy boundary states, they may have duals in the $k=8$ parafermion theory. If so, the existence of these solutions also suggests that there are unknown boundary states in the parafermion theory.

Finally, we notice that table \ref{tab:solutions exotic 2} includes two pseudo-real solutions (solution number 4491 for $J=3/2$ and number 6354 for $J=2$) which have similar energy and $\theta$-independent invariants as the real solutions presented here. Analogously to Cardy solutions and B-branes, we believe that the pseudo-real solutions are connected to the real ones by complex marginal deformations.

%4669 3/2
\begin{table}[t!]
\centering
\begin{tabular}{|l|llllll|}\hline
Level    & Energy  & $E_{0,0}$ & $E_{1/2,1/2}       $ & $\ps E_{1,1}            $ & $\ps E_{1,0} $ & $\ps \Delta _S$ \\\hline
4        & 1.16165 & 1.14432   & $0.66963+0.010145 i$ & $   -0.372793+0.078599 i$ & $   -0.147291$ & $   -0.0018960$ \\
5        & 1.16131 & 1.14445   & $0.66779+0.010206 i$ & $   -0.371460+0.079163 i$ & $   -0.134488$ & $   -0.0018266$ \\
6        & 1.15993 & 1.14961   & $0.66315+0.008729 i$ & $   -0.372276+0.074751 i$ & $   -0.095785$ & $   -0.0010325$ \\
7        & 1.15982 & 1.14958   & $0.66255+0.009130 i$ & $   -0.371323+0.074747 i$ & $   -0.097006$ & $   -0.0010147$ \\
8        & 1.15926 & 1.15187   & $0.66100+0.008633 i$ & $   -0.372584+0.072576 i$ & $   -0.077517$ & $   -0.0007109$ \\
9        & 1.15921 & 1.15185   & $0.66071+0.008807 i$ & $   -0.372224+0.072782 i$ & $   -0.075520$ & $   -0.0007032$ \\
10       & 1.15891 & 1.15306   & $0.65989+0.008561 i$ & $   -0.372362+0.072160 i$ & $   -0.068430$ & $   -0.0005426$ \\\hline
\end{tabular}
\caption{Selected observables of exotic solution number 4669 in the $k=8$ model with $J=\frac{3}{2}$ boundary conditions. We omitted data from levels 2 and 3 because the solution is complex at these levels.} \label{tab:sol 8 3/2 exotic data}
\vspace{7mm}
\begin{tabular}{|c|lllll|}\cline{1-4}
          & $\ps $Energy         & $\ps E_{0,0}        $ & $\ps \Delta_S      $ & \mc{2}{|c}{}                            \\ \cline{1-4}
$\inf$    & $\ps 1.1580        $ & $\ps 1.157          $ & $   -0.0002        $ & \mc{2}{|c}{}                            \\
$\sigma$  & $\ps 0.0004        $ & $\ps 0.005          $ & $\ps 0.0003        $ & \mc{2}{|c}{}                            \\\cline{1-4}
          & $\ps J_{+-}        $ & $\ps J_{-+}         $ & $\ps J_{33}        $ & \mc{2}{|c}{}                            \\\cline{1-4}
$\inf$    & $\ps 0.73+0.02    i$ & $\ps 0.73-0.02     i$ & $\ps 1.157         $ & \mc{2}{|c}{}                            \\
$\sigma$  & $\ps 0.08+0.01    i$ & $\ps 0.08+0.01     i$ & $\ps 0.005         $ & \mc{2}{|c}{}                            \\\cline{1-4}
          & $\ps E_{1/2,1/2}   $ & $\ps E_{1/2,-1/2}   $ & \mc{3}{|c}{}                                                   \\\cline{1-3}
$\inf$    & $\ps 0.655+0.008  i$ & $   -0.655+0.008   i$ & \mc{3}{|c}{}                                                   \\
$\sigma$  & $\ps 0.005+0.001  i$ & $\ps 0.005+0.001   i$ & \mc{3}{|c}{}                                                   \\\cline{1-4}
          & $\ps E_{1,1}       $ & $\ps E_{1,0}        $ & $\ps E_{1,-1}      $ & \mc{2}{|c}{}                            \\\cline{1-4}
$\inf$    & $   -0.3723+0.069 i$ & $   -0.03           $ & $   -0.3723+0.069  i$ & \mc{2}{|c}{}                            \\
$\sigma$  & $\ps 0.0008+0.004 i$ & $\ps 0.04           $ & $\ps 0.0008+0.004  i$ & \mc{2}{|c}{}                            \\\cline{1-5}
          & $\ps E_{3/2,3/2}   $ & $\ps E_{3/2,1/2}    $ & $\ps E_{3/2,-1/2}  $ & $\ps E_{3/2,-3/2}   $ & \mc{1}{|c}{}    \\\cline{1-5}
$\inf$    & $   -0.683+0.26   i$ & $\ps 0.53+0.39     i$ & $   -0.53+0.39    i$ & $\ps 0.683+0.26    i$ & \mc{1}{|c}{}    \\
$\sigma$  & $\ps 0.005+0.02   i$ & $\ps 0.02+0.03     i$ & $\ps 0.02+0.03    i$ & $\ps 0.005+0.02    i$ & \mc{1}{|c}{}    \\\cline{1-6}
          & $\ps E_{2,2}       $ & $\ps E_{2,1}        $ & $\ps E_{2,0}       $ & $\ps E_{2,-1}       $ & $\ps E_{2,-2} $ \\\cline{1-6}
$\inf$    & $\ps 0.37         i$ & $\ps 0.32+0.56     i$ & $   -0.39          $ & $\ps 0.32-0.56     i$ & $   -0.37    i$ \\
$\sigma$  & $\ps 0.02         i$ & $\ps 0.03+0.05     i$ & $\ps 0.05          $ & $\ps 0.03+0.05     i$ & $\ps 0.02    i$ \\\cline{1-6}
\end{tabular}
\caption{Extrapolations of gauge invariants of exotic solution number 4669 in the $k=8$ model with $J=\frac{3}{2}$ boundary conditions. Due to large number of invariants, we decided to show invariants only up to $j=2$ to fit a page.} \label{tab:sol 8 3/2 exotic extrapolation}
\end{table}

%6534 2
\begin{table}[t]
\begin{tabular}{|l|llllll|}\hline
Level    & Energy  & $E_{0,0}$ & $\ps E_{1/2,1/2}        $ & $\ps E_{1,1}            $ & $E_{1,0} $ & $\ps \Delta _S$ \\\hline
4        & 1.25683 & 1.20475   & $   -0.274082+0.018397 i$ & $   -0.737259-0.277261 i$ & $0.251618$ & $   -0.0064416$ \\
5        & 1.25319 & 1.20724   & $   -0.260742+0.027474 i$ & $   -0.746221-0.262986 i$ & $0.291839$ & $   -0.0052499$ \\
6        & 1.24806 & 1.21918   & $   -0.251575+0.024688 i$ & $   -0.743583-0.248075 i$ & $0.337671$ & $   -0.0032275$ \\
7        & 1.24701 & 1.21939   & $   -0.247524+0.028925 i$ & $   -0.747284-0.244595 i$ & $0.334880$ & $   -0.0029470$ \\
8        & 1.24498 & 1.22457   & $   -0.243820+0.027853 i$ & $   -0.747330-0.237617 i$ & $0.357334$ & $   -0.0021747$ \\
9        & 1.24449 & 1.22458   & $   -0.241864+0.029755 i$ & $   -0.748414-0.235977 i$ & $0.362410$ & $   -0.0020537$ \\
10       & 1.24339 & 1.22734   & $   -0.240052+0.029199 i$ & $   -0.747833-0.233558 i$ & $0.370721$ & $   -0.0016452$ \\\hline
\end{tabular}
\caption{Selected observables of exotic solution number 6534 in the $k=8$ model with $J=2$ boundary conditions.} \label{tab:sol 8 2 exotic data}
\vspace{7mm}
\begin{tabular}{|c|lllll|}\cline{1-4}
          & $\ps $Energy         & $\ps E_{0,0}        $ & $\ps \Delta_S      $ & \mc{2}{|c}{}                            \\ \cline{1-4}
$\inf$    & $\ps 1.239         $ & $\ps 1.237          $ & $   -0.0004        $ & \mc{2}{|c}{}                            \\
$\sigma$  & $\ps 0.002         $ & $\ps 0.012          $ & $\ps 0.0010        $ & \mc{2}{|c}{}                            \\\cline{1-4}
          & $\ps J_{+-}        $ & $\ps J_{-+}         $ & $\ps J_{33}        $ & \mc{2}{|c}{}                            \\\cline{1-4}
$\inf$    & $\ps 0.5+0.14     i$ & $\ps 0.5-0.14      i$ & $\ps 1.237         $ & \mc{2}{|c}{}                            \\
$\sigma$  & $\ps 0.2+0.06     i$ & $\ps 0.2+0.06      i$ & $\ps 0.012         $ & \mc{2}{|c}{}                            \\\cline{1-4}
          & $\ps E_{1/2,1/2}   $ & $\ps E_{1/2,-1/2}   $ & \mc{3}{|c}{}                                                   \\\cline{1-3}
$\inf$    & $   -0.23+0.034   i$ & $\ps 0.23+0.034    i$ & \mc{3}{|c}{}                                                   \\
$\sigma$  & $\ps 0.02+0.004   i$ & $\ps 0.02+0.004    i$ & \mc{3}{|c}{}                                                   \\\cline{1-4}
          & $\ps E_{1,1}       $ & $\ps E_{1,0}        $ & $\ps E_{1,-1}      $ & \mc{2}{|c}{}                            \\\cline{1-4}
$\inf$    & $   -0.753-0.21   i$ & $\ps 0.42           $ & $   -0.753+0.21   i$ & \mc{2}{|c}{}                            \\
$\sigma$  & $\ps 0.004+0.02   i$ & $\ps 0.06           $ & $\ps 0.004+0.02   i$ & \mc{2}{|c}{}                            \\\cline{1-5}
          & $\ps E_{3/2,3/2}   $ & $\ps E_{3/2,1/2}    $ & $\ps E_{3/2,-1/2}  $ & $\ps E_{3/2,-3/2}   $ & \mc{1}{|c}{}    \\\cline{1-5}
$\inf$    & $\ps 0.14+0.38    i$ & $   -0.40+0.73     i$ & $\ps 0.40+0.73    i$ & $   -0.14+0.38     i$ & \mc{1}{|c}{}    \\
$\sigma$  & $\ps 0.01+0.04    i$ & $\ps 0.03+0.08     i$ & $\ps 0.03+0.08    i$ & $\ps 0.01+0.04     i$ & \mc{1}{|c}{}    \\\cline{1-6}
          & $\ps E_{2,2}       $ & $\ps E_{2,1}        $ & $\ps E_{2,0}       $ & $\ps E_{2,-1}       $ & $\ps E_{2,-2} $ \\\cline{1-6}
$\inf$    & $\ps 0.56          $ & $   -0.43-0.50     i$ & $\ps 0.16          $ & $   -0.43+0.50     i$ & $\ps 0.56     $ \\
$\sigma$  & $\ps 0.01          $ & $\ps 0.06+0.08     i$ & $\ps 0.10          $ & $\ps 0.06+0.08     i$ & $\ps 0.01     $ \\\cline{1-6}
\end{tabular}
\caption{Extrapolations of gauge invariants of exotic solution number 6534 in the $k=8$ model with $J=2$ boundary conditions.} \label{tab:sol 8 2 exotic extrapolation}
\end{table}

\FloatBarrier

\FloatBarrier
\subsection{Survey of solutions}\label{sec:exotic:other}
Finally, let us briefly discuss the remaining exotic solutions. We provide a survey of these solutions up to $k=8$ in tables \ref{tab:solutions exotic 1} and \ref{tab:solutions exotic 2}. There are only few real and well-behaved solutions. They include the B-branes, which we discussed in subsection \ref{sec:exotic:B-branes}, one marginal solution for $k=4$ discussed in subsection \ref{sec:exotic:k=4} and two solutions for $k=8$ discussed in subsection \ref{sec:exotic:k=8}, which suggest existence of unknown symmetry-breaking boundary states.

We unfortunately cannot say much about most of the remaining solutions. They are mostly pseudo-real with relatively generic properties, so we can deduce only that they probably represent some symmetry-breaking \SL boundary states. Furthermore, they often suffer from numerical instabilities or they are fully complex at low levels, which makes their analysis even more problematic.

There is one interesting property of exotic solutions, which we noticed when going over the extrapolated energies in tables \ref{tab:solutions exotic 1} and \ref{tab:solutions exotic 2}. Energies some of these solutions match each other or some combination Cardy boundary states (these matches are mentioned in the notes). The extrapolated energies have only a limited precision, so some of the cases could be just coincidences, but they often share other properties (most notably, they often have similar invariants $E_{j,0}$ like the two B-brane solutions in table \ref{tab:sol 6 3/2 B-brane extrapolation}) and coincidental agreements of multiple invariants have only a low probability.
A more likely explanation is therefore that such solutions belong to some broader family of boundary states connected by marginal deformations.

In particular, when we choose the setting with $k=8$ and $J=2$, there are several related solutions with energy similar to three 0-branes. These solutions are pseudo-real, so they cannot describe \SU Cardy branes, but they could potentially describe combinations of \SL boundary states. We tried to identify them using the $R^2$ minimization procedure, but we found no parameters that would match their invariants well enough. Therefore we placed these solutions in this section among exotic solutions.

\begin{table}[!]
\centering
\begin{tabular}{|l|l|l|l|l|l|l|}\hline
$k$           & $J$               & No.  & $E^{(\inf)}$ & $E_J$               & Real & Notes                          \\\hline
\mr{1}{*}{3}  & \mr{1}{*}{$\dt1$} & 169  & 1.0562       & \mr{1}{*}{0.986534} & yes  & R(4), B-brane                  \\\hline\hline
\mr{3}{*}{4}  & \mr{3}{*}{1}      & 697  & 1.07457      & \mr{3}{*}{1.074570} & no   & I, exactly marginal            \\
              &                   & 745  & 1.07457      &                     & yes  & energy similar to 1-brane      \\
              &                   & 783  & 1.0747       &                     & no   & energy similar to 1-brane      \\\hline\hline
\mr{7}{*}{5}  & \mr{1}{*}{$\dt1$} & 215  & 0.96         & \mr{1}{*}{0.867780} & no   & I(3)                           \\\cline{2-7}
              & \mr{6}{*}{1}      & 1833 & 1.0756       & \mr{6}{*}{1.082104} & no   & I(3), pseudo-real B-brane      \\
              &                   & 1861 & 0.97         &                     & no   & I                              \\
              &                   & 1887 & 1.082        &                     & no   & I, energy similar to 1-brane   \\
              &                   & 1957 & 1.089        &                     & no   & I, energy similar to 1-brane   \\
              &                   & 2007 & 1.076        &                     & yes  & R(6), B-brane                  \\
              &                   & 2461 & 0.98         &                     & no   & I                              \\\hline\hline
\mr{10}{*}{6} & \mr{1}{*}{1}      & 2590 & 0.892        & \mr{1}{*}{1.056040} & no   & I                              \\\cline{2-7}
              & \mr{9}{*}{$\dt3$} & 18   & 1.07149      & \mr{9}{*}{1.143050} & yes  & B-brane                        \\
              &                   & 2235 & 0.786        &                     & no   & I                              \\
              &                   & 2785 & 0.70         &                     & no   & I                              \\
              &                   & 3061 & 1.07149      &                     & no   & pseudo-real B-brane            \\
              &                   & 3243 & 1.10228      &                     & no   &                                \\
              &                   & 3383 & 1.10217      &                     & no   & energy similar no. 3243        \\
              &                   & 3591 & 1.10247      &                     & no   & energy similar no. 3243        \\
              &                   & 3779 & 1.079        &                     & no   & I                              \\
              &                   & 4393 & 1.114        &                     & no   & I                              \\\hline
\end{tabular}
\caption{List of exotic solutions in the SU(2)$_k$ WZW model with $k\leq 8$, part 1. The structure of the table and notes follows tables \ref{tab:solutions regular 1} and \ref{tab:solutions SL 1}. We added a column with solution numbers (which come from ordering of seeds by the homotopy continuation algorithm and which do not have any deeper meaning) so that we can refer to them and a column which denotes whether the solutions are real or pseudo-real. Energies are rounded roughly according to error estimates of our extrapolations, but in case of solutions which are complex at low levels or which have numerical instabilities, error estimates are often just a guesswork. We removed the column with expected energies because we do not know them for most solutions.} \label{tab:solutions exotic 1}
\end{table}

\begin{table}[!]
\centering
\begin{tabular}{|l|l|l|l|l|l|l|}\hline
$k$           & $J$                & No.   & $E^{(\inf)}$ & $E_J$                & Real & Notes                                 \\\hline
\mr{13}{*}{7} & \mr{1}{*}{1}       & 2491  & 1.26         & \mr{1}{*}{1.016721}  & no   & I                                     \\\cline{2-7}
              & \mr{12}{*}{$\dt3$} & 2256  & 0.750        & \mr{12}{*}{1.156172} & no   & energy similar to $\dt1$-brane        \\
              &                    & 2691  & 0.804        &                      & no   & I                                     \\
              &                    & 2917  & 0.745        &                      & no   & I                                     \\
              &                    & 3501  & 1.062        &                      & no   & I(3), pseudo-real B-brane             \\
              &                    & 3607  & 1.062        &                      & yes  & R(6), B-brane                         \\
              &                    & 4640  & 1.19815$^*$  &                      & no   & R(9)                                  \\
              &                    & 4740  & 1.07         &                      & no   & R(4), I                               \\
              &                    & 4812  & 0.96         &                      & no   & R(6)                                  \\
              &                    & 4988  & 1.19800$^*$  &                      & no   & R(9)                                  \\
              &                    & 5364  & 1.20515$^*$  &                      & yes  & R(9)                                  \\
              &                    & 5655  & 1.19368$^*$  &                      & no   & R(10)                                 \\
              &                    & 7090  & 1.14         &                      & no   & I                                     \\\hline\hline
\mr{26}{*}{8} & \mr{1}{*}{1}       & 4471  & 1.19         & \mr{1}{*}{0.973249}  & no   & I(3)                                  \\\cline{2-7}
              & \mr{11}{*}{$\dt3$} & 1679  & 0.705        & \mr{11}{*}{1.144123} & no   & energy similar to $\dt1$-brane        \\
              &                    & 3581  & 1.051        &                      & no   & I(3,5), pseudo-real B-brane           \\
              &                    & 4383  & 1.16         &                      & no   & R(5), I                               \\
              &                    & 4491  & 1.1576       &                      & no   & energy similar no. 4669               \\
              &                    & 4663  & 1.35         &                      & no   & I                                     \\
              &                    & 4669  & 1.1580       &                      & yes  & R(4)                                  \\
              &                    & 5335  & 1.12         &                      & no   & R(6), I                               \\
              &                    & 5343  & 1.058        &                      & no   & R(4), I, pseudo-real B-brane?         \\
              &                    & 6281  & 1.075        &                      & no   & R(5)                                  \\
              &                    & 6285  & 1.27027$^*$  &                      & yes  & R(10)                                 \\
              &                    & 6449  & 1.24984$^*$  &                      & no   & R(8)                                  \\\cline{2-7}
              & \mr{14}{*}{2}      & 10    & 1.05148      & \mr{14}{*}{1.203002} & yes  & B-brane                               \\
              &                    & 2581  & 0.72         &                      & no   & I                                     \\
              &                    & 2969  & 1.08591$^*$  &                      & no   & R(10)                                 \\
              &                    & 3461  & 1.05149      &                      & no   & pseudo-real B-brane                   \\
              &                    & 3695  & 1.1119       &                      & no   & energy similar three 0-branes         \\
              &                    & 3707  & 1.1128       &                      & no   & energy similar three 0-branes         \\
              &                    & 3859  & 1.109        &                      & no   & energy similar three 0-branes         \\
              &                    & 4255  & 1.1122       &                      & no   & energy similar three 0-branes         \\
              &                    & 5068  & 1.12         &                      & no   & I(3,5), energy similar three 0-branes \\
              &                    & 5806  & 1.06         &                      & no   & I                                     \\
              &                    & 6354  & 1.234        &                      & no   & R(4), energy similar no. 6534         \\
              &                    & 6534  & 1.239        &                      & yes  & R(4)                                  \\
              &                    & 9339  & 1.117        &                      & no   & I, energy similar three 0-branes      \\
              &                    & 10419 & 1.04         &                      & no   & I                                     \\\hline
\end{tabular}
\caption{List of exotic solutions in the SU(2)$_k$ WZW model with $k\leq 8$, part 2.} \label{tab:solutions exotic 2}
\end{table}

\FloatBarrier
\section{Summary and future directions}\label{sec:discussion}
In this paper, we described many OSFT solutions in the \SUk WZW model which correspond to several types of boundary states. So let us review the results and discuss some possible future directions of this research.

We divide solutions in this model into three groups. The first group includes real solutions describing \SU Cardy boundary states. These represent the basic type of solutions which is expected based on background independence of OSFT. From the OSFT perspective, solutions of this type have similar properties as lump solutions in free boson theories and solutions in Virasoro minimal models \cite{KudrnaThesis}. We reproduce the expected $g$-functions and boundary states components with decent precision and we are able to uniquely identify most of the solutions. We have found just two peculiar properties of these solutions. They have 'accidental' symmetries of certain invariants, which may be connected to absence of the marginal field in these solutions. We also noticed that some of their invariants converge better than we expected based on their conformal weights, which makes their identification easier.

When we consider these solutions in the context of boundary conformal field theory, it is most important to know what are their parameters $J$ and $\theta$, which characterize Cardy boundary states in the \SUk WZW model restricted to our ansatz. The parameter $J$ does not seem to follow any special selection rules, but (as in other OSFT models) it is restricted by the fact that numerical OSFT solutions typically have lower energy than the background, so we mostly encounter solutions with low values of $J$. We found only few solutions with energy higher than the background and they mostly represent two 0-branes.

Rules governing the angle $\theta$ are more interesting. Out of the continuous family of boundary states, we find only a finite number and their parameters satisfy the following relation:
\begin{equation}\label{theta 3}
\theta=\pm(J_f-J_i+n)\frac{2\pi}{k},\quad n\in \mathbb{Z}.
\end{equation}
That means that the angle is given by an integer multiple of $\frac{\pi}{k}$ and it is related to the difference between the initial and final value of $J$.
In addition to this, branes described by our solutions lie in the segment given by the reference brane (see figure \ref{fig:Branes k=9}) and there is a preference for branes that touch the reference brane at one point (which have $n=0$). We also noticed that solutions with $|n|>0$ are more difficult to find and they converge slower.

The structure of branes described by the equation (\ref{theta 3}) is in fact the same as structure of branes in the parafermion theory. Therefore it seems likely that the condition $J_0^3|\Psi\ra=0$ and the fact that these solutions do not excite the marginal field turn off the free boson degrees of freedom and effectively reduce the \SUk WZW model OSFT to parafermion OSFT. We have seen that $k=2,3$ solutions have duals in minimal models (which are isomorphic to the parafermion theories) and we expect that similar dualities hold even for higher $k$.

It would be to interesting to see whether whether the selection rule (\ref{theta 3}) is specific to OSFT or whether it also applies to boundary states that can be obtained by other RG flow methods. One possibility would be to use boundary conformal field theory perturbation techniques (see for example \cite{RecknagelSchomerus}, chapter 5). This method requires a boundary field with conformal dimension close to 1, so it would work only on backgrounds which include such boundary field, but even such limited results would be useful.
Another method which could be used for investigating RG flow in this model  is the truncated conformal space approach (TCSA) \cite{TCSAYurov}\cite{TCSAGraham}\cite{TCSADorey}\cite{TCSATakacs}.

The second group of solutions does not describe Cardy boundary states in the \SUk WZW model, but in the \SLk WZW model.
The \SL group is a complexification of the \SU group, so it is not surprising that these solutions are complex (their invariants do not satisfy the \SU reality condition (\ref{reality})), although they have some reality properties, like real action.
They generally do not have any special symmetries and they often suffer from numerical instabilities, so their properties are generally worse than properties of \SU Cardy solutions.

As before, we find mainly solutions with low values of the parameter $J$, mostly 0-branes. They are characterized by a complex angle $\theta-i\log \rho$. Its real part follows the equation (\ref{theta 3}) with $n=0$, while the other parameter $\rho$ seems to be quite generic. For a given value of $\theta$, there are always two solutions with $\rho$ and $1/\rho$. In case of solutions that describe two 0-branes, which is the most common type \SL solutions, we find $\theta_1=-\theta_2$ and their $\rho$ parameters follow one of three options: $\rho_1=\rho_2$, $\rho_1=1/\rho_2$ or one of the two parameters is close to one. The third option means that such solutions describe combinations of \SU and \SL boundary states.

Finally, there is a third group of exotic solutions, which do not fit the previous two groups. These solutions are potentially the most interesting, because they most likely describe symmetry-breaking boundary states either in the \SUk or the \SLk WZW model (depending on whether they are real or not). We have found quite a large number of exotic solutions, but, unfortunately, only few of them are real. The rest are pseudo-real and they often suffer from various numerical problems, so we are not able to say much about their physical properties.
When it comes to real exotic solutions, we identified most of them as the B-brane boundary states from \cite{WZW B-branes}. Their invariants match the expected values quite well, so we are essentially sure about the identification. We have also found some pseudo-real B-branes, which are probably related to real B-branes by complex marginal deformations.
Apart from B-branes, we have found only three nice real solutions. The first one appears in the $k=4$ model with $J=1$ boundary conditions, which has additional marginal operators. It probably describes a marginal deformation of the initial 1-brane by the field $\phi_{2,0}$. We checked whether it belongs to the first group of symmetry-breaking states from \cite{WZW Symmetry-breaking}, but it seems likely that it does not. The other two real solutions that we were not able identify are presented in section \ref{sec:exotic:k=8} and they probably describe yet unknown boundary states in the $k=8$ model with $g$-functions approximately 1.1580 and 1.239.
From the OSFT point of view, behavior of real exotic solutions is analogous to solutions describing \SU Cardy boundary states, they have the same symmetries and similar precisions. It is quite likely that real exotic solutions also have duals in parafermion theories, although we are not able to verify this directly with the exception of the $k=3$ B-brane solution.

We were surprised that the number of real exotic solutions is quite low, especially when compared with free boson theory on torus \cite{KudrnaThesis}, which admits more exotic solutions. A possible cause is the condition (\ref{J03 psi}), which we used to fix the \SU symmetry of OSFT equations. It imposes the condition (\ref{J03 B}) on boundary states that correspond to our solutions, which seems to be a big restriction for new symmetry-breaking boundary states.

In a future work, it would be interesting to try to relax the condition $J_0^3 |\Psi\ra=0$, which may help us find more exotic solutions. However, doing so would require significant modifications of the numerical approach because we would need to fix the \SU symmetry in some alternative way. The simplest option seems to be to take inspiration from the free boson theory and to remove states that are odd with respect to some $\mathbb{Z}_2$ symmetry of the action. In analog with the symmetry $X\rar -X$, we consider transformations that flip signs of two of the three currents (because we need to preserve the \SU algebra), for example $J^{1,2}\rar -J^{1,2}$. In our basis, the three transformations act as
\begin{eqnarray}
&J^3\rar J^3,&\quad J^\pm\rar -J^\pm,\\
&J^3\rar -J^3,&\quad J^\pm\rar J^\mp, \\
&J^3\rar -J^3,&\quad J^\pm\rar -J^\mp.
\end{eqnarray}
However, imposing even parity with respect to one of these symmetries does not fully fix the SU(2) symmetry, so we have impose two of them, which automatically implies the third one. Unfortunately, these symmetries are not compatible with our basis and it would be necessary to abandon splitting of the Hilbert space according to $J_0^3$ eigenvalues and use the basis generated by $J^{1,2,3}$ instead.

This ansatz would lead both to some simplifications and complications in the numerical approach. Algorithms to compute the action and Ellwood invariants would be simpler because we would not need to worry about $J_0^3$ eigenvalues, but there would be an overall increase in the number of states at a given level, which would increase time and memory requirements of the calculations, and more Ellwood invariants would be nonzero. A Cardy boundary state in this ansatz would be characterized by all three angles instead of just one like now, so identification of solutions would be much more complicated, especially for solutions describing more than one D-brane.

\begin{figure}
   \centering
   \includegraphics[width=8cm]{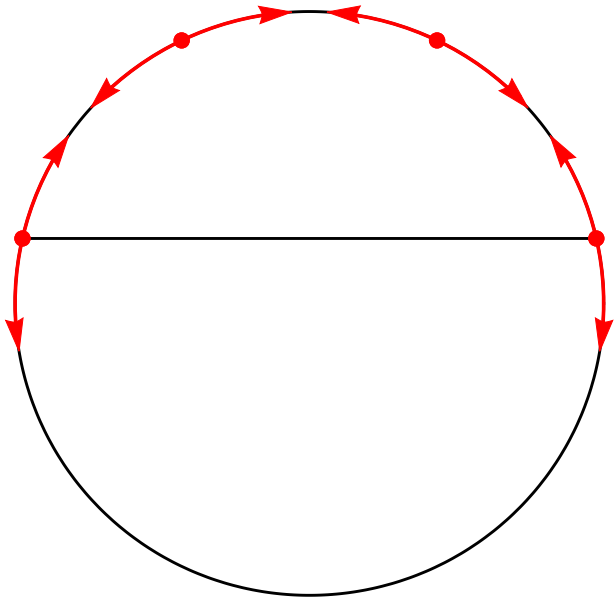}
   \caption{Schematic depiction of the expected mechanism of covering the moduli space of 0-branes in the $k=7$ model. The dots represent the solutions found in this paper using the traditional approach, while the red arrows show how the solutions can be moved by marginal deformations. The actual length of the arrows is unknown and they may or may not overlap.}
   \label{fig:Branes k=7 mar}
\end{figure}

Another topic that should be explored in the future is how to restore the D-brane modulus. The approach used in this paper leads only to discrete number of solutions, whose parameters follow (\ref{theta 3}). In order to find a continuum of solutions, we propose an approach inspired by marginal deformations \cite{MarginalSen}\cite{MarginalKMOSY}\cite{MarginalTachyonKM}\cite{KudrnaThesis}. We can fix the value of the marginal field, remove the equation corresponding to the marginal field and try to solve this reduced system of equations. However, unlike in the traditional marginal approach, we do not suggest to look for solutions that preserve the energy, but for solutions that describe lower energy D-branes. In general, we expect that these solutions will cover the moduli space as depicted in figure \ref{fig:Branes k=7 mar}. Solutions found in this paper will work as seeds and there will be an area covered by marginal deformations around each of them.
We tried a quick low level test of this approach for $k=2$ and it seems to work at least for small values of the marginal parameter. But the precision goes quickly down with increasing value of the marginal parameter and the branch of solutions possibly goes off-shell. Therefore it is unclear how much of the moduli space can be actually covered by this approach. It is possible that the areas around seed solutions will overlap and the entire moduli space will be covered, but it seems more likely that there will be some gaps and we will recover only part of the moduli space. The later option is supported by the fact that we typically find solutions only in one section of the circle (given by the initial D-brane), so covering the remaining part without any seed solutions seems to be difficult. We would like to explore this approach in one of our future works.

This approach should also work for complex marginal deformations, which generate \SL boundary states. Specifically, if we consider purely imaginary marginal field, we should be able to take the \SU solutions from section \ref{sec:regular} and continuously deform them until we get the \SL solutions from section \ref{sec:SL} with the same values of $J$ and $\theta$.

\section*{Acknowledgements}

\noindent
We would like to thanks Carlo Maccaferri, Martin Schnabl and Jakub Vo\v{s}mera for valuable comments and useful discussions.

This research has been supported by the GACR grant 20-25775X.

Computational resources were supplied by the project "e-Infrastruktura CZ" (e-INFRA CZ LM2018140 ) supported by the Ministry of Education, Youth and Sports of the Czech Republic.

\begin{appendix}

\section{Complex conjugation}\label{app:complex}
In this appendix, we will discuss complex conjugation of \SU primary states.

Complex conjugation in string field theory is defined as a combination of BPZ and Hermitian conjugations. Modes of the \SU currents satisfy
\begin{equation}
{\rm bpz} (J_n^a)=(-1)^{n+1}J_{-n}^a
\end{equation}
and
\begin{eqnarray}
(J_n^3)^\dagger=J_{-n}^3,\\
(J_n^\pm)^\dagger=J_{-n}^\mp.
\end{eqnarray}
By combining the two conjugations, we find
\begin{eqnarray}
(J_n^3)^\ast=(-1)^{n+1}J_{n}^3,\\
(J_n^\pm)^\ast=(-1)^{n+1}J_{n}^\mp.
\end{eqnarray}

We also need to know how the complex conjugation acts on primary states, which is more difficult to derive. For consistency, we must have
\begin{equation}\label{complex 1}
(J_0^a |j,m\ra)^\ast=(J_0^a)^\ast |j,m\ra^\ast.
\end{equation}
By choosing $a=3$, we find $|j,m\ra^\ast\sim |j,-m\ra$ and choosing $a=\pm$ gives us
\begin{equation}
|j,m\ra^\ast =(-1)^m f_1(j) |j,-m\ra,
\end{equation}
where $f_1(j)$ is a function that equals $\pm 1$ for each $j$. Similarly, bulk states satisfy an analogous relation
\begin{equation}
|j,m,n\ra^\ast =(-1)^{m-n} f_2(j) |j,-m,-n\ra.
\end{equation}
So far, we have determined complex conjugation of primary states up to two unknown functions $f_1(j)$ and $f_2(j)$. To fix these functions, we will analyze reality of boundary states. We would like SU(2) boundary states to be real under our reality condition.
Complex conjugation of the Ishibashi states (\ref{Ishibashi}) gives us
\begin{equation}
|j,g\rra^\ast=\sum_{m,n} (-1)^{j-m}\left(D^j_{-m,n}(g)\right)^\ast (-1)^{m-n}f_2(j)|j,-m,-n\ra+\dots.
\end{equation}
Using a formula for complex conjugation of $D^j_{m,n}$ \cite{GaberdielSU2}
\begin{equation}
\left(D^j_{m,n}(g)\right)^\ast=(-1)^{m-n}D^j_{-m,-n}(g),
\end{equation}
we get
\begin{eqnarray}
|j,g\rra^\ast &=&\sum_{m,n} (-1)^{j-m-2n} D^j_{m,-n}(g)\, f_2(j)|j,-m,-n\ra+\dots\\
              &=&\sum_{m,n} (-1)^{j-m} D^j_{-m,n}(g)\, f_2(j)|j,m,n\ra+\dots,
\end{eqnarray}
where we changed the summation variables as $m\rar -m$, $n\rar -n$ and used that $(-1)^{2(m+n)}=1$ because $m+n$ is always an integer. The condition $|j,g\rra^\ast=|j,g\rra$ then leads to a simple result $f_2(j)=1$.

Next, we consider a bulk-boundary correlation function $\la \phi_{j,m,n}(z) \psi_{J,-m-n}^{(aa)}(w)\ra$.
Dressed bulk-boundary structure constants of these correlators (based on (\ref{OPE bulk-bound})) are
\begin{equation}\label{conj1}
^{(a)}B_{(j,m,n)(J,-m-n)}= (-1)^J\,\Cbb ajJ \Cbound aaaJJ0
\left(\begin{array}{ccc}
j & j & J \\
m & n & -m-n
\end{array}\right)\la \Id\ra^{(a)}
\end{equation}
By complex conjugation of the r.h.s. of this equation we get
\begin{equation}\label{conj2}
^{(a)}B_{(j,m,n)(J,-m-n)}^\ast=(-1)^{2J}\,\Cbb ajJ \Cbound aaaJJ0
\left(\begin{array}{ccc}
j & j & J \\
m & n & -m-n
\end{array}\right)\la \Id\ra^{(a)}
\end{equation}
because bulk-boundary structure constants (\ref{C bulk-bound}) satisfy $(\Cbb ajJ)^\ast=(-1)^J\,\Cbb ajJ$ and the remaining terms are real. By complex conjugation of the fields in the correlator, we get
\begin{eqnarray}\label{conj3}
^{(a)}B_{(j,m,n)(J,-m-n)}^\ast&=&(-1)^{m-n}(-1)^{-m-n}f_1(J)\,^{(a)}B_{(j,-m,-n)(J,m+n)} \nn \\
&=&(-1)^{J-2n}f_1(J)\Cbb ajJ \Cbound aaaJJ0
\left(\begin{array}{ccc}
j & j & J \\
-m & -n & m+n
\end{array}\right)\la \Id\ra^{(a)}  \nn\\
&=&(-1)^{2J}f_1(J)\Cbb ajJ \Cbound aaaJJ0
\left(\begin{array}{ccc}
j & j & J \\
m & n & -m-n
\end{array}\right)\la \Id\ra^{(a)},
\end{eqnarray}
where we used a symmetry of the 3-j symbols
\begin{equation}
\left(\begin{array}{ccc}
j_1 & j_2 & j_3 \\
m_1 & m_2 & m_3
\end{array}\right)
=(-1)^{j_1+j_2+j_3}
\left(\begin{array}{ccc}
j_1 & j_2 & j_3 \\
-m_1 & -m_2 & -m_3
\end{array}\right)
\end{equation}
and that $(-1)^{2(j-n)}=1$. The expressions (\ref{conj2}) and (\ref{conj3}) are equal if we set $f_1(j)=1$. The final results are therefore simply
\begin{eqnarray}
|j,m\ra^\ast &=&(-1)^m |j,-m\ra, \\
|j,m,n\ra^\ast &=&(-1)^{m-n} |j,-m,-n\ra.
\end{eqnarray}

These rules for complex conjugation of boundary fields tell us that string fields describing Cardy boundary states are real, which is the expected result.

\section{Ishibashi states}\label{app:Ishibashi}
In this appendix, we will derive an explicit form of SU(2) Ishibashi states. A more abstract discussion of this topic can be found in \cite{BCRational}\cite{RecknagelSchomerus}.

We consider a chiral basis of a spin $j$ representation denoted as $|n\ra$. These states do not have to be orthogonal or have some other special properties.
Using this basis, we parameterize Ishibashi states as
\begin{equation}
|j,g\rra=\sum_{n_1,n_2} M_{n_1n_2}^j(g)|n_1\ra\overline{|n_2\ra}.
\end{equation}

Let us start with the lowest level elements and the trivial gluing condition. One-point function of a bulk primary operator (with unit normalization) equals to
\begin{equation}
|z-\bar z|^{2h_j}\la \phi_{j,m_1,m_2}\ra^{(a)}=\Cbb aj0 \la j,m_1,j,m_2|0,0\ra N_j^{bulk} \la \Id\ra^{(a)}=B_a^j (-1)^{j-m_1} \delta_{m_1,-m_2},
\end{equation}
where we used (\ref{Cardy BS}), (\ref{norm bulk}) and $\Cbb aj0=\frac{S_a^{\ j}}{S_a^{\ 0}}$.
Alternatively, we can express the same correlator as contraction of a bulk primary and a boundary state
\begin{equation}
\la j,m_1,m_2\ww a\rra=\sum_{n_1,n_2} B_a^j \la j,m_1,m_2| M^j_{n_1n_2}(\Id) |j,n_1,n_2\ra=B_a^j (-1)^{2j-m_1-m_2}M_{-m_1-m_2}^j(\Id).
\end{equation}
The comparison of these two equations gives us the first elements of the matrix $M^j(\Id)$
\begin{equation}
M_{m_1m_2}^j(\Id)=(-1)^{j-m_1}\delta_{m_1,-m_2}.
\end{equation}
Therefore the Ishibashi states are
\begin{equation}\label{Ishibashi lev0 id}
|j,\Id\rra=\sum_m (-1)^{j-m} |j,m,-m\ra+\dots.
\end{equation}
It is easy to check that this expression solves the gluing condition (\ref{Gluing J}) for $g=\Id$.

Now, let us consider the full gluing conditions. We need a solution to
\begin{equation}
(J^a_n+\Omega^a_{\ b}(g)\bar J^b_{-n})\sum_{n_1,n_2} M_{n_1n_2}(g)|n_1\ra\overline{|n_2\ra}=0.
\end{equation}
We contract this condition with arbitrary states $|m_1\ra,\overline{|m_2\ra}$ and we get
\begin{eqnarray}
0&=&\la m_1|\overline{\la m_2|} (J^a_n+\Omega^a_{\ b}(g)\bar J^b_{-n})\sum_{n_1,n_2} M_{n_1n_2}(g)|n_1\ra\overline{|n_2\ra} \\
&=&\sum_{n_1,n_2} (G_{m_1 n_1}G_{m_2 (g(J^a_{-n}) n_2)}+(-1)^{1-n}G_{(J^a_{-n}m_1) n_1}G_{m_2 n_2})M_{n_1n_2}(g),\label{gluing aux1}
\end{eqnarray}
where we use Gram matrix of BPZ products $G_{mn}=\la m|n\ra$, $G_{m(J^a_{-n}n)}$ represents $\la m |J^a_{-n} |n\ra$ and $g(J^a_{-n})$ is an abbreviation for $\Omega^a_{\ b}(g)J^b_{-n}$. Similarly, we define $G_{m g(n)}=\la m |g(n)\ra=\sum_k D_{nk}(g)\la m |k\ra$, where $D(g)$ is a representation (generally reducible) of $g$ on our basis.

The solution to (\ref{gluing aux1}) is given by
\begin{equation}\label{Ishibashi full}
M_{n_1n_2}(g)=(-1)^{2j+|n_2|}G^{-1}_{n_1g(n_2)},
\end{equation}
where $|n_2|$ is the eigenvalue of the number operator and the inverse matrix $G^{-1}_{n_1g(n_2)}$ should be understood as
\begin{equation}
G^{-1}_{n_1g(n_2)}=\sum_k G^{-1}_{k n_2}D_{kn_1}(g^{-1})=\sum_k G^{-1}_{n_1 k}D_{k n_2}(g).
\end{equation}
To show that (\ref{Ishibashi full}) is indeed the correct solution, we substitute this expression into (\ref{gluing aux1}) and we get
\begin{eqnarray}
&&\sum_{n_1,n_2,k}(-1)^{2j+|n_2|} G_{m_1 n_1}G_{m_2 (g(J^a_{-n}) n_2)}G^{-1}_{n_1 k}D_{k n_2}(g) \nn \\
&+&\sum_{n_1,n_2,k}(-1)^{2j+1-n+|n_2|}G_{(J^a_{-n}m_1) n_1}G_{m_2 n_2}G^{-1}_{k n_2}D_{kn_1}(g^{-1}) \nn \\
&=&\sum_{n_2,k} (-1)^{2j+|n_2|}\delta_{m_1 k}G_{m_2 (g(J^a_{-n}) n_2)}D_{k n_2}(g) \nn\\
&+&\sum_{n_1,k}(-1)^{1-n+|m_2|}G_{(J^a_{-n}m_1) n_1}\delta_{m_2k}D_{kn_1}(g^{-1}) \\
&=& (-1)^{2j+|m_1|}G_{m_2 (g(J^a_{-n})g(m_1))}+(-1)^{1-n+|m_2|}G_{(J^a_{-n}m_1) g^{-1}(m_2)} \nn \\
&=& (-1)^{|m_1|}G_{g(J^a_{-n}m_1)m_2}-(-1)^{|m_1|}G_{(J^a_{-n}m_1) g^{-1}(m_2)} \nn \\
&=&0,\nn
\end{eqnarray}
where we used that $|m_2|=|m_1|+n$ and that the Gram matrix has a symmetry $G_{mn}=(-1)^{2j} G_{nm}$ for spin $j$ representation.
In the last step, we used $G_{mn}=G_{g(m)g(n)}$. This equation means
\begin{equation}
G_{mn}=\sum_{k,l}D_{mk}(g)D_{nl}(g)G_{kl}.
\end{equation}
This relation looks a bit unusual, but it is correct. We can see that by decomposing our basis into irreducible representations. Then it simplifies to
\begin{equation}
(-1)^{j-m}\delta_{m,-n}=\sum_{k,l}D_{mk}^j(g)D_{nl}^j(g)(-1)^{j-k}\delta_{k,-l},
\end{equation}
which is satisfies thanks to the relation
\begin{equation}
D^j_{mn}(g^{-1})=(-1)^{n-m}D_{-n-m}^j(g),
\end{equation}
which hold for $g\in$\SL \cite{GaberdielSU2}.

For $g=\Id$, we find that (\ref{Ishibashi full}) matches (\ref{Ishibashi lev0 id}) at the lowest level, which justifies that factor $(-1)^{2j}$ which was added to (\ref{Ishibashi full}) by hand.

For generic gluing conditions, we find from (\ref{Ishibashi full}) that the lowest level elements of Ishibashi states are
\begin{equation}\label{Ishibashi lev0 gen}
|j,g\rra=\sum_{m,n} (-1)^{j-m} D^j_{-mn}(g) |j,m,n\ra+\dots.
\end{equation}

%(auxiliary equations)
%\begin{equation}
%M_{mn}|j,m,n\ra=(-1)^{j-m}\delta_{m,-k}D^j_{kn}|j,m,n\ra=(-1)^{j-m}D^j_{-mn}|j,m,n\ra
%\end{equation}
%\begin{equation}
%\la j,m_1,-m_1|(-1)^{j-m}D^j_{-mn}|j,m,n\ra=(-1)^{j+m}\delta_{m_1,-m}\delta_{-m_1,-n}D^j_{-m,n}=(-1)^{j-m_1}D^j_{m_1m_1}
%\end{equation}

\section{F-matrices and sewing relations}\label{app:sewing}
In this appendix, we discuss properties of F-matrices and B-matrices in the \SUk WZW theory, which are needed for structure constants, and we sketch how to derive the formulas for structure constants using sewing relations. The derivation is analogous to Virasoro minimal models \cite{Runkel1}\cite{Runkel2}\cite{Runkel3}, but one has to make corrections because some equations include additional signs. We also provide an explicit expression for the fusion matrix.

\subsection{Fusion matrices}\label{app:sewing:fusion}
Let us begin with F-matrices. A formula for the F-matrix in the \SUk WZW model can be found in \cite{RecknagelSchomerus}, but the F-matrix in this form does not lead to the correct structure constants\footnote{We found some typos in the formula in \cite{RecknagelSchomerus}, which can be fixed by requiring that the F-matrix has the correct symmetries (\ref{F1}) and that it satisfies the pentagon identity (\ref{pentagon}).}. The reason is that the F-matrix admits a 'gauge' transformations \cite{MooreSeiberg}\cite{TopologicalDefectsOSFT} of the form
\begin{equation}
\Fmatrix pqijkl \rar \frac{\Lambda(j,k,q)\Lambda(i,l,q)}{\Lambda(i,j,p)\Lambda(k,l,p)}\Fmatrix pqijkl.
\end{equation}
This transformation preserves the pentagon identity (\ref{pentagon}), but it changes structure constants and therefore it leads to a different theory. The 'gauge' can be fixed by requiring that the F-matrix leads to correct bulk structure constants (see (\ref{C bulk})), which can be found in \cite{GaberdielWZW}\cite{RecknagelSchomerus}.
%\footnote{The paper \cite{ZamolodchikovFateev} has a different expression for bulk structure constants. When we properly normalize all operators, the results do not seem to be equivalent. One can check which results are correct using the duality between $k=3$ WZW model and Potts model times free boson. By comparing some three-point functions in the two models, we determined that the results from \cite{GaberdielWZW} are correct.}.
Once we include the 'gauge' correction, the final result for the F-matrix is
\begin{eqnarray}\label{F matrix}
\Fmatrix pqijkl &=& \frac{\Lambda(j,k,q)\Lambda(i,l,q)}{\Lambda(i,j,p)\Lambda(k,l,p)}\Delta(i,j,p)\Delta(k,l,p)\Delta(j,k,q)\Delta(i,l,q) \\
\nonumber && \times (-1)^{i+j+k+l}\sqrt{\qsin{2p+1}\qsin{2q+1}}\, \sum_s (-1)^s\qfac{s+1} \\
\nonumber && \times \Big ( \qfac{s-i-j-p}\qfac{s-k-l-p}\qfac{s-j-k-q}\qfac{s-i-l-q}  \\
\nonumber && \times \qfac{i+j+k+l-s}\qfac{i+k+p+q-s}\qfac{j+l+p+q-s}\Big)^{-1},
\end{eqnarray}
where the sum over $s$ goes over such values that all arguments of $\qfac .$ are nonnegative and where we use the following definitions:
\begin{equation}
\Delta(i,j,k)=\sqrt{\frac{\qfac{i+j-k}\qfac{i+k-j}\qfac{j+k-i}}{\qfac{i+j+k+1}}},
\end{equation}
\begin{eqnarray}
\qsin n&=&\frac{\sin\frac{n\pi}{k+2}}{\sin\frac{\pi}{k+2}},\\
\qfac n&=&\prod_{i=1}^n \qsin i,\\
\qfac 0&=&1,
\end{eqnarray}
\begin{equation}
\Lambda(i,j,k)=\left(\frac{a_i a_j}{a_k}\prod_{l=1}^{i+j-k}\frac{\gamma(l)\gamma(2i-l+1)}{\gamma(2k-l+1)\gamma(2j-l+1)}\right)^{\frac{1}{2}},
\end{equation}
\begin{equation}
\gamma(n)=\frac{\Gamma\left(\frac{n}{k+2}\right)}{\Gamma\left(-\frac{n}{k+2}\right)},
\end{equation}
\begin{equation}
a_n=\prod_{i=1}^{2n}\sqrt{\frac{\gamma(i)}{\gamma(i+1)}}.
\end{equation}

Next, we list several relations satisfied by the F-matrices which are useful during manipulations with sewing relations and structure constants:
\begin{equation}\label{F1}
\Fmatrix pqijkl=\Fmatrix pqjilk=\Fmatrix pqklij,
\end{equation}

\begin{equation}\label{F2}
\sum_r \Fmatrix prijkl \Fmatrix rqilkj=\delta_{pq},
\end{equation}

\begin{equation}\label{F3}
\Fmatrix 00iiii=(-1)^{2i}\frac{S_0^{\ 0}}{S_0^{\ i}},
\end{equation}

\begin{equation}\label{F4}
\Fmatrix k0jiij=(-1)^{2i}\frac{S_0^{\ k}}{S_0^{\ j}}\Fmatrix j0ikki,
\end{equation}

\begin{equation}\label{F5}
\Fmatrix 0kiijj \Fmatrix k0jiij=\frac{S_0^{\ 0}S_0^{\ k}}{S_0^{\ i}S_0^{\ j}},
\end{equation}

\begin{equation}\label{F6}
\Fmatrix pinjkl \Fmatrix n0liil=\Fmatrix nklijp \Fmatrix p0lkkl,
\end{equation}

\begin{equation}\label{pentagon}
\sum_s \Fmatrix qspjkb \Fmatrix plaisb \Fmatrix srlijk= \Fmatrix praijq \Fmatrix qlarkb,
\end{equation}
Note that these relations are meaningful only if the labels satisfy the required \SUk fusion rules and that the equations (\ref{F3}) and (\ref{F4}) differ from the minimal model case.

\subsection{Braiding matrices}\label{app:sewing:braiding}
The relation between F-matrices and braiding matrices is the \SUk WZW model is different from Virasoro minimal models. To see why, we consider the limit $k\rar \inf$. In this limit, F-matrices are essentially just the usual SU(2) 6j-symbols
\begin{equation}
\lim_{k\rar\inf}\Fmatrix pqijkl=\sqrt{(2p+1)(2q+1)}(-1)^{i+j+k+l}
\left\{\begin{array}{ccc}
i & j & p \\
k & l & q \\
\end{array}\right\}.
\end{equation}
Furthermore, weights of primary operators become zero and conformal blocks become just products of Clebsch-Gordan coefficients. Therefore we get
\begin{equation}
\la i,m_i,j,m_j|p,m_p\ra \la p,m_p,k,m_k|l,m_l\ra=\sum_q \Fmatrix pqijkl \la j,m_j,k,m_k|q,m_q\ra \la i,m_i,q,m_q|l,m_l\ra
\end{equation}
and
\begin{equation}
\la i,m_i,j,m_j|p,m_p\ra \la p,m_p,k,m_k|l,m_l\ra=\sum_q \BFmatrix pqijkl \la i,m_i,k,m_k|q,m_q\ra \la q,m_q,j,m_j|l,m_l\ra.
\end{equation}
Thanks to properties of Clebsch-Gordan coefficients, we find the F-matrix and the B-matrix in this limit differ by a sign
\begin{equation}
\BFmatrix pqijkl=(-1)^{i+l-p-q} \Fmatrix pqijlk.
\end{equation}

Going back to finite $k$, the relation between the F-matrix and the B-matrix changes to \cite{BCRational}
\begin{equation}\label{F/B matrix relation}
\BFmatrix pqijkl=e^{i\pi (\Delta_i+\Delta_l-\Delta_p-\Delta_q)} \Fmatrix pqijlk,
\end{equation}
where $\Delta_i\equiv h_i-i$.

\subsection{Sewing relations and structure constants}\label{app:sewing:sewing}
Structure constants in Virasoro minimal models can be derived using sewing relations \cite{Runkel1}. We would like to derive structure constants in the \SUk WZW model in the same way. However, some sewing relation in the \SUk WZW model are slightly different because there are some additional signs. The reason is the formula for B-matrix (\ref{F/B matrix relation}), which appears at intermediate steps of derivation of sewing relations. Unfortunately, we cannot take the relations from \cite{LewellenSewing}\cite{Runkel1} and simply replace $h_i$ by $\Delta_i$. Therefore, we decided to rederive the sewing relations using the new expression for the B-matrix. Our results seem to be equivalent to \cite{BCRational}, this reference however uses different conventions and therefore there are some differences in complex phases.

Sewing relations for bulk or boundary fields only obviously remain the same because the do not involve the B-matrix:
\begin{equation}
\Cbulk jkq\, \Cbulk iql\, \Fmatrix qpilkj=\Cbulk ijp\, \Cbulk pkl\, \Fmatrix pqijkl,
\end{equation}
\begin{equation}
\Cbound bcd jkq\, \Cbound abd iql=\sum_p \Cbound abc ijp\, \Cbound acd pkl\, \Fmatrix pqijkl.
\end{equation}
Differences appear in sewing relations that involve both bulk and boundary fields.
By considering a correlator of one bulk field and two boundary fields $\la \phi_i(z) \psi_p^{(ab)}(x_1) \psi_q^{(ba)}(x_2)\ra$, we get the following sewing relation
\begin{eqnarray}
\Cbb bil\ \Cbound abb plq\, \Cbound aba qq0 &=& \sum_{k,m} \Cbb aik\, \Cbound aba pqk\, \Cbound aaa kk0\, \Fmatrix kmpqii \Fmatrix mlpiiq \\
&&\times e^{i\pi(2\Delta_m-2\Delta_i-\Delta_p-\Delta_q+\Delta_k-h_k/2+h_l/2)}  \nn
\end{eqnarray}
Another sewing relation follows from a correlator of two bulk fields and one boundary field $\la \phi_k(z_1) \phi_l(z_1) \psi_i^{(aa)}(x)\ra$:
\begin{eqnarray}
\Cbb akq\ \Cbb alt\, \Cbound aaa qti &=&  \sum_{p,r} \Cbulk klp\ \Cbb api\, \Fmatrix prpilk \Fmatrix pqklrk \Fmatrix rtqlli \\
&&\times  e^{i\pi(\Delta_k+\Delta_r-\Delta_p-\Delta_q-h_i/2+h_p+h_q/2-h_k+h_t/2-h_l)}  \nn
\end{eqnarray}
After substituting the expression for $\Delta$, the two equations above change to
\begin{eqnarray}\label{sewing 1}
\Cbb bil\, \Cbound abb plq\, \Cbound aba qq0 &=& \sum_{k,m} \Cbb aik\, \Cbound aba pqk\, \Cbound aaa kk0\, \Fmatrix kmpqii \Fmatrix mlpiiq \\
&&\times e^{i\pi(2h_m-2h_i-h_p-h_q+h_k/2+h_l/2)}(-1)^{p-q+k}  \nn
\end{eqnarray}
and
\begin{eqnarray}\label{sewing 2}
\Cbb akq\ \Cbb alt\, \Cbound aaa qti &=&  \sum_{p,r} \Cbulk klp\ \Cbb api\, \Fmatrix prpilk \Fmatrix pqklrk \Fmatrix rtqlli \\
&&\times  e^{i\pi(h_r-h_q/2-h_i/2+h_t/2-h_l)}(-1)^{k+r-p-q}  \nn
\end{eqnarray}
By setting $i=0$ in the second equation, we get also get a simpler sewing relation for just two bulk fields
\begin{equation}\label{sewing 3}
\Cbb akq\ \Cbb alq\, \Cbound aaa qq0 =  \sum_{p} \Cbulk klp\ \Cbb ap0\,  \Fmatrix pqkllk  (-1)^{k+l-p-q}
\end{equation}

Now we can find a solution for structure constants following \cite{Runkel1}. Boundary structure constants remain the same:
\begin{equation}\label{C bound}
\Cbound abc ijk = \Fmatrix bkiacj.
\end{equation}
Bulk structure constants can be obtained by setting $a=0$ in (\ref{sewing 3}):
\begin{equation}\label{C bulk}
\Cbulk ijk =(-1)^{i+j-k} \left(\Fmatrix k0jiij \right)^{-1}.
\end{equation}
This equation includes an additional sign compared to the minimal model solution.
Finally, bulk-boundary structure constants follow from (\ref{sewing 1}), where we set $a=0$, $p=q=b$. Although the sewing relation is different, the solution is surprisingly the same as in Virasoro minimal models:
\begin{eqnarray}\label{C bulk-bound}
\Cbb bil =  \frac{S_i^{\ 0}}{S_0^{\ 0}} \sum_m e^{i \pi (2h_m-2h_b-2h_i+h_l/2)} \Fmatrix 0mbbii \Fmatrix mlbiib
\end{eqnarray}
However, unlike in the minimal model case, some of these structure constants are purely imaginary.

\section{Numerical methods}\label{app:Numerics}
In this appendix, we would like to describe several numerical algorithms which we use to compute the OSFT action and Ellwood invariants. We will focus on the \SUk WZW model sector, because most calculations can be factorized and algorithms in the ghost sector and in the universal part of the matter sector are the same as in \cite{KudrnaThesis}.

The general framework of our algorithms (representation of the string field, matrix representations of various objects, conservation laws, etc.) is described in \cite{KudrnaThesis}, but we have to adapt the algorithms because the \SU state space is generated by three non-commuting currents. The situation is further complicated by the fact that we divide the state space according to $J_0^3$ eigenvalues and we make restriction to states that belong to $\hh^{(0)}$ and $\hh^{(\pm 1)}$. This leads to algorithms which have a structure somewhat similar to algorithms for ghost theory in the $bc$ basis, where the $J_0^3$ eigenvalue plays a role analogous to the ghost number.

Our calculations are executed in Mathematica (symbolic manipulations) and C++ (time-consuming numerical calculations).

\subsection{Gram matrix}\label{app:Numerics:Gram matrix}
First, we will focus on the Gram matrix, which is the key ingredient of the quadratic term in the OSFT action. In the \SU sector, we need to compute the matrix of BPZ products $G_{ij}=\la i^{(0)}|j^{(0)}\ra$, where the states $|i^{(0)}\ra$ form a basis of the space $\hh^{(0)}$. As described in \cite{KudrnaThesis}, we compute the Gram matrix using a recursive algorithm which expresses a given element as a sum of elements at lower levels. In this way, we reduce level of elements until we get to a BPZ product of two primary operators, which we know explicitly.

In order to compute an element $G_{ij}$, we first separate the first current operator from the state $|i\ra$ as $|i\ra=J_{-n}^a|\hat i\ra$ and move this operator to the other side of the BPZ product:
\begin{equation}
G_{ij}=\la J_{-n}^a \hat i |j\ra=(-1)^{n+1}\la \hat i |J_{n}^a j\ra.
\end{equation}
The next steps depends on the value of $a$. If $a=3$, there are no problems and we can compute $G_{ij}$ as
\begin{equation}\label{Gram matrix 0}
G_{ij}=(-1)^{n+1} \sum_k\mm(J_n^3)_{jk}^0 G_{\hat i k}.
\end{equation}
The object $\mm(J_n^3)_{jk}^0$ is a matrix representation of the current operator $J_n^3$ acting on the state space, see \cite{KudrnaThesis} for more details. The upper index $0$ denotes that the operator $J_n^3$ acts on states from $\hh^{(0)}$.
If $a=\pm$, the algorithm becomes more complicated because the states $|\hat i\ra$ and $J_n^a|j\ra$ belong to $\hh^{(\pm 1)}$. Therefore we have to define an auxiliary Gram matrix $G^{aux}_{ij}=\la i^{(-1)}|j^{(+1)}\ra$. Then we get either
\begin{equation}
G_{ij}=(-1)^{n+1}\sum_k \mm(J_n^+)_{jk}^0 G_{\hat i k}^{aux}
\end{equation}
or
\begin{equation}
G_{ij}=(-1)^{n+1}\sum_k \mm(J_n^-)_{jk}^0 G_{k \hat i}^{aux}
\end{equation}
depending on whether the first operator in $|i\ra$ is $J^+_{-n}$ or $J^-_{-n}$.

The auxiliary matrix $G^{aux}$ can be computed analogously, but there are few complications. Once again, we would like to separate one operator from the state $\la i|$ and act with it on $|j\ra$. The case $a=3$ is simple and the recursive step is very similar to (\ref{Gram matrix 0}):
\begin{equation}
G_{ij}^{aux}=(-1)^{n+1} \sum_k\mm(J_n^3)_{jk}^{1} G_{\hat i k}^{aux}.
\end{equation}
The case $a=-1$ also proceeds without any issues, but there is a problem if $a=+1$. We cannot remove a $J^+$ operator because that would lead to a state from $\hh^{(-2)}$, which we want to avoid. States with $J^+$ at the first position typically have $J^-$ at some later position\footnote{We canonically order the current operators in states so that $J^+$ are first, $J^-$ are second and finally $J^3$ are at last positions. For this particular step, it would be more convenient to change the order, but similar problems would appear elsewhere.}, that is $|i\ra=J_{-n_1}^+\dots J_{-n_2}^- \dots|j,m\ra$. In such cases, we can commute the first $J_{-n}^-$ operator to the first position through operators in front of it, but we have to take into account that the commutators produce new states. It can also happen that there are no explicit $J^-$ operators present, for example in states of the form $J_{-n}^+|j,-2\ra$. In such cases, we have to remove one $J_0^-$ from the primary state and commute it forward. Therefore we can write in general (including the trivial case when $J_n^-$ is at the first position)
\begin{equation}
|i^{(-1)}\ra=C_i J_{-n}^-|\hat i^{(0)}\ra+\sum_k C_{i,k}|\hat i_k^{(-1)}\ra,
\end{equation}
where $C_i$ equals 1 if $n>0$ and $1/\alpha_{j,m+1}$ if $n=0$. The second part includes constants $C_{i,k}$ and states $|\hat i_k^{(-1)}\ra$ which come from commuting $J_{-n}^-$ through the operators in front of it. Therefore the general formula for the recursive step for $J_n^-$ reads
\begin{equation}
G_{ij}^{aux}=(-1)^{n+1}C_i \sum_k\mm(J_n^-)_{jk}^1 G_{\hat i k}+\sum_k  C_{i,k} G_{\hat i_k j}^{aux}.
\end{equation}
There is no risk of infinite loops (which we will have to deal with later) because commutators with $J_{-n}^-$ tend to reduce the number of operators and change $J^+$ to $J^3$, so we eventually always get to states which have $J^3_{-n}$ or $J^-_{-n}$ with $n>0$ at the first position.

Another issue that plays a role in this theory is the existence of null states. Especially for low levels $k$, there is a large number of null states, so removing them is essential. To do so, we use the approach described in \cite{KudrnaThesis}. We find null states by analyzing the Gram matrix, then we choose a basis that represents the irreducible part of the Hilbert space and remove representatives of null states from the string field. This must be accompanied by modification of matrix representations. Null states from the Hilbert space $\hh^{(0)}$ are encoded in the basic Gram matrix $G$ and null states from the auxiliary spaces $\hh^{(\pm 1)}$ in the matrix $G^{aux}$. We first compute the Gram matrices in Mathematica to get exact expressions for null states and later in C++ we compute the Gram matrices only for the irreducible part of the state space.

\subsection{Cubic vertex}\label{app:Numerics:vertices}
When it comes to the cubic vertex, the situation gets even more complicated because there are three states involved. Therefore we will describe our recursive algorithm only schematically, precise formulas involving matrix representations can be obtained by making similar modifications as in the previous subsection to the formulas from \cite{KudrnaThesis}.
The main problem we face during these calculations is a risk of the recursive algorithm going into infinite loops when using certain conservation laws.

For reference, let us define a matrix of 'physical' vertices $V_{ijk}=\la V_3| i^{(0)}\ra|j^{(0)}\ra|k^{(0)}\ra$ and a matrix of auxiliary vertices $V_{ijk}^{aux}=\la V_3| i^{(-1)}\ra|j^{(0)}\ra|k^{(1)}\ra$.
The general idea of computing the cubic vertex is the same as for the Gram matrix, we construct a recursive algorithm that expresses a given element as a sum of elements at lower levels. In this way, we eventually get to the basic vertex for three primary operators.
There are no issues when computing elements of the physical part of the vertex. We can choose an arbitrary operator from any of the three entries and use the corresponding conservation law. If we pick $J^3$, we stay within the physical part of the vertex, but if we pick $J^\pm$, the recursive formula also involves the auxiliary part of the vertex.

The real problem lies in the auxiliary vertex. We want to avoid states from $\hh^{(\pm 2)}$, so we have to select such conservation laws that we stay in the desired part of the Hilbert space. Therefore the only options are to pick either $J_{-n}^3$ operator from an arbitrary entry of the vertex, $J_{-n}^-$ from the first entry or $J_{-n}^+$ from the third entry. If $n>0$, we are always safe because the corresponding conservation law reduces level of the elements. However, we are sometimes forced to use $J_0$ conservation laws which do not decrease level, so, if done incorrectly, their application can lead to infinite loops.

As a simple example, consider the element $\la V_3|2,-1\ra |2,0\ra |k^{(1)}\ra$, where $ |k^{(1)}\ra$ is an arbitrary state. If we use $J_0^-$ conservation law on the first entry, the first state changes to $|2,0\ra$ and the second state to $|2,-1\ra$ (plus there are other terms that come from $J_0^-$ acting on the last state, but these are not important now). Once we exchange the order of the two states to match the definition of $V^{aux}$, we notice that the recursive algorithm points to the same element again and therefore it enters an infinite loop. There are also two-step infinite loops. For example, consider the element $\la V_3|J_{-m}^+|2,-2\ra J_{-n}^+|2,-1\ra |k^{(1)}\ra$. In the first step, we apply the $J_0^-$ conservation law on the first state and the result includes $\la V_3|J_{-n}^+|2,-2\ra J_{-m}^+|2,-1\ra |k^{(1)}\ra$, in the second step, we use $J_0^-$ conservation law again we get back to $\la V_3|J_{-m}^+|2,-2\ra J_{-n}^+|2,-1\ra |k^{(1)}\ra$.

Such infinite loops can be avoided by choosing which conservation laws we use following the steps below. There, $m_j$ denotes spin projection of the primary in the state $|j^{(0)}\ra$ and $N_i$ and $N_k$ are number operator eigenvalues of the states $|i^{(-1)}\ra$ and $|k^{(1)}\ra$ respectively.

\begin{enumerate}
  \item If there are any operators $J^3_{-n}$ or $J^-_{-n}$ with $n>0$ in the state $|i^{(-1)}\ra$, we remove them using the corresponding conservation laws.
  \item Similarly, if there are any operators $J^3_{-n}$ or $J^+_{-n}$ with $n>0$ in the state $|k^{(1)}\ra$, we remove them.
  \item Next, we check $m_j$. If $m_j>0$, we use $J^-_0$ conservation law on $|i^{(-1)}\ra$.
  \item If $m_j<0$, we use $J^+_0$ conservation law on $|k^{(1)}\ra$.
  \item Finally, if $m_j=0$, we check $N_i$ and $N_k$. If $N_i>0$ and $N_k=0$, we use $J^+_0$ conservation law on the state $|k^{(1)}\ra$.
  \item If $N_k>0$ and $N_i=0$,  we use $J^-_0$ conservation law on the state $|i^{(-1)}\ra$.
  \item In other cases, it is save to use both of these options.
\end{enumerate}

We have explicitly checked for all types of problematic vertices that an algorithm following these rules does not lead to infinite loops.

Next, let us make few comments regarding the time requirements of this algorithm. In \cite{KudrnaThesis}, we stated that computing solutions using Newton's method takes more time than evaluation of the cubic vertex. However, that is not true in this model and computing the cubic vertex is typically the most time consuming task. The asymptotic estimates from \cite{KudrnaThesis} are no longer valid for two reasons.

The first reason is that evaluation of one element of the cubic vertex requires a lot of operations. Matrix representation of operators like $L_n$ or $\alpha_n$ are typically sparse, so their application requires only few operations. In our model, there are however many nontrivial elements because most commutators of current modes are nonzero and because we have to make replacements for null states. Therefore the scaling of time needed for the cubic vertex is no longer $O(N^3)$, but somewhere between $O(N^3)$ and $O(N^4)$.

Second, there is a huge asymmetry between the constituent models. In other OSFT settings, the constituent models typically have more or less the same number of states in the truncated Hilbert space. In \SUk WZW model, the Hilbert space is generated by three currents, which means that it is much larger than state spaces of the universal matter and ghost sectors. If we denote the number of states in a Virasoro Verma module up a given level as $N_L$, than we can roughly estimate the number of states in the \SU Hilbert space (including null states) as $N_L^{\sqrt 3}$ and number of states in the full state space as $N_L^{\sqrt 5}$. So the number of states in the \SU space is much closer to number of states in the full state space than in other models.

When combined, these two reasons mean that evaluation of the cubic vertex typically takes much more time than computing one solution using Newton's method. So, unless one wants to compute a large number of solutions, the cubic vertex is the main restriction that decides the available level and the results are worth saving for possible future calculations.

\subsection{Ellwood invariants}\label{app:Numerics:Ellwood}
Our strategy to compute Ellwood invariants again follows the approach described in \cite{KudrnaThesis}, which involves a recursive algorithm based on conservation laws.

First, let us have a look at structure of Ellwood invariant conservation laws for SU(2) currents, which take schematic form:
\begin{equation}\label{Elw 1}
\la E[\vv]|(J^a_n+(-1)^n J^a_{-n})|\Psi\ra= \la E[ J^a (\vv)]|\Psi\ra.
\end{equation}
The left hand side includes the usual combination of modes, which allows us to trade a creation operator for an annihilation operator. Furthermore, this part does not lead to change of the $J_0^3$ eigenvalue of the string field. The right hand side, which we will describe in more detail later, is determined by the OPE between the current $J^a$ and the vertex operator $\vv$. If we choose $a=\pm$, this part changes $J_0^3$ eigenvalues and it leads to states from the auxiliary spaces $\hh^{(\pm 1)}$. Therefore we have to apply conservation laws also on the these auxiliary states, which involves the same complications as in section \ref{app:Numerics:Gram matrix} (restricted choice of conservation laws to avoid states from $\hh^{(\pm 2)}$, commutation of $J^-_{-n}$ through operators in front of it). We deal with these problems similarly as before.

When it comes to action of conservation laws on the Ellwood state, we decided to implement the approach described towards the end of section 2.5.1 in \cite{KudrnaThesis}. The reason is that bulk vertex operators that define Ellwood invariants can include both \SU primaries and current descendants. The invariants considered in this paper are quite simple, so we could deal with them using other methods, but we would like have the option of adding more complicated invariants in future works, so we decided for this approach. The full expression for conservation laws in this approach takes the form
\begin{equation}
\la E[\vv]|(J^a_n+(-1)^nJ^a_{-n})|\Psi\ra=\sum_m \ee_m^+(J_{-n}^a) \la E[J_m^a \vv]|\Psi\ra+\sum_m \ee_m^-(J_{-n}^a) \la E[\bar J_m^a \vv]|\Psi\ra,
\end{equation}
where $\ee_m^\pm(J_{-n}^a)$ are constants that follow from expansion of functions $v_n(z)$ (which generate conservation laws) around the points $\pm i$, see equations (2.5.237) and (2.5.244) in \cite{KudrnaThesis}. We represent bulk vertex operators as products of two chiral Hilbert spaces (with unrestricted $J_0^3$ eigenvalue), so the expressions $J_m^a \vv$ and $\bar J_m^a \vv$ represent action of current modes on the left and right part of the bulk state, which can be written in terms of matrix representation.

To write an explicit recursive algorithm, we choose a basis $\vv_{ij}$ of bulk vertex operators and we define $E_{ijk}=\la E[\vv_{ij}]|k^{(0)}\ra$ (we need also auxiliary objects $E_{ijk}^\pm=\la E[\vv_{ij}]|k^{(\pm1)}\ra$). The recursive algorithm then takes the schematic form
\begin{equation}
E_{ijk}=(-1)^n\left(-\sum_l \mm(J^a_n)_{\hat k l} E_{ijl}+\sum_{m,l}  \ee_m^+(J_{-n}^a) \mm(J^a_m)_{i l} E_{lj\hat k}
+\sum_{m,l} \ee_m^-(J_{-n}^a) \mm(J^a_m)_{j l} E_{il\hat k}\right).
\end{equation}
To complete it, we need to make modifications for the individual currents so that it correctly switches between $E_{ijk}$ and $E_{ijk}^\pm$ and sometimes for commuting a required operator $J^a_{-n}$ to the first position. These modifications once again follow section \ref{app:Numerics:Gram matrix}, so we will not go into details.

Conservation laws allow us to remove all currents from the string field. Then we are left we expressions of the form
\begin{equation}
\la E[\vv]|j,m\ra=2^{h_j}\la \vv(i,-i) \phi_{j,m}(0)\ra.
\end{equation}
These are ordinary bulk-boundary correlators, which can be computed using standard CFT methods.

\FloatBarrier
\section{More numerical data}\label{app:Data}
%\subsection{$k=5$ solutions}\label{app:Data:k=5}
In this appendix, we provide several examples of solutions describing two 0-branes. We choose $k=5$ and $J=1$ as the background. The first solution represents an \SU Cardy boundary state, the other three describe \SL boundary states, which illustrate the three types of two 0-brane solutions mentioned in section \ref{sec:SL:other}. Since this background includes a large number of Ellwood invariants, we provide only their infinite level extrapolations to keep the amount of data reasonable. All extrapolations are based on level 10 results.

The data given in table \ref{tab:sol 5 1 SU extrapolation} describe a real solution corresponding to an \SU boundary state. We can easily identify it as two 0-branes with $\theta_1=-\theta_2=\frac{2\pi}{k}$. Apart from the fact that it describes two branes, it has very similar properties to the examples of solutions in section \ref{sec:regular}. In particular, we observe that it has the same symmetries and the extrapolations have similar precision.

\begin{table}[!t]
\begin{tabular}{|c|llllll|}\cline{1-4}
          & $\ps $Energy      & $\ps E_{0,0}     $ & $\ps \Delta_S    $ & \mc{3}{|c}{}                                                 \\\cline{1-4}
$\inf$    & $\ps 0.9646     $ & $\ps 0.951       $ & $   -0.0013      $ & \mc{3}{|c}{}                                                 \\
$\sigma$  & $\ps 0.0004     $ & $\ps 0.003       $ & $\ps 0.0003      $ & \mc{3}{|c}{}                                                 \\
Exp.      & $\ps 0.963163   $ & $\ps 0.963163    $ & $\ps 0           $ & \mc{3}{|c}{}                                                 \\\cline{1-4}
          & $\ps J_{+-}     $ & $\ps J_{-+}      $ & $\ps J_{33}      $ & \mc{3}{|c}{}                                                 \\\cline{1-4}
$\inf$    & $   -0.84       $ & $   -0.84        $ & $\ps 0.951       $ & \mc{3}{|c}{}                                                 \\
$\sigma$  & $\ps 0.05       $ & $\ps 0.05        $ & $\ps 0.003       $ & \mc{3}{|c}{}                                                 \\
Exp.      & $   -0.779215   $ & $   -0.779215    $ & $\ps 0.963163    $ & \mc{3}{|c}{}                                                 \\\cline{1-4}
          & $\ps E_{1/2,1/2}$ & $\ps E_{1/2,-1/2}$ & \mc{4}{|c}{}                                                                      \\\cline{1-3}
$\inf$    & $\ps 0.397      $ & $   -0.397       $ & \mc{4}{|c}{}                                                                      \\
$\sigma$  & $\ps 0.001      $ & $\ps 0.001       $ & \mc{4}{|c}{}                                                                      \\
Exp.      & $\ps 0.399532   $ & $   -0.399532    $ & \mc{4}{|c}{}                                                                      \\\cline{1-4}
          & $\ps E_{1,1}    $ & $\ps E_{1,0}     $ & $\ps E_{1,-1}    $ & \mc{3}{|c}{}                                                 \\\cline{1-4}
$\inf$    & $   -1.162      $ & $   -1.46        $ & $   -1.162       $ & \mc{3}{|c}{}                                                 \\
$\sigma$  & $\ps 0.004      $ & $\ps 0.02        $ & $\ps 0.004       $ & \mc{3}{|c}{}                                                 \\
Exp.      & $   -1.16804    $ & $   -1.44377     $ & $   -1.16804     $ & \mc{3}{|c}{}                                                 \\\cline{1-5}
          & $\ps E_{3/2,3/2}$ & $\ps E_{3/2,1/2} $ & $\ps E_{3/2,-1/2}$ & $\ps E_{3/2,-3/2}$ & \mc{2}{|c}{}                            \\\cline{1-5}
$\inf$    & $   -1.161      $ & $   -0.47        $ & $\ps 0.47        $ & $\ps 1.161       $ & \mc{2}{|c}{}                            \\
$\sigma$  & $\ps 0.004      $ & $\ps 0.03        $ & $\ps 0.03        $ & $\ps 0.004       $ & \mc{2}{|c}{}                            \\
Exp.      & $   -1.16804    $ & $   -0.446151    $ & $\ps 0.446151    $ & $\ps 1.16804     $ & \mc{2}{|c}{}                            \\\cline{1-6}
          & $\ps E_{2,2}    $ & $\ps E_{2,1}     $ & $\ps E_{2,0}     $ & $\ps E_{2,-1}    $ & $\ps E_{2,-2}    $ & \mc{1}{|c}{}       \\\cline{1-6}
$\inf$    & $\ps 0.397      $ & $\ps 1.07        $ & $\ps 1.22        $ & $\ps 1.07        $ & $\ps 0.397       $ & \mc{1}{|c}{}       \\
$\sigma$  & $\ps 0.001      $ & $\ps 0.03        $ & $\ps 0.28        $ & $\ps 0.03        $ & $\ps 0.001       $ & \mc{1}{|c}{}       \\
Exp.      & $\ps 0.399532   $ & $\ps 1.04599     $ & $\ps 1.29291     $ & $\ps 1.04599     $ & $\ps 0.399532    $ & \mc{1}{|c}{}       \\\cline{1-7}
          & $\ps E_{5/2,5/2}$ & $\ps E_{5/2,3/2} $ & $\ps E_{5/2,1/2} $ & $\ps E_{5/2,-1/2}$ & $\ps E_{5/2,-3/2}$ & $\ps E_{5/2,-5/2}$ \\\cline{1-7}
$\inf$    & $\ps 0.951      $ & $\ps 0.84        $ & $\ps 0.1         $ & $   -0.1         $ & $   -0.84        $ & $   -0.951       $ \\
$\sigma$  & $\ps 0.003      $ & $\ps 0.05        $ & $\ps 0.8         $ & $\ps 0.8         $ & $\ps 0.05        $ & $\ps 0.003       $ \\
Exp.      & $\ps 0.963163   $ & $\ps 0.779215    $ & $\ps 0.297634    $ & $   -0.297634    $ & $   -0.779215    $ & $   -0.963163    $ \\\cline{1-7} \end{tabular}
\caption{Extrapolations of observables of a solution describing two \SU Cardy 0-branes with $\theta_1=-\theta_2=\frac{2\pi}{k}$ in the $k=5$ model with $J=1$ boundary conditions. }
\label{tab:sol 5 1 SU extrapolation}
\end{table}

The next solution given in table \ref{tab:sol 5 1 SL1 extrapolation} describes an \SL boundary state. We notice that all of its invariants are real numbers, although the solution itself is not because it violates the condition (\ref{reality}). This property of invariants suggests a special relation between parameters of the two branes. By the $R^2$ minimization procedure, we found that $\theta_1=-\theta_2=\frac{2\pi}{k}$, which is typical for two 0-brane solutions, and that the two $\rho$ parameters are equal, $\rho_1=\rho_2\approx0.50$. Their value is quite close to $\dt1$, but that seems to be just a coincidence. The parameters $\rho$ are determined mainly from $E_{j,m}$ invariants with negative $m$. Invariants with positive $m$ have very small expected values and most of them have relative errors over 100\%, which means that these invariants do not give us any meaningful information. If we make a comparison of invariants which do not depend on $\rho$ with the \SU solution above, we find that the \SL solution is surprisingly more precise, although it has larger error estimates.

\begin{table}[!t]
\begin{tabular}{|c|llllll|}\cline{1-4}
          & $\ps $Energy      & $\ps E_{0,0}     $ & $\ps \Delta_S    $ & \mc{3}{|c}{}                                                 \\\cline{1-4}
$\inf$    & $\ps 0.9636     $ & $\ps 0.963       $ & $   -0.0003      $ & \mc{3}{|c}{}                                                 \\
$\sigma$  & $\ps 0.0005     $ & $\ps 0.003       $ & $\ps 0.0009      $ & \mc{3}{|c}{}                                                 \\
Exp.      & $\ps 0.963163   $ & $\ps 0.963163    $ & $\ps 0           $ & \mc{3}{|c}{}                                                 \\\cline{1-4}
          & $\ps J_{+-}     $ & $\ps J_{-+}      $ & $\ps J_{33}      $ & \mc{3}{|c}{}                                                 \\\cline{1-4}
$\inf$    & $   -0.21       $ & $   -3.11        $ & $\ps 0.96        $ & \mc{3}{|c}{}                                                 \\
$\sigma$  & $\ps 0.65       $ & $\ps 1.15        $ & $\ps 1.25        $ & \mc{3}{|c}{}                                                 \\
Exp.      & $   -0.19       $ & $   -3.12        $ & $\ps 0.963163    $ & \mc{3}{|c}{}                                                 \\\cline{1-4}
          & $\ps E_{1/2,1/2}$ & $\ps E_{1/2,-1/2}$ & \mc{4}{|c}{}                                                                      \\\cline{1-3}
$\inf$    & $\ps 0.204      $ & $   -0.800       $ & \mc{4}{|c}{}                                                                      \\
$\sigma$  & $\ps 0.010      $ & $\ps 0.016       $ & \mc{4}{|c}{}                                                                      \\
Exp.      & $\ps 0.200      $ & $   -0.799       $ & \mc{4}{|c}{}                                                                      \\\cline{1-4}
          & $\ps E_{1,1}    $ & $\ps E_{1,0}     $ & $\ps E_{1,-1}    $ & \mc{3}{|c}{}                                                 \\\cline{1-4}
$\inf$    & $   -0.23       $ & $   -1.444       $ & $   -4.67        $ & \mc{3}{|c}{}                                                 \\
$\sigma$  & $\ps 0.11       $ & $\ps 0.060       $ & $\ps 0.26        $ & \mc{3}{|c}{}                                                 \\
Exp.      & $   -0.29       $ & $   -1.44377     $ & $   -4.67        $ & \mc{3}{|c}{}                                                 \\\cline{1-5}
          & $\ps E_{3/2,3/2}$ & $\ps E_{3/2,1/2} $ & $\ps E_{3/2,-1/2}$ & $\ps E_{3/2,-3/2}$ & \mc{2}{|c}{}                            \\\cline{1-5}
$\inf$    & $\ps 0.04       $ & $   -0.28        $ & $\ps 0.90        $ & $\ps 9.42        $ & \mc{2}{|c}{}                            \\
$\sigma$  & $\ps 0.39       $ & $\ps 0.17        $ & $\ps 0.16        $ & $\ps 0.40        $ & \mc{2}{|c}{}                            \\
Exp.      & $   -0.15       $ & $   -0.22        $ & $\ps 0.89        $ & $\ps 9.34        $ & \mc{2}{|c}{}                            \\\cline{1-6}
          & $\ps E_{2,2}    $ & $\ps E_{2,1}     $ & $\ps E_{2,0}     $ & $\ps E_{2,-1}    $ & $\ps E_{2,-2}    $ & \mc{1}{|c}{}       \\\cline{1-6}
$\inf$    & $\ps 0.16       $ & $\ps 0.56        $ & $\ps 1.1         $ & $\ps 4.13        $ & $\ps 6.63        $ & \mc{1}{|c}{}       \\
$\sigma$  & $\ps 0.31       $ & $\ps 0.74        $ & $\ps 1.1         $ & $\ps 0.67        $ & $\ps 0.05        $ & \mc{1}{|c}{}       \\
Exp.      & $\ps 0.02       $ & $\ps 0.26        $ & $\ps 1.29291     $ & $\ps 4.18        $ & $\ps 6.39        $ & \mc{1}{|c}{}       \\\cline{1-7}
          & $\ps E_{5/2,5/2}$ & $\ps E_{5/2,3/2} $ & $\ps E_{5/2,1/2} $ & $\ps E_{5/2,-1/2}$ & $\ps E_{5/2,-3/2}$ & $\ps E_{5/2,-5/2}$ \\\cline{1-7}
$\inf$    & $\ps 0.76       $ & $\ps 0.58        $ & $\ps 1.01        $ & $\ps 0.42        $ & $   -6.23        $ & $   -30.97       $ \\
$\sigma$  & $\ps 1.54       $ & $\ps 1.57        $ & $\ps 2.84        $ & $\ps 2.59        $ & $\ps 1.44        $ & $\ps  0.68       $ \\
Exp.      & $\ps 0.03       $ & $\ps 0.10        $ & $\ps 0.15        $ & $   -0.60        $ & $   -6.23        $ & $   -30.82       $ \\\cline{1-7} \end{tabular}
\caption{Extrapolations of observables of a solution describing two \SL 0-branes with $\theta_1=\frac{2\pi}{k}$, $\theta_2=\frac{-2\pi}{k}$ and $\rho_1=\rho_2\approx0.50$ in the $k=5$ model with $J=1$ boundary conditions.}
\label{tab:sol 5 1 SL1 extrapolation}
\end{table}

The solution in table \ref{tab:sol 5 1 SL2 extrapolation} has somewhat different properties. Invariants of this solution are generic complex numbers, but we notice that they have a symmetry given by:
\begin{equation}
E_{j,m}=(-1)^{2j}E_{j,-m}.
\end{equation}
Although similar, this symmetry is not the same as the reality condition (\ref{reality}), so the solution is only pseudo-real. However, the symmetry implies a relation between parameters of the two 0-branes. The parameters $\theta$ take the usual values $\theta_1=-\theta_2=\frac{2\pi}{k}$ and $\rho$ parameters are inversely proportional: $\rho_1\approx0.49$ and $\rho_2=1/\rho_1\approx2.05$. The solution suffers from the odd level instability, therefore we have data from only 5 levels, which leads to lesser precision of extrapolations.

\begin{table}[!t]
\begin{tabular}{|c|llllll|}\cline{1-4}
          & $\ps $Energy       & $\ps E_{0,0}     $ & $\ps \Delta_S    $ & \mc{3}{|c}{}                                                \\\cline{1-4}
$\inf$    & $\ps 0.962       $ & $\ps 0.964       $ & $\ps 0.0002      $ & \mc{3}{|c}{}                                                \\
$\sigma$  & $\ps -           $ & $\ps 0.002       $ & $\ps -           $ & \mc{3}{|c}{}                                                \\
Exp.      & $\ps 0.963163    $ & $\ps 0.963163    $ & $\ps 0           $ & \mc{3}{|c}{}                                                \\\cline{1-4}
          & $\ps J_{+-}      $ & $\ps J_{-+}      $ & $\ps J_{33}      $ & \mc{3}{|c}{}                                                \\\cline{1-4}
$\inf$    & $   -1.68-0.91 i $ & $   -1.68-0.91  i$ & $   -0.5         $ & \mc{3}{|c}{}                                                \\
$\sigma$  & $\ps 0.46+0.44 i $ & $\ps 0.46+0.44  i$ & $\ps 0.5         $ & \mc{3}{|c}{}                                                \\
Exp.      & $   -1.73-1.12 i $ & $   -1.73-1.12  i$ & $\ps 0.963163    $ & \mc{3}{|c}{}                                                \\\cline{1-4}
          & $\ps E_{1/2,1/2} $ & $\ps E_{1/2,-1/2}$ & \mc{4}{|c}{}                                                                     \\\cline{1-3}
$\inf$    & $\ps 0.507-0.982i$ & $   -0.507+0.982i$ & \mc{4}{|c}{}                                                                     \\
$\sigma$  & $\ps 0.002+0.003i$ & $\ps 0.002+0.003i$ & \mc{4}{|c}{}                                                                     \\
Exp.      & $\ps 0.507-0.960i$ & $   -0.507+0.960i$ & \mc{4}{|c}{}                                                                     \\\cline{1-4}
          & $\ps E_{1,1}     $ & $\ps E_{1,0}     $ & $\ps E_{1,-1}    $ & \mc{3}{|c}{}                                                \\\cline{1-4}
$\inf$    & $   -2.51-1.67  i$ & $   -1.6         $ & $   -2.51-1.67  i$ & \mc{3}{|c}{}                                                \\
$\sigma$  & $\ps 0.09+0.06  i$ & $\ps 0.1         $ & $\ps 0.09+0.06  i$ & \mc{3}{|c}{}                                                \\
Exp.      & $   -2.59-1.68  i$ & $   -1.44377     $ & $   -2.59-1.68  i$ & \mc{3}{|c}{}                                                \\\cline{1-5}
          & $\ps E_{3/2,3/2} $ & $\ps E_{3/2,1/2} $ & $\ps E_{3/2,-1/2}$ & $\ps E_{3/2,-3/2}$ & \mc{2}{|c}{}                           \\\cline{1-5}
$\inf$    & $   -5.0+3.6    i$ & $   -0.64+0.9   i$ & $\ps 0.64-0.9   i$ & $\ps 5.0-3.6    i$ & \mc{2}{|c}{}                           \\
$\sigma$  & $\ps 0.8+0.4    i$ & $\ps 0.04+0.7   i$ & $\ps 0.04+0.7   i$ & $\ps 0.8+0.4    i$ & \mc{2}{|c}{}                           \\
Exp.      & $   -5.1+3.6    i$ & $   -0.57+1.1   i$ & $\ps 0.57-1.1   i$ & $\ps 5.1-3.6    i$ & \mc{2}{|c}{}                           \\\cline{1-6}
          & $\ps E_{2,2}     $ & $\ps E_{2,1}     $ & $\ps E_{2,0}     $ & $\ps E_{2,-1}    $ & $\ps E_{2,-2}    $ & \mc{1}{|c}{}      \\\cline{1-6}
$\inf$    & $\ps 3.6+10.9   i$ & $\ps 1.8+1.0    i$ & $\ps 0.9         $ & $\ps 1.8+1.0    i$ & $\ps 3.6+10.9   i$ & \mc{1}{|c}{}      \\
$\sigma$  & $\ps 0.5+ 2.2   i$ & $\ps 0.1+1.1    i$ & $\ps 2.1         $ & $\ps 0.1+1.1    i$ & $\ps 0.5+ 2.2   i$ & \mc{1}{|c}{}      \\
Exp.      & $\ps 3.5+10.8   i$ & $\ps 2.3+1.5    i$ & $\ps 1.29291     $ & $\ps 2.3+1.5    i$ & $\ps 3.5+10.8   i$ & \mc{1}{|c}{}      \\\cline{1-7}
          & $\ps E_{5/2,5/2} $ & $\ps E_{5/2,3/2} $ & $\ps E_{5/2,1/2} $ & $\ps E_{5/2,-1/2}$ & $\ps E_{5/2,-3/2}$ & $\ps E_{5/2,-5/2}$\\\cline{1-7}
$\inf$    & $\ps 18.1        $ & $\ps 2.7-1.74   i$ & $   -0.7+0.6    i$ & $\ps 0.7-0.6    i$ & $   -2.7+1.74   i$ & $   -18.1       $ \\
$\sigma$  & $\ps 3.1         $ & $\ps 1.1+0.04   i$ & $\ps 4.6+0.7    i$ & $\ps 4.6+0.7    i$ & $\ps 1.1+0.04   i$ & $\ps 3.1        $ \\
Exp.      & $\ps 17.4        $ & $\ps 3.4-2.41   i$ & $\ps 0.4-0.7    i$ & $   -0.4+0.7    i$ & $   -3.4+2.41   i$ & $   -17.4       $ \\\cline{1-7} \end{tabular}
\caption{Extrapolations of observables of a solution describing two \SL 0-branes with $\theta_1=\frac{2\pi}{k}$, $\theta_2=\frac{-2\pi}{k}$, $\rho_1\approx0.49$ and $\rho_2=1/\rho_1\approx2.05$ in the $k=5$ model with $J=1$ boundary conditions. Extrapolations are done only using even level data due to odd level instabilities.}
\label{tab:sol 5 1 SL2 extrapolation}
\end{table}

Extrapolations of invariants of the final two 0-brane solution are given in table \ref{tab:sol 5 1 SL3 extrapolation}. Most of the invariants are generic complex numbers and they do not have any symmetries or special properties. The solution has odd level instabilities, which leads to lesser precision of extrapolations. We determined parameters of the two 0-branes it describes to be $\theta_1=-\theta_2=\frac{2\pi}{k}$, $\rho_1\approx0.99$ and $\rho_2\approx1.82$. One of the $\rho$ parameters is quite generic, but the other is close to 1, which means that this solution most likely describes a combination of an \SU brane and an \SL brane.

\begin{table}[!t]
\footnotesize
\begin{tabular}{|c|llllll|}\cline{1-4}
          & $\ps $Energy       & $\ps E_{0,0}     $ & $\ps \Delta_S    $ & \mc{3}{|c}{}                                                \\\cline{1-4}
$\inf$    & $\ps 0.9641      $ & $\ps 0.957       $ & $   -0.0007      $ & \mc{3}{|c}{}                                                \\
$\sigma$  & $\ps -           $ & $\ps 0.001       $ & $\ps -           $ & \mc{3}{|c}{}                                                \\
Exp.      & $\ps 0.963163    $ & $\ps 0.963163    $ & $\ps 0           $ & \mc{3}{|c}{}                                                \\\cline{1-4}
          & $\ps J_{+-}      $ & $\ps J_{-+}      $ & $\ps J_{33}      $ & \mc{3}{|c}{}                                                \\\cline{1-4}
$\inf$    & $   -1.85-0.66  i$ & $   -0.54-0.27  i$ & $\ps 0.77        $ & \mc{3}{|c}{}                                                \\
$\sigma$  & $\ps 0.08+0.15  i$ & $\ps 0.19+0.03  i$ & $\ps 0.09        $ & \mc{3}{|c}{}                                                \\
Exp.      & $   -1.68-0.65  i$ & $   -0.51-0.20  i$ & $\ps 0.963163    $ & \mc{3}{|c}{}                                                \\\cline{1-4}
          & $\ps E_{1/2,1/2} $ & $\ps E_{1/2,-1/2}$ & \mc{4}{|c}{}                                                                     \\\cline{1-3}
$\inf$    & $\ps 0.5638-0.542i$ & $   -0.310+0.289i$ & \mc{4}{|c}{}                                                                    \\
$\sigma$  & $\ps 0.0004+0.009i$ & $\ps 0.001+0.003i$ & \mc{4}{|c}{}                                                                    \\
Exp.      & $\ps 0.5633-0.504i$ & $   -0.310+0.277i$ & \mc{4}{|c}{}                                                                    \\\cline{1-4}
          & $\ps E_{1,1}     $ & $\ps E_{1,0}     $ & $\ps E_{1,-1}    $ & \mc{3}{|c}{}                                                \\\cline{1-4}
$\inf$    & $   -2.55-1.03  i$ & $   -1.46        $ & $   -0.759-0.316i$ & \mc{3}{|c}{}                                                \\
$\sigma$  & $\ps 0.03+0.04  i$ & $\ps 0.04        $ & $\ps 0.017+0.009i$ & \mc{3}{|c}{}                                                \\
Exp.      & $   -2.52-0.98  i$ & $   -1.44377     $ & $   -0.760-0.296i$ & \mc{3}{|c}{}                                                \\\cline{1-5}
          & $\ps E_{3/2,3/2} $ & $\ps E_{3/2,1/2} $ & $\ps E_{3/2,-1/2}$ & $\ps E_{3/2,-3/2}$ & \mc{2}{|c}{}                           \\\cline{1-5}
$\inf$    & $   -4.14+2.2   i$ & $   -0.66+0.64  i$ & $\ps 0.36-0.32  i$ & $\ps 0.678-0.40 i$ & \mc{2}{|c}{}                           \\
$\sigma$  & $\ps 0.14+0.1   i$ & $\ps 0.06+0.07  i$ & $\ps 0.04+0.10  i$ & $\ps 0.004+0.02 i$ & \mc{2}{|c}{}                           \\
Exp.      & $   -4.10+2.1   i$ & $   -0.63+0.56  i$ & $\ps 0.35-0.31  i$ & $\ps 0.681-0.35 i$ & \mc{2}{|c}{}                           \\\cline{1-6}
          & $\ps E_{2,2}     $ & $\ps E_{2,1}     $ & $\ps E_{2,0}     $ & $\ps E_{2,-1}    $ & $\ps E_{2,-2}    $ & \mc{1}{|c}{}      \\\cline{1-6}
$\inf$    & $\ps 2.38+6.2   i$ & $\ps 2.21+0.86  i$ & $\ps 1.39        $ & $\ps 0.76+0.41  i$ & $\ps 0.23+0.60  i$ & \mc{1}{|c}{}      \\
$\sigma$  & $\ps 0.11+0.4   i$ & $\ps 0.06+0.28  i$ & $\ps 0.57        $ & $\ps 0.26+0.16  i$ & $\ps 0.02+0.14  i$ & \mc{1}{|c}{}      \\
Exp.      & $\ps 2.39+6.1   i$ & $\ps 2.26+0.88  i$ & $\ps 1.29291     $ & $\ps 0.68+0.27  i$ & $\ps 0.22+0.56  i$ & \mc{1}{|c}{}      \\\cline{1-7}
          & $\ps E_{5/2,5/2} $ & $\ps E_{5/2,3/2} $ & $\ps E_{5/2,1/2} $ & $\ps E_{5/2,-1/2}$ & $\ps E_{5/2,-3/2}$ & $\ps E_{5/2,-5/2}$\\\cline{1-7}
$\inf$    & $\ps 9.7         $ & $\ps 2.9-1.3    i$ & $\ps 0.1+0.01   i$ & $   -0.8+0.3    i$ & $   -0.59+0.35  i$ & $   -0.56       $ \\
$\sigma$  & $\ps 0.8         $ & $\ps 0.4+0.2    i$ & $\ps 1.1+0.09   i$ & $\ps 0.2+1.7    i$ & $\ps 0.20+0.23  i$ & $\ps 0.04       $ \\
Exp.      & $\ps 10.1        $ & $\ps 2.7-1.4    i$ & $\ps 0.4-0.38   i$ & $   -0.2+0.2    i$ & $   -0.45+0.24  i$ & $   -0.51       $ \\\cline{1-7} \end{tabular}
\caption{Extrapolations of observables of a solution describing probably an \SU 0-brane with $\theta_1=\frac{2\pi}{k}$ ($\rho_1\approx0.99$) and \SL 0-brane with $\theta_2=\frac{-2\pi}{k}$ and $\rho_2\approx1.82$ in the $k=5$ model with $J=1$ boundary conditions. Extrapolations are done only using even level data due to odd level instabilities.}
\label{tab:sol 5 1 SL3 extrapolation}
\end{table}

\FloatBarrier
\end{appendix}

\end{document}